\documentclass[preprint,aps,pre,amsmath,amssymb,longbibliography]{revtex4-2}
\usepackage{graphicx}
\usepackage{amsmath,amssymb,bm,amsthm}
\usepackage{rotating}
\def\beq{\begin{equation}}
\def\eeq{\end{equation}}

\def\<{\langle}
\def\>{\rangle}

\def\Tr{{\rm Tr}}

\begin{document}

\title{Effective length scales, dispersion relations, and discrete densities of states 
for Laplacian eigenvectors on complex  networks}

\author{Per Arne Rikvold}
\email{p.a.rikvold@fys.uio.no}
\email{prikvold@fsu.edu}
\affiliation{PoreLab, NJORD Centre, Department of Physics, University of Oslo, Oslo, Norway.\\
Department of Physics, Florida State University, Tallahassee, FL 32306-4350, USA\\ 
}

\begin{abstract}
To construct dispersion relations for diffusion or oscillation processes on random 
networks, it is necessary to obtain effective length scales for the eigenvectors of 
a graph Laplacian matrix, whose eigenvalues represent inverse time scales. 
For this purpose, we adapt a method originally introduced in 
condensed-matter physics to estimate correlation lengths for disordered  
materials as the ratio of volume to interface area [P.\ Debye,  H.~R.\ Anderson 
and H. Brumberger, J.\ Appl.\ Phys.\ {\bf 28}, 679 (1957)]. 
In a graph setting of vertices connected by edges, 
we interpret this as the ratio of twice the total number of edges 
to the number of edges connecting vertices bearing values of different sign 
on the particular eigenvector. After describing the method and the necessary concepts 
in pedagogical detail, we apply it to nine different graphs representing  
natural and artificial networks, including two tree graphs without and with random 
shortcuts, the nervous system of a roundworm, 
a food web, a social network of dolphins, 
an electrical power grid, and a model porous material. 
The results identify both distributed and localized eigenvectors. They 
are given in graphical format showing example eigenvectors, 
dispersion relations, and discrete densities of states, as well as tables summarizing 
the main numerical results.
\end{abstract}

\maketitle

\section{Introduction}
\label{sec:I}

Relationships between time scales of diffusion or oscillations and spatial length scales are 
generally known as {\it dispersion relations}. For systems with global symmetries, 
for instance 
under translation or rotation, such relations can be obtained by well-known mathematical 
techniques, such as Fourier transforms using complex exponentials as 
basis functions \cite{MONA99}.
For systems without such obvious, global symmetries, this task becomes much harder 
as both spatial dimension and distance become more ambiguous. 
Nevertheless, many such systems can be given a discrete description as {\it graphs}: 
collections of $N$ points, known as vertices or nodes, 
connected by $M$ line segments, known as edges or bonds. 
In these cases, significant progress 
has recently been made with techniques often called ``Fourier transforms 
on graphs'' \cite{SHUM13,SHUM16,RICA19,MACM22}, 
in which functions defined on the vertices are expanded in terms of the 
eigenvectors of a matrix known as the {\it graph Laplacian}, $\bf L$.
The corresponding eigenvalues form the {\it spectrum} of the Laplacian, which can be 
characterized by its {\it density of states} (DOS). 
In order to produce proper dispersion relations for the Laplacian eigenvectors, 
we must associate effective {\it length scales} with the eigenvalues. 
We will achieve this by adapting a method originally introduced in condensed-matter physics 
by Debye {\it et al.}~\cite{DEBY57}. In a graph setting,  
it amounts to counting the proportion of edges that
connect vertices bearing values of different sign in that particular eigenvector.
More detailed definitions of these concepts and methods 
are given in Sec.~\ref{sec:II}.

Graph theory was first introduced by Euler in 
his discussion of the famous problem of the Seven Bridges of K\"onigsberg \cite{EULER1736}.
Later applications of the abstraction of systems into vertices connected by edges 
to transportation systems such as 
electrical circuits \cite{KIRCH1845}, power grids \cite{KINN05} 
and other geographically constrained transport systems \cite{ZHU22,ALLH22} are numerous,  
as well as to a wider range of problems including flow in porous media \cite{REIS25}, 
fluid dynamics \cite{MACM22}, chemistry \cite{LEIT24}, neuronal networks \cite{ARNA18},
ecology \cite{ROSS13}, linguistics \cite{KUHL16}, computer science \cite{SHUM13}, 
and many more. 

In this paper we concentrate on the structures of Laplacian eigenvectors 
for graph representations of several different networks, both artificial and natural, and we 
use these structures to construct corresponding dispersion relations and DOS. 
The remainder of the paper is organized as follows. 
Section~\ref{sec:II} contains an introductory summary of the concepts 
and methods used. These are  
graphs and their matrix representations, including the graph Laplacian and 
some of its normalized forms  (\ref{sec:IIa}). Special emphasis is given to the 
walk-normalized form, which is particularly well suited 
for the study of diffusion and conduction processes  (\ref{sec:IIb}).
Our proposed method to estimate a length scale associated with a particular Laplacian 
eigenvector from the number of edges connecting vertices bearing values of different sign, 
$+$, $0$ or $-$,  
is described in detail in Sec.~\ref{sec:IId}.  
Some global measures that describe a graph as a whole, rather than its individual
eigenvectors, are discussed in  Sec.~\ref{sec:GGM}.
In Sec.~\ref{sec:III}, we present applications to nine different example graphs: 
a line graph (\ref{Sec:Lin64}) and a small-world graph produced by adding random shortcuts 
to the line graph (\ref{Sec:Lin64L}). These are followed by a Cayley tree without 
(\ref{Sec:Cayley40}) 
and with random shortcuts (\ref{Sec:Cayley40L}), 
the nervous system of a roundworm (\ref{Sec:CE277}), 
a food web (\ref{Sec:SM48}), a social group of dolphins (\ref{Sec:Dol62}),
an electric power grid (\ref{Sec:FLpow}), and a graph extracted from a random pack 
of glass beads (\ref{sec:IV}). 
A summary and conclusions are given in Sec.~\ref{sec:V}.
Additional mathematical details are given in the Appendix.

\section{Concepts and Methods}
\label{sec:II}

\subsection{Graphs and their matrix representations}
\label{sec:IIa}
As mentioned in Sec.~\ref{sec:I}, 
a graph is defined as a collection of $N$ vertices, connected by 
$M$ edges. For all the vertices to be connected by edges, 
$N-1 \le M \le N(N-1)/2$, where the lower limit characterizes {\it trees}, and the 
upper limit represents a fully connected graph, known as {\it the complete graph}. 
The simplest, but nevertheless quite useful, kind of graph is an 
undirected, unweighted graph. This is the only type of graph considered here.
It is defined by the symmetric, $N \times N$ 
{\it adjacency matrix} $\bf A$ with entries $a_{ij} = a_{ji} = 1$ if the distinct vertices, 
$v_i$ and $v_j$ with $i \neq j$, are connected by an edge, and $0$ otherwise. 

Closely related to $\bf A$ is the {\it graph Laplacian},
\beq 
{\bf L} = {\bf D} - {\bf A} ,
\eeq
 where {\bf D} is the diagonal matrix of {\it vertex degrees} with elements 
$d_i = \sum_j a_{ij}$ \cite{NEWM10}. ${\bf L}$ is the discrete analog of the negative of the 
diffusion operator in continuous, Euclidean space. 
Since $\bf L$ is symmetric and has vanishing row (and column) sums, it is singular and 
has a uniform eigenvector with eigenvalue zero (the Perron-Frobenius vector). 
Assuming the network is connected, all the other $N-1$ eigenvalues are positive. 

Two normalized versions of the graph Laplacian are commonly used. The first is the 
{\it symmetrically normalized} Laplacian, 
\beq
{\mathcal L} = {\bf D}^{-1/2} {\bf L}{\bf D}^{-1/2} \;.
\eeq 
The second is the {\it walk-normalized} Laplacian, 
\beq
{\bf W} = {\bf D}^{-1} {\bf L} \; . 
\label{eq:W}
\eeq
Acting to the right, it represents the transition rates for a random walk 
between the vertices, following the edges \cite{MASU17}. 
It is related to ${\mathcal L}$
by the similarity transformation, ${\mathcal L} = {\bf D}^{1/2} {\bf W}{\bf D}^{-1/2}$, 
and the two normalized forms therefore have the same set of $N$ eigenvalues, 
$\{ \lambda_n \}_{n=1}^N$, confined to the interval $ [0,2]$. 
(Note that the index $n$ refers to the eigenvectors, while the index $i$, used above 
in the definition of $\bf A$, refers 
to the vertices. Both $n$ and $i$ run between 1 and $N$.)
Further basic properties of the 
normalized spectrum for a connected graph with $N$ vertices are listed 
in Lemma~1.7 of Ref.~\cite{CHUN96}. We cite those here in the Appendix \ref{sec:APP}, 
with the only difference that 
the eigenvalues are listed in descending order, from $\lambda_1$ for the maximum to 
$\lambda_N = 0$ for the minimum. 
Since all the row sums of ${\bf W}$ (and also of ${\bf L}$) vanish, the right (column)
eigenvector $| 0 \rangle$, corresponding 
to $\lambda_N = 0$, is the Perron-Frobenius vector, which has all elements equal. 

\subsection{The walk-normalized graph Laplacian}
\label{sec:IIb}
In this paper, we concentrate on the walk-normalized Laplacian, 
${\bf W} = {\bf D}^{-1} {\bf L}$. 
Since $\bf W$ is not symmetric, its left (row) eigenvectors, $\langle \lambda_n |$, 
are different from 
the simple transpose of the corresponding right eigenvectors, $| \lambda_n \rangle$. 
Instead, the $i$th elements of the two vectors, $q_i(\lambda_n)$ and $w_i(\lambda_n)$, 
respectively, are related as  $q_i(\lambda_n) = d_i w_i(\lambda_n)$. 
This normalization of $\bf L$ is particularly useful in the study of conduction 
or diffusion on graphs, as the time evolution of the potential distribution over the vertices 
of an isolated circuit of Ohmic conductors is given by the matrix differential equation, 
\beq
\frac{{\rm d}}{{\rm d} t}  | w_i \rangle = - {\bf W} | w_i \rangle \;. 
\label{eq:DE0}
\eeq
Its solution is the approach of a nonuniform, initial potential distribution 
toward the stationary, uniform distribution proportional to the eigenfunction $| 0 \rangle$.  
The time evolution of the set of vertex values, $w_i(t)$, is given by the matrix exponential 
function $\bf exp$ \cite{MOLE03,ARNA20} as 
\beq
| w_i (t) \rangle  = {\bf exp}(- {\bf W} t) | w_i (0) \rangle \;.
\label{eq:Sol}
\eeq
This matrix function is defined by the same power series in the argument, as its scalar analog:
\beq
{\bf exp} ({\bf  M}) = {\bf 1} + \sum_{j=1}^\infty \frac{{\bf M}^j}{j!} .
\label{eq:mexp1}
\eeq
Alternatively, it can be expressed as an expansion in the eigenvectors and eigenvalues 
of $\bf M$,
\beq
{\bf exp} ({\bf  M}) = \sum_{n=1}^N 
\frac{ |\lambda_n \rangle e^{\lambda_n} \langle \lambda_n | }
{\langle \lambda_n | \lambda_n\rangle}\; ,
\label{eq:mexp2}
\eeq
where $\lambda_n$ are the eigenvalues, and 
the left and right eigenvectors, the row vector $\langle \lambda_n |$ and 
the column vector $| \lambda_n \rangle$, 
are assumed to have identical sets of eigenvalues, and to be mutually orthonormalized. 
For extensive details about the matrix exponential and its various representations, 
see Ref.~\cite{MOLE03}.
Although we will not explicitly consider diffusion or conduction on networks in this paper, 
the results in Eqs.~(\ref{eq:DE0}--\ref{eq:mexp2}) provide a strong argument for the 
importance of a detailed study of the eigenvectors and eigenvalues of $\bf W$.


The spectra of the walk-normalized Laplacian $\bf W$ and the correspondingly 
normalized adjacency matrix, 
 ${\bf D}^{-1} {\bf A}$, are closely related as 
 ${\bf W} = {\bf 1} - {\bf D}^{-1} {\bf A}$, where $\bf 1$ is the identity matrix. 
 The eigenvalues $\mu_i$ of ${\bf D}^{-1} {\bf A}$ obey the following conditions. 
 \beq
1 = \mu_N \ge \mu_{N-1} \ge \mu_{N-2} \; ... \ge \mu_{1} \ge -1 \;,
\label{eq:avals}
\eeq
confined to [$ -1,+1 $].
Since all the row sums of ${\bf D}^{-1} {\bf A}$ equal unity, the Perron-Frobenius 
eigenvector also corresponds to its eigenvalue, $ \mu_N = 1$. 

The relation between the spectra of these two operators is simply
 \beq
\lambda_n = 1- \mu_n \;, \{ n = 1 \; ... (N-1) \} \; , {\rm and} \; \lambda_N = 1- \mu_N = 0 \;, 
\label{eq:wvals}
\eeq
confined to [$0,2$] \cite{BOLL15}. 
This further leads to the conclusion that $\lambda_n = 1 \Leftrightarrow \mu_n = 0$ or, in other 
words, that $\bf W$ has a space of eigenvectors with eigenvalue $\lambda = 1$ 
{\it if and only if} $\bf A$ is singular. 

The adjacency matrix $\bf A$ may or may not be singular. 
Typically, singularity is caused by 
one or more small subgraphs, called {\it motifs} \cite{BANE08B,BANE08,BANE09}
or {\it singular cores} \cite{SCIR07}, connected to  
other vertices of degree at least two. 
Such configurations can be created by motif doubling or addition 
\cite{BANE08B}, and they 
are often found in ``boundary'' regions where the graph being studied 
has been separated from a larger graph. 
The resulting eigenvector has only nonzero elements corresponding to the motif vertices, 
and exactly zero on all the other vertices in the graph. It is 
therefore strongly {\it localized}, in contrast to the eigenvectors that do not belong to the 
nullspace of $\bf A$, which are generally {\it extended}. 
In addition to eigenvalues with $\lambda =1$, such doubling or joining of motifs 
may also create pairs of localized eigenvectors that are positioned symmetrically 
about $\lambda =1$ \cite{BANE08B,BANE08,BANE09}, or sometimes about other 
rational numbers. Several related theorems and examples are found in
Refs.~\cite{BANE08B,BANE08,BANE09,SCIR07}. 
Concrete examples will be discussed in detail in connection with some of the 
applications studied in Sec.~\ref{sec:III}.

\subsection{Estimating length scales for Laplacian eigenvectors}
\label{sec:IId}

For systems defined on a regular lattice, correlation lengths can be obtained from spatial 
correlation functions, or by Fourier transformation. However, these methods rely on 
translational or other, global symmetries of the lattice. 
When such global symmetries cannot be identified, as in 
many random networks with varying local vertex degrees and/or ``small-world'' shortcuts 
\cite{WS98,MONA99,ROSA22}, length scales must be estimated by alternative means. 

A practical method to estimate an effective correlation length in a random 
two-phase medium, without 
relying on global symmetries, was introduced in condensed-matter physics 
by Debye {\it et al.}~\cite{DEBY57}. 
Generalized to Laplacian eigenvectors $| \lambda_n \rangle$ of unweighted, undirected graphs, 
this method consists in measuring the ``area'' of the 
interface separating vertices $v_i$ bearing values $w_i $ of different sign by counting 
the number of the  
corresponding edges,  or ``broken bonds'', $M_{\rm b}(\lambda_n)$
 \cite{SHUM13,SHUM16,RICA19}. 
Defining the ``volume'' of the network as twice the number of edges, $2M$ \cite{GRAD10}, 
we see that 
 \beq
\xi(\lambda_n) = 2M / M_{\rm b}(\lambda_n) \in [ 2, 2M ] 
\label{eq:Debye}
\eeq
has the dimension of a length. It can be considered as an effective length scale
for $| \lambda_n \rangle$ 
\cite{DEBY57,RIKV85}, 
measured in edges along shortest paths (geodesics) and averaged over the whole system. 
In fact, volume and interface area are the two first Minkowski functionals from integral geometry 
\cite{ARMS19}, here implemented in a graph setting. 

As discussed above, some graphs may have localized eigenvectors that contain 
vertices bearing the value of exactly zero, $w_i = 0$. We therefore generalize the Sign 
function used to detect broken bonds to include the three possibilities, $+$, $-$, and $0$. 
This will be seen in some of the applications discussed in Sec.~\ref{sec:III} below. 

A {\it dispersion relation} is usually presented as a relation between an 
inverse characteristic length, such as $\xi^{-1}$, 
and a corresponding rate or frequency, such as the eigenvalues $\lambda$ of $\bf W$. 
For this purpose one conventionally uses the {\it wavenumber}, 
$k(\lambda_n) = 2 \pi / \xi(\lambda_n)$. 
Here, we will express the wavenumber in units of $\pi$, as 
\beq
K(\lambda_n) =  \frac{2}{\xi(\lambda_n)} = \frac{M_{\rm b}(\lambda_n)}{M} \;.
\label{eq:DebyeK}
\eeq 
We note that $K(\lambda_n)$ is simply the ratio of the number of ``broken'' edges in the 
eigenvector $| \lambda_n \rangle$ to the total number of edges in the graph. 

For a periodic lattice of linear size $M$ edges and free boundary conditions 
with unit lattice constant, $K$ will vary between a 
minimum of $1/M$ and a maximum of $1$, corresponding to a maximum $\xi = 2M$ 
and a minimum $\xi = 2$. In this way, our definition agrees with the analytic 
dispersion relation for a linear chain graph, 
   \beq
\lambda(K) = 1 - \cos(K \pi) \;\; {\rm with} \;\; K \in [1/M, 1] \;.
\label{eq:lin}
\eeq
Since the graph Laplacian is a generalization of a spatial second derivative, we expect 
its extended eigenvectors to retain the convex form for small values of $K$, even in the 
more complicated networks considered below. We return briefly to this point in 
Sec.~\ref{sec:V}. 

The inverse of Eq.~(\ref{eq:lin}) is  
   \beq
K(\lambda)  =\frac{1}{\pi} \arccos(1 - \lambda) \;,
\label{eq:linv}
\eeq
which yields the discrete, normalized {\it density of states} (DOS) as 
\beq
{\rm DOS}( \lambda) = \frac{{\rm d} K}{{\rm d} \lambda} 
=\frac{1}{\pi \sqrt {\lambda (2- \lambda)}} \;.
\label{eq:DOS}
\eeq
In the figures that follow, the discrete, $(N-1)$-point DOS will be defined as 
$\frac{1}{(N-1)(\lambda_{n-1} - \lambda_{n})}$ and shown vs 
$(\lambda_{n-1} + \lambda_{n})/2$, with $n$ running from $N$ to $2$ in order of increasing 
$\lambda_n$. 
Our choice of a discrete DOS was made to emphasize 
local divergencies, maxima, and minima.  
It contrasts with the locally smoothed forms used, e.g., in Refs. \cite{MONA99,BANE08,BANE09}. 
Numerical results for the dispersion relation and DOS for a line graph with $N=64$ 
are included in Fig.~\ref{fig:disprelsL64} and discussed in Sec.~\ref{Sec:Lin64}.

\subsection{Global graph measures and small-world property}
\label{sec:GGM}

In addition to our focus on the 
spectral graph measures discussed above, we will also determine some 
commonly used measures that describe the global organization of a graph, without reference 
to a specific set of basis vectors. Beyond the obvious numbers, $N$ and $M$, they are: 
the {\it mean degree}, the {\it edge density}, the {\it global clustering coefficient}, 
the {\it graph diameter}, and the {\it mean pair separation}. 

The mean degree is simply 
\beq
\langle d \rangle  = \frac{2M}{N} \;,
\label{eq:MD}
\eeq
and the edge density is 
\beq
\rho  = \frac{2M}{N(N-1)} = \frac{\langle d \rangle}{N-1}\;.
\label{eq:MD}
\eeq

The global clustering parameter can be defined as \cite{NEWM10}
\beq
C = \frac{3 \times {\rm (number \; of \; triangles)}}
{{\rm (number \; of \; connected \; vertex \; triples)}} \;.
\label{eq:GCP}
\eeq
It can be calculated exactly in terms of powers of the adjacency matrix $\bf A$ as
\beq
C = 
\frac{ {\rm \Tr} {\bf A}^3}
{2 \sum_{i=1}^{N-1} \sum_{j=i+1}^N  {{(\bf A}^2)}_{i,j}}
\;,
\label{eq:GCPe}
\eeq
where the double sum in the denominator runs over the upper triangle of 
the symmetric ${\bf A}^2$.
Note that this is different from the {\it mean} clustering coefficient used in Ref.~\cite{WS98}. 

Both the graph diameter, $D$, and the mean pair separation, $L$, are obtained from the 
symmetric {\it distance matrix}. Each of its elements, $D_{ij}$, is the 
shortest graph distance (geodesic) between the two vertices, $i$ and $j$. Simply,
\beq
D = {\rm Max}(D_{ij}) \;,
\label{eq:D}
\eeq
and 
\beq
L = \langle D_{ij} \rangle 
= \frac{2 \sum_{i=1}^{N-1} \sum_{j=i+1}^N D_{ij}}{N(N-1)}
 \;.
\label{eq:L}
\eeq
The computational work involved is that of determining all the $D_{ij}$. For unweighted, 
undirected graphs, this can be accomplished with a breadth-first search \cite{NEWM10}. 

A {\it small-world} graph is loosely defined as one with $D$ and $L$ of 
order $\ln (N) $ or smaller, while also satisfying $O(N^{-1}) << C << 1$ \cite{WS98,NEWM00}. 
However, since some of the applications discussed here are 
based on trees, which have  
$C = 0$, this definition is inconvenient in our case. We will therefore simply consider the ratio 
of $L$ to its value for an Erd{\"o}sz-R{\'e}nyi random graph 
with the same values of 
$N$ and $\langle d \rangle$, $L_{\rm rand} = \ln N /\ln \langle d \rangle$  \cite{NEWM00} 
and view the graphs studied here as small-world if $L/L_{\rm rand} < 1.5$.    


\section{Applications}
\label{sec:III}


The nine applications presented here have been chosen to be sufficiently small that 
their eigenvalues and eigenvectors can be numerically (and sometimes exactly) 
determined using commercial software on a workstation or laptop computer. 
By restricting the network sizes this way, we also wish to ensure that the structures 
of the eigenvectors should be visually understandable. 

\clearpage

\subsection{Linear chain}
\label{Sec:Lin64}

In order to calibrate our proposed 
method for extracting effective length scales from the individual 
Laplacian eigenvectors for a graph, Eq.~(\ref{eq:Debye}), 
we start with a simple, unweighted line graph with 
$N=64$ and $M=63$. 
Its global parameters are $\langle d \rangle \approx 1.9688$, 
$\rho \approx 0.03125$, $C=0$, $D=63$, and $L \approx 21.6667$. 
As $L/L_{\rm rand} \approx 3.5292 > 1.5$, this is not a small-world graph 
by our definition. 
We further note that a line graph with even $N$ does not have any vertex values exactly 
equal to zero. 

\begin{figure}[h]
\begin{center}
\hspace*{-1.5truecm}
\includegraphics[angle=0,width=.35\textwidth]{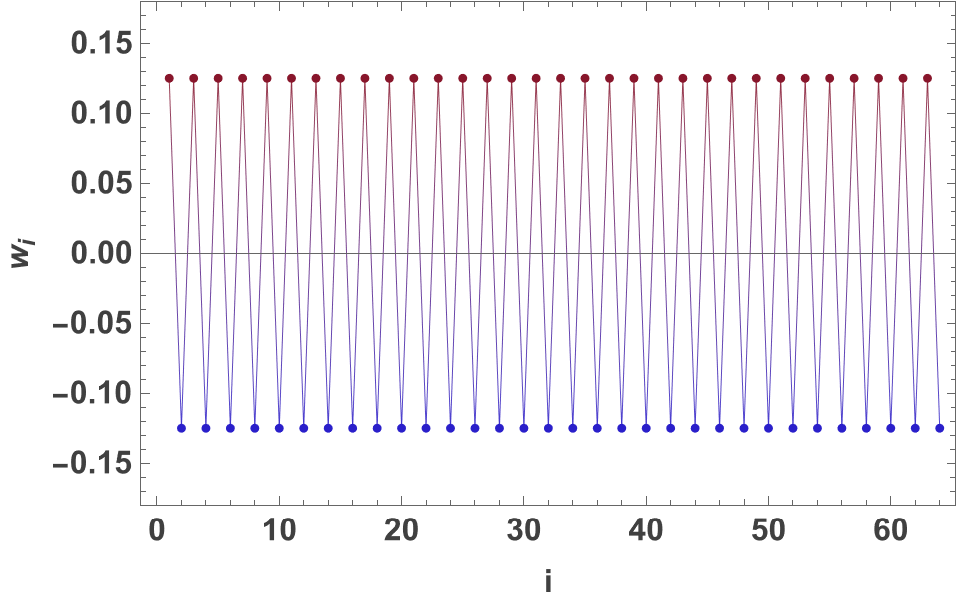}
\includegraphics[angle=0,width=.35\textwidth]{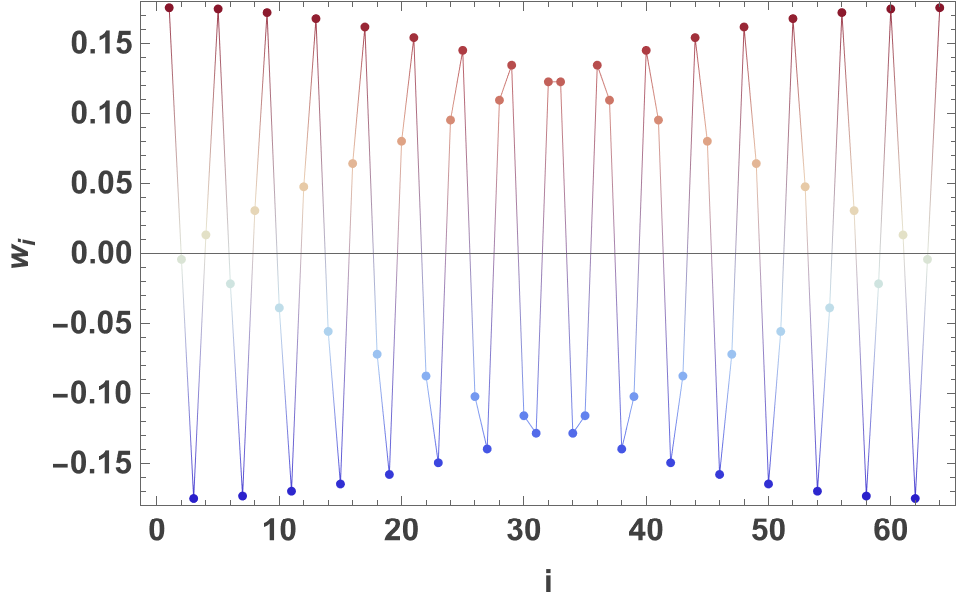}
\includegraphics[angle=0,width=.35\textwidth]{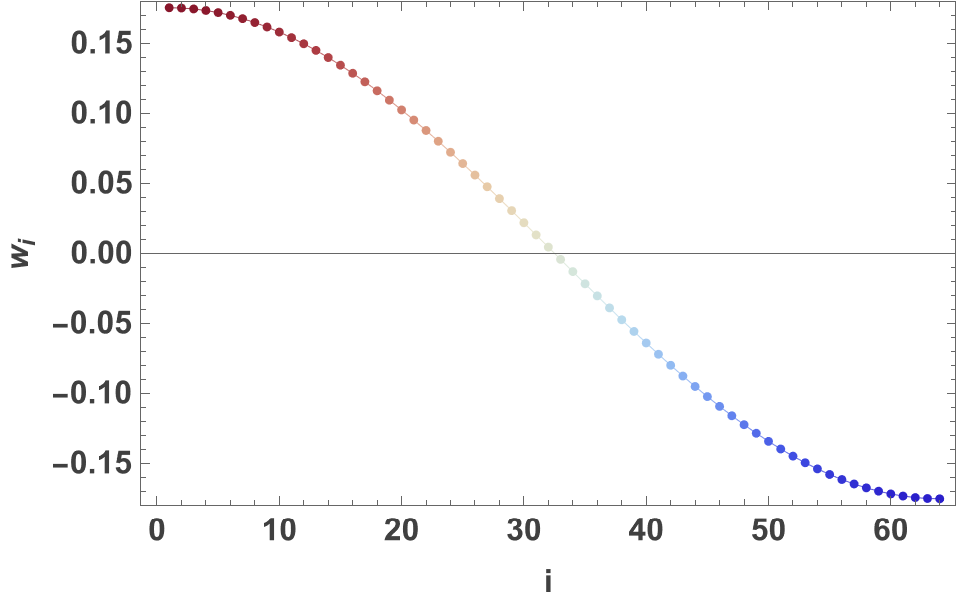}
(a)\;\;\;\;\;\;\;\;\;\;\;\;\;\;\;\;\;\;\;\;\;\;\;\;\;\;\;\;\;\;\;\;\;\;\;\;\;\;\;\;\;\;\; (b) \;\;\;\;\;\;\;\;\;\;\;\;\;\;\;\;\;\;\;\;\;\;\;\;\;\;\;\;(c)
\vspace*{-0.1truecm}
\includegraphics[angle=0,width=.47\textwidth]{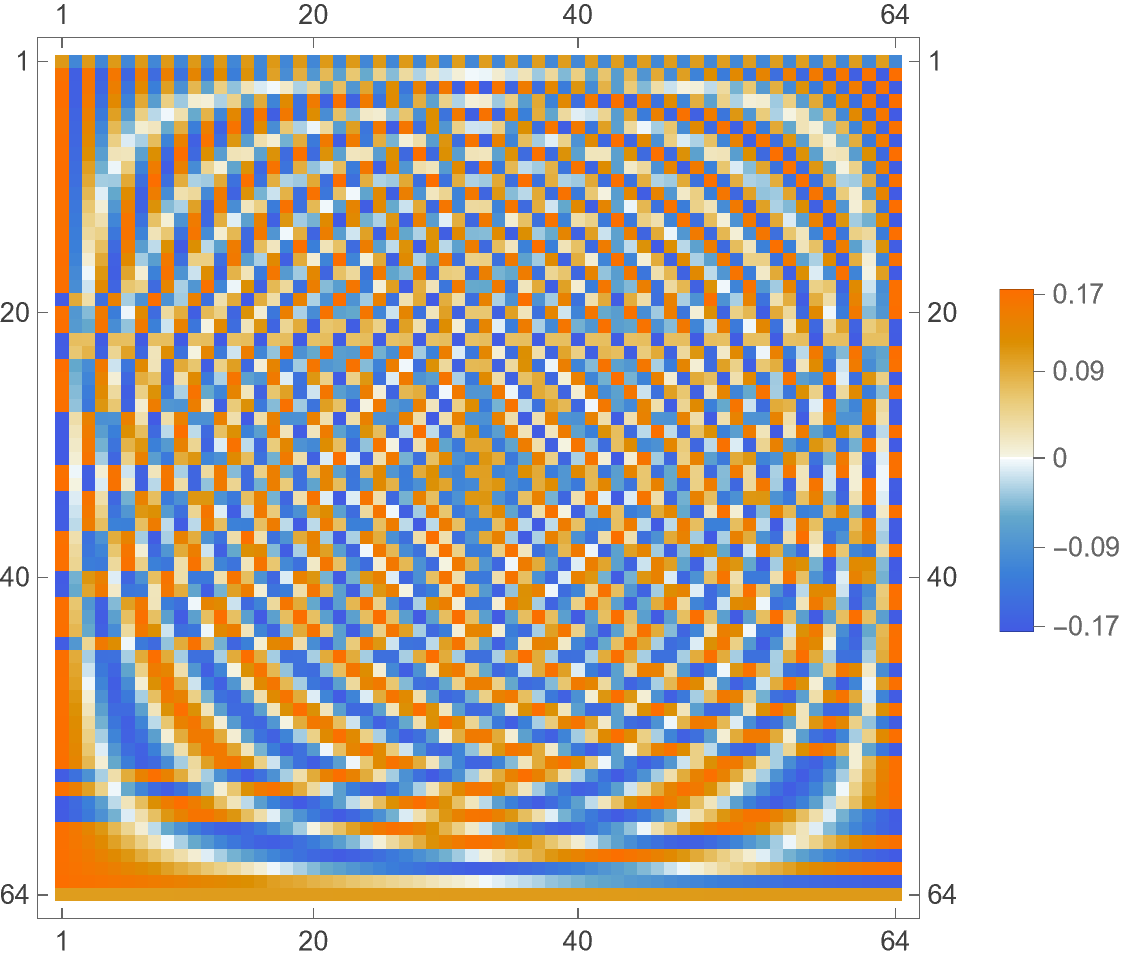} 
\includegraphics[angle=0,width=.5\textwidth]{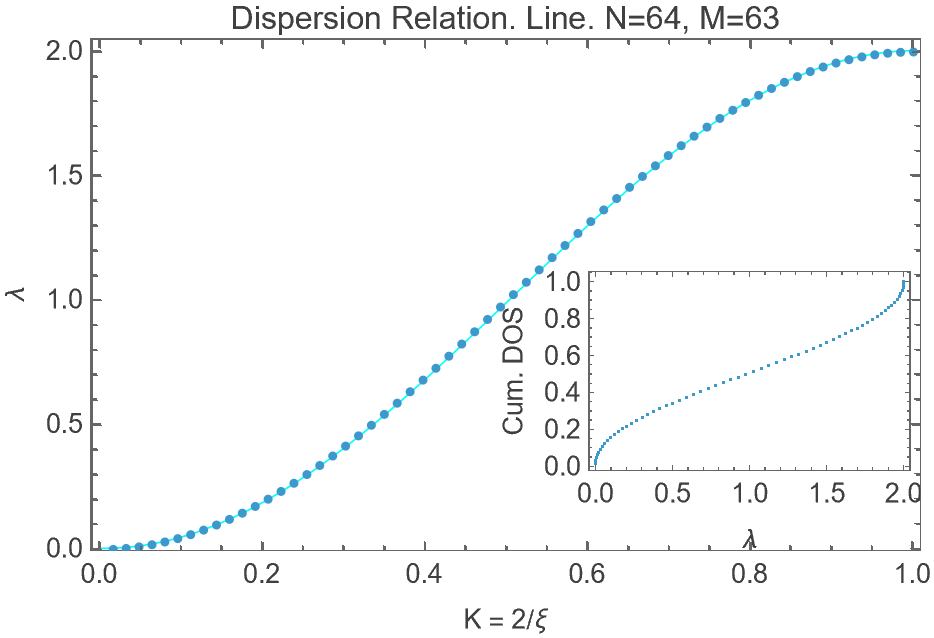} 
(d) \;\;\;\;\;\;\;\;\;\;\;\;\;\;\;\;\;\;\;\;\;\;\;\;\;\;\;\;\;\;\;\;\;\;\;\;\;\;\;\;\;\;\;\;\;\;\;\;\;\;\;\;\;\; (e)\\
\includegraphics[angle=0,width=.47\textwidth]{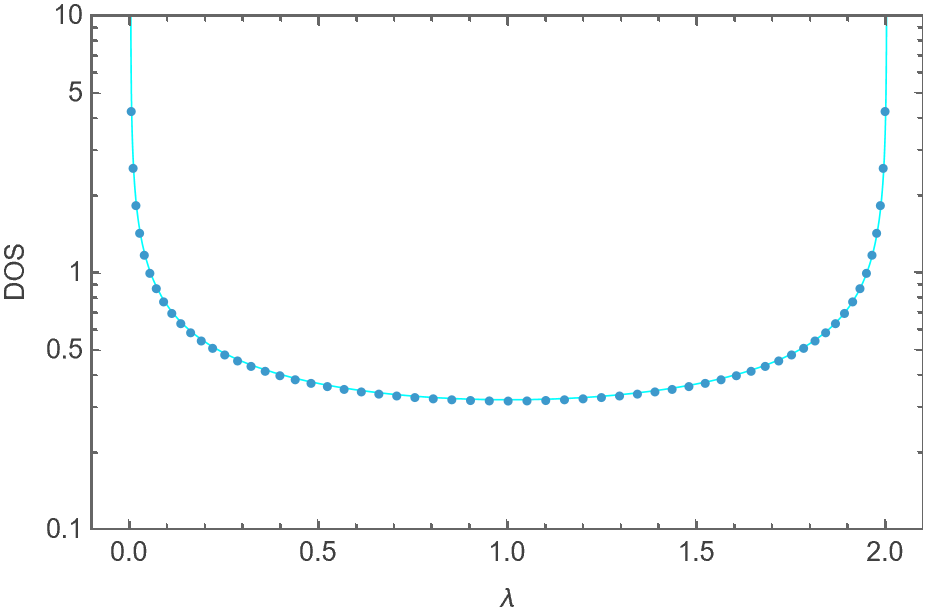} 
(f)
\end{center}
\vspace*{-0.5truecm}
\caption[]{
Unweighted, undirected line graph with $N=64$ and $M=63$. 
(a) The eigenvector  $| \lambda_{1} \rangle$ with 
$\lambda = 2$ and $K = 63/63 =1$.
(b) $| \lambda_{32} \rangle$ with $\lambda \approx 1.02493$ and 
$K = 32/63 \approx 0.508$. 
(c)  The Fiedler eigenvector,  $| \lambda_{63} \rangle$ with 
$\lambda \approx 0.001243$ and $K = 1/63 \approx 0.01587$.
(d) Vertex values for all the $64$ eigenvectors. The top row is
$| \lambda_1 \rangle$, and the bottom row is the uniform $| \lambda_N \rangle$. 
The vertex values in parts (a)--(d) are shown on a color scale, with dark blue for the 
largest negative and dark red for the largest positive values. This convention is 
maintained in all the remaining figures in this paper. 
(e) Dispersion relation.
The inset shows the cumulative DOS for the eigenvalues.  
(f)
Discrete density of states (DOS) as defined in the text after Eq.~(\ref{eq:DOS}).
See details in Sec.~\ref{Sec:Lin64} and numerical data in the tables. 
Continuous curves in the background in parts (e) and (f) are exact, analytical results. 
}
\label{fig:disprelsL64}
\end{figure}

The three representative eigenvectors, the bipartite (``antiferromagnetic'') 
$| \lambda_{1} \rangle$ with $\lambda = 2$ and $K = 63/63 =1$,
$| \lambda_{32} \rangle$ with $\lambda \approx 1.0249$ and $K = 32/63 \approx 0.5079$,
and the Fiedler vector (the eigenvector with the smallest nonzero eigenvalue, 
which bisects the graph into two parts of similar size)  
$| \lambda_{63} \rangle$ with 
$\lambda \approx 0.001243$ and $K = 1/63 \approx 0.01587$
are shown as $w_i$ vs $i$ in Fig.~\ref{fig:disprelsL64} (a), (b), and (c), respectively. 
The vertex values, 
$w_i$, are shown by color from dark red for large positive to dark blue for large negative 
values. This color coding for vertex values is maintained throughout the paper, including 
Fig.~\ref{fig:disprelsL64}(d), which shows the vertex values for all the 64 eigenvectors. 
The top row is $| \lambda_1 \rangle$, and the bottom row is the uniform $| \lambda_N \rangle$. 
The dispersion relation, $\lambda (K)$, is shown in Fig.~\ref{fig:disprelsL64}(e). The data 
points are obtained from the number of broken bonds in each eigenvector via 
Eq.~(\ref{eq:DebyeK}), while the curve in the background is 
the analytic dispersion relation, Eq.~(\ref{eq:linv}). 
The agreement is excellent. 
The inset in part (e) shows the cumulative DOS for the eigenvalues.  
The discrete density of states (DOS) as defined in the text after Eq.~(\ref{eq:DOS}) is 
shown in Fig.~\ref{fig:disprelsL64}(f). It is the numerical derivative of the cumulative DOS. 
The curve in the background is the analytical result, Eq.~(\ref{eq:DOS}). 
Again, the agreement between the numerical results obtained by the broken-bond method 
and the analytical curve is excellent for this calibration case. 
The numerical results for this and all the other application networks 
are summarized in Tables~\ref{tab:table} and \ref{tab:table2}. 

\clearpage

\clearpage

\subsection{Linear chain with random shortcuts}
\label{Sec:Lin64L}

Next we consider a network generated from the $N=64$ 
line graph of Fig.~\ref{fig:disprelsL64} by adding 22 random shortcuts between 
non-neighbor vertices \cite{NEWM99,NEWM00}, $N=64$ and $M=85$. 
Its global parameters are $\langle d \rangle \approx 2.6563$, 
$\rho \approx 0.04216$, $C=0$, $D=10$, and 
$L \approx 4.5987$. The significant reductions in $D$ and $L$ 
from those of the unmodified line graph qualifies this as a small-world network \cite{WS98} . 
Consistent with this, $L/L_{\rm rand} \approx 1.0803 < 1.5$.

\begin{figure}[h]
\begin{center}
\vspace*{-0.1truecm}

\hspace*{-1.5truecm}
\includegraphics[angle=0,width=.35\textwidth]{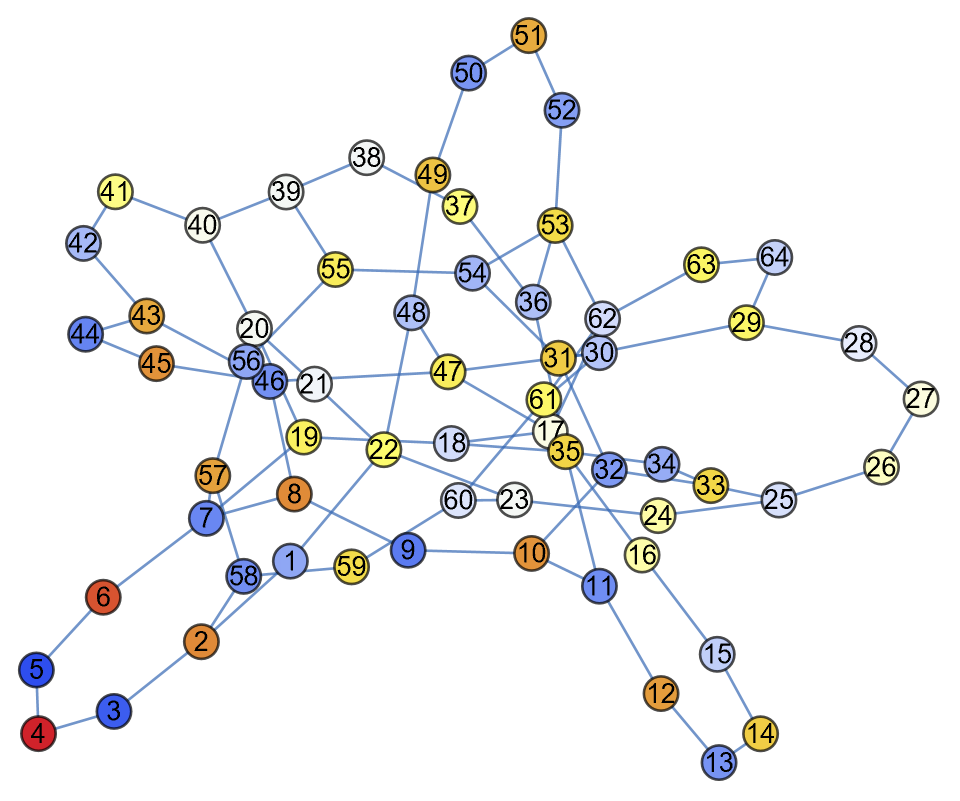}
\includegraphics[angle=0,width=.35\textwidth]{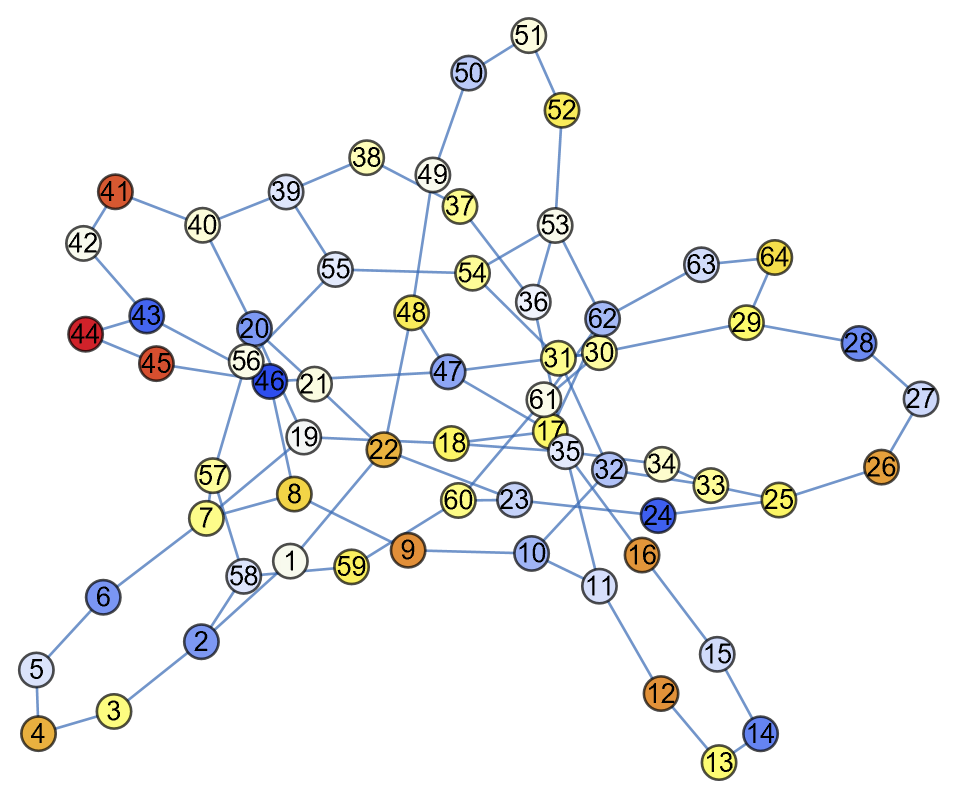}
\includegraphics[angle=0,width=.35\textwidth]{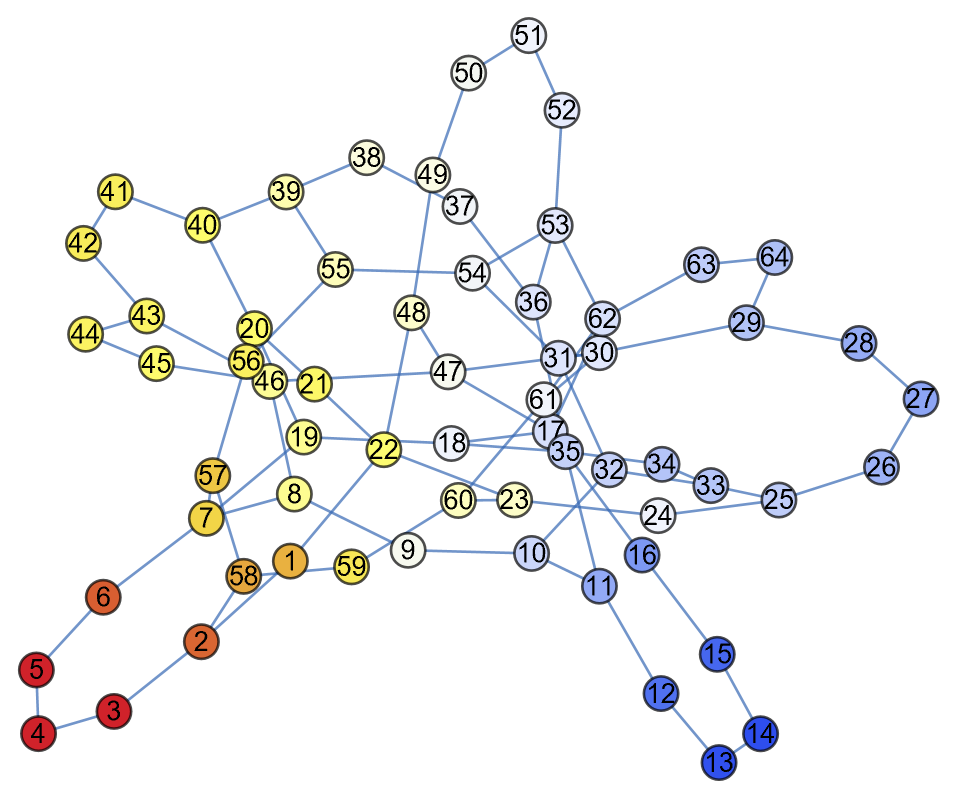}
(a)\;\;\;\;\;\;\;\;\;\;\;\;\;\;\;\;\;\;\;\;\;\;\;\;\;\;\;\;\;\;\;\;\;\;\;\;\;\;\;\;\;\;\; (b) \;\;\;\;\;\;\;\;\;\;\;\;\;\;\;\;\;\;\;\;\;\;\;\;\;\;\;\;(c)
\hspace*{1.0truecm}
\includegraphics[angle=0,width=.45\textwidth]{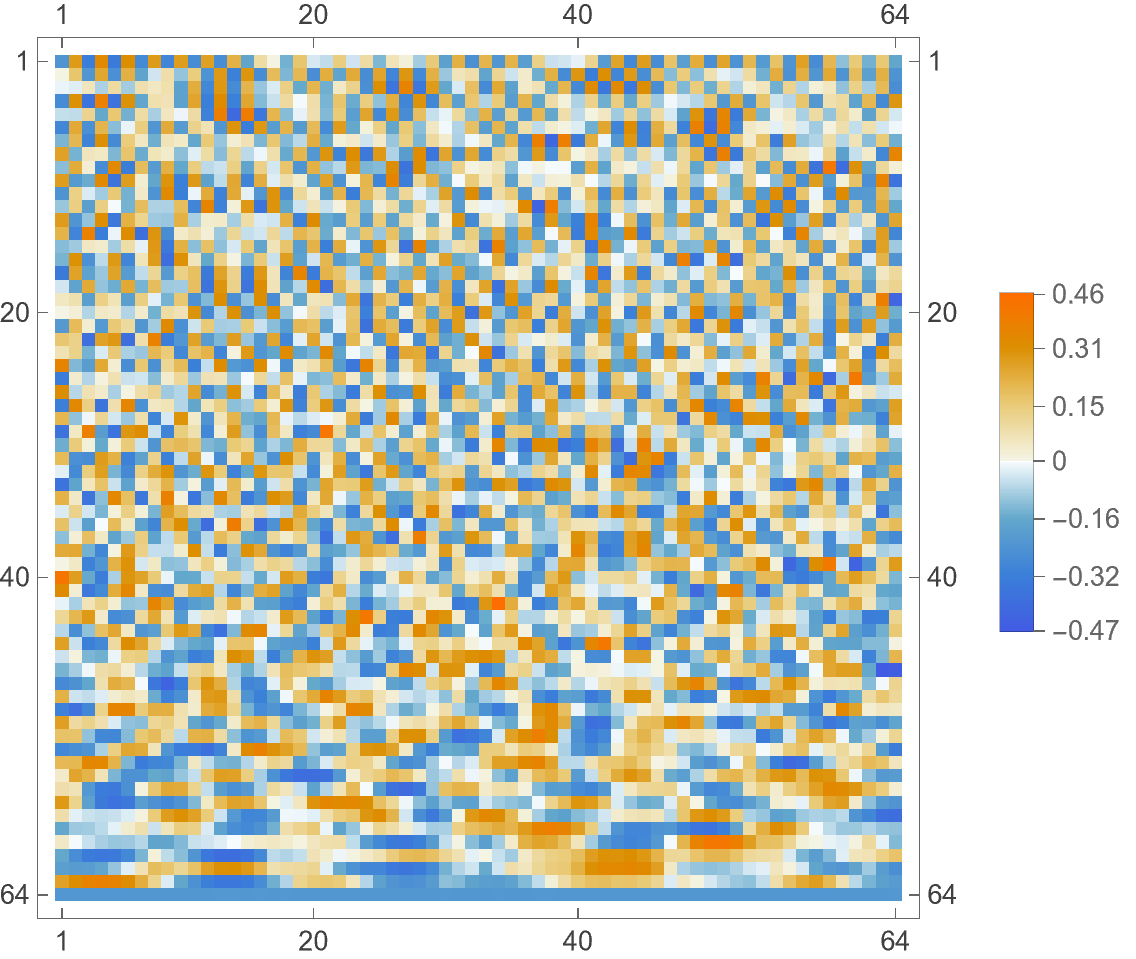} 
\includegraphics[angle=0,width=.45\textwidth]{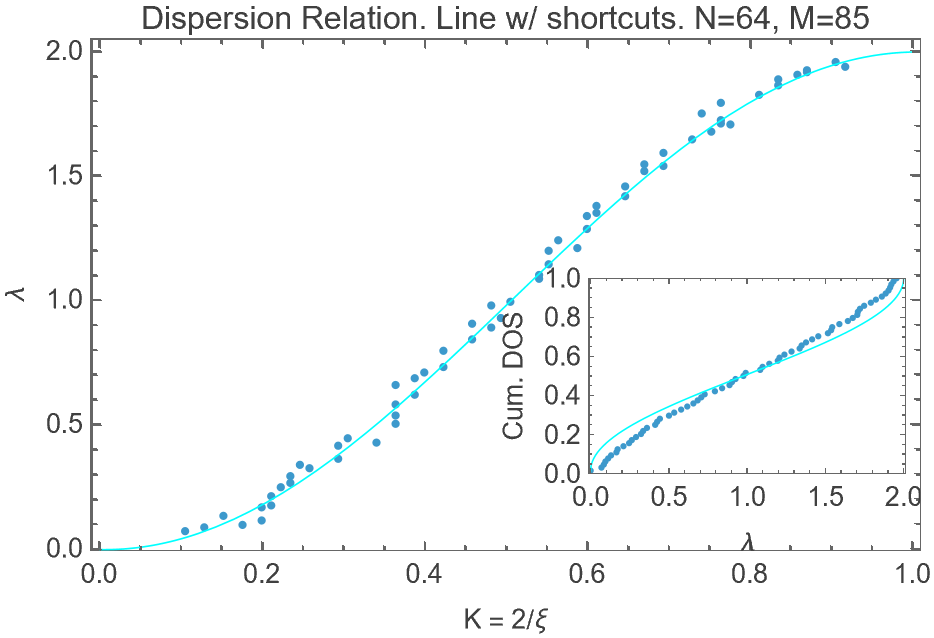} 
(d)\;\;\;\;\;\;\;\;\;\;\;\;\;\;\;\;\;\;\;\;\;\;\;\;\;\;\;\;\;\;\;\;\;\;\;\;\;\;\;\;\;\;\; (e)\\
\includegraphics[angle=0,width=.35\textwidth]{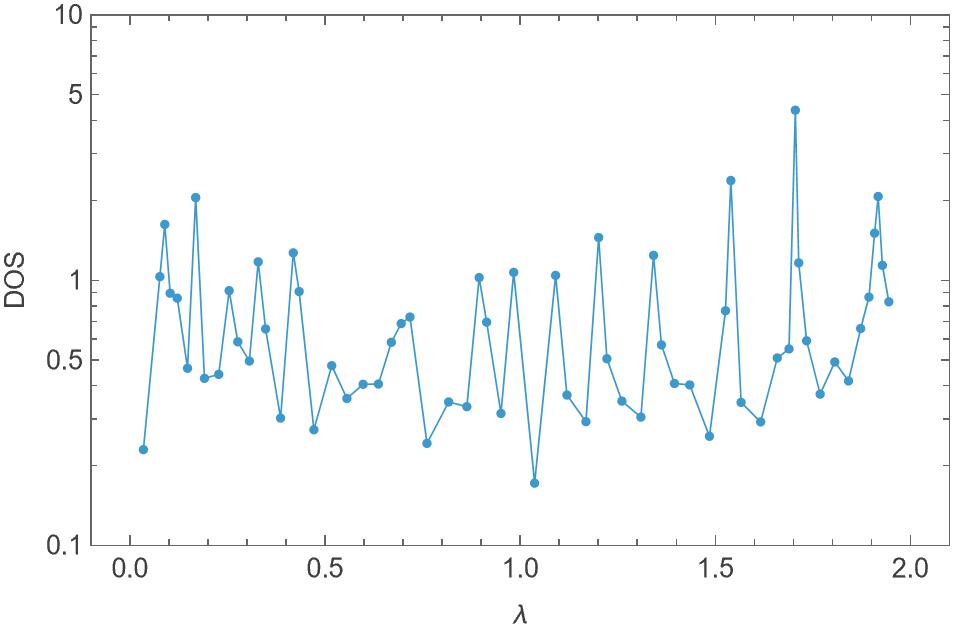} 
(f)
\end{center}
\vspace*{-0.5truecm}
\caption[]{
A small-world network generated from the $N=64$ line graph of Fig.~\ref{fig:disprelsL64} 
by adding 22 random shortcuts, $N=64$ and $M=85$. 
(a) $| \lambda_{1} \rangle$ with 
$\lambda \approx 1.9545$ and $K = 77/85 \approx 0.9059$.
(b) $| \lambda_{32} \rangle$ with $\lambda \approx 0.9909$ 
 and $K = 43/85 \approx 0.5059$. 
(c)  The Fiedler eigenvector  $| \lambda_{63} \rangle$ with 
$\lambda \approx 0.06909$  and $K = 9/85 \approx 0.1059$.
(d) Vertex values for the $N=64$ eigenvectors. 
(e) Dispersion relation.
The inset shows the cumulative DOS for the eigenvalues.  
In both main figure and inset, the corresponding results for the unmodified line graph
are shown as cyan curves.
(f) Discrete density of states (DOS). 
The maximum peak corresponds to the nearly degenerate pair, 
$| \lambda_{12} \rangle$ with $\lambda \approx 1.7071$
and 
$| \lambda_{13} \rangle$ with $\lambda \approx 1.7035$ .
}
\label{fig:SmallWorld64}
\end{figure}

The three representative eigenvectors, 
$| \lambda_{1} \rangle$ with $\lambda = 1.9545$ and $K =77/85 \approx 0.9059$,
$| \lambda_{32} \rangle$ with $\lambda \approx 1.0249$ and $K = 32/63 \approx 0.5079$,
and the Fiedler vector  $| \lambda_{63} \rangle$ with 
$\lambda \approx 0.06909$ and $K = 9/85 \approx 0.1059$
are shown in a spring-electric layout in Fig.~\ref{fig:SmallWorld64} (a), (b), and (c), respectively. 
As $\lambda_1 < 2$, $| \lambda_{1} \rangle$ is not bipartite. However, 
$K$ can be read as the percentage of broken bonds, which remains about $91 \%$ in (a). 
As always, the Fiedler vector bisects the graph into two parts of similar size. 

Vertex values for all the $N=64$ eigenvectors are shown in part (d). 
As in Fig.~\ref{fig:disprelsL64} and all the following figures, the top row is 
$| \lambda_1 \rangle$, and the bottom row is the 
uniform $| \lambda_N \rangle$ with $\lambda = 0$. There are no vertex values exactly 
equal to 0. 

The dispersion relation is shown in part (e) with the cumulative DOS as an inset.  
In both main figure and inset, the corresponding analytical 
results for the unmodified line graph are shown as continuous curves in cyan. 
The ranges of the nonzero $\lambda$ and $K$ are smaller than for the pure line graph, 
as shown in Tables~\ref{tab:table} and~\ref{tab:table2}, and 
within this reduced range the data points are scattered 
closely around the curve representing the unmodified line graph.

The discrete density of states (DOS) is shown on a logarithmic scale in part (f). 
It is quite different from the smooth curve with divergences at 1 and 2 shown in 
Fig.~\ref{fig:disprelsL64}(f). 
The global maximum corresponds to the nearly degenerate pair, 
$| \lambda_{12} \rangle$ and $| \lambda_{13} \rangle$with $\lambda \approx 1.7$.

\clearpage 

\clearpage

\subsection{Cayley tree}
\label{Sec:Cayley40}

\begin{figure}[h]
\begin{center}
\vspace*{-0.1truecm}
\includegraphics[angle=0,width=.3\textwidth]{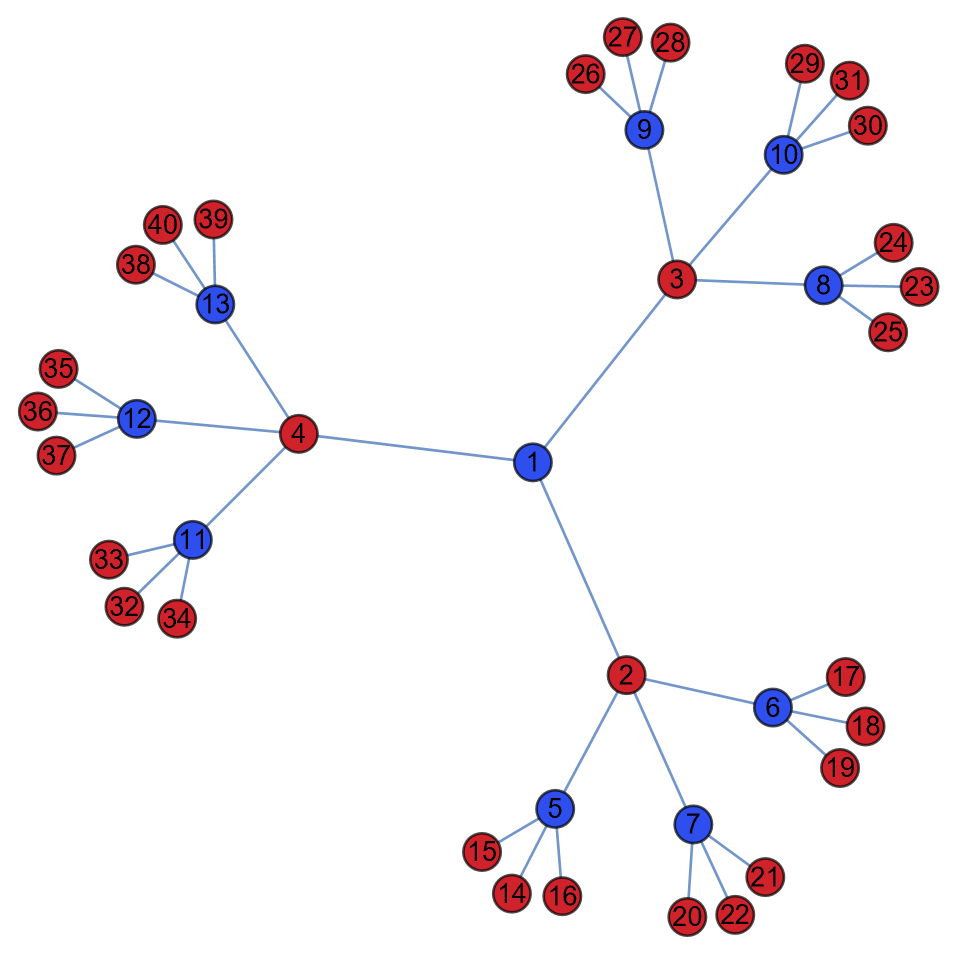}
\includegraphics[angle=0,width=.3\textwidth]{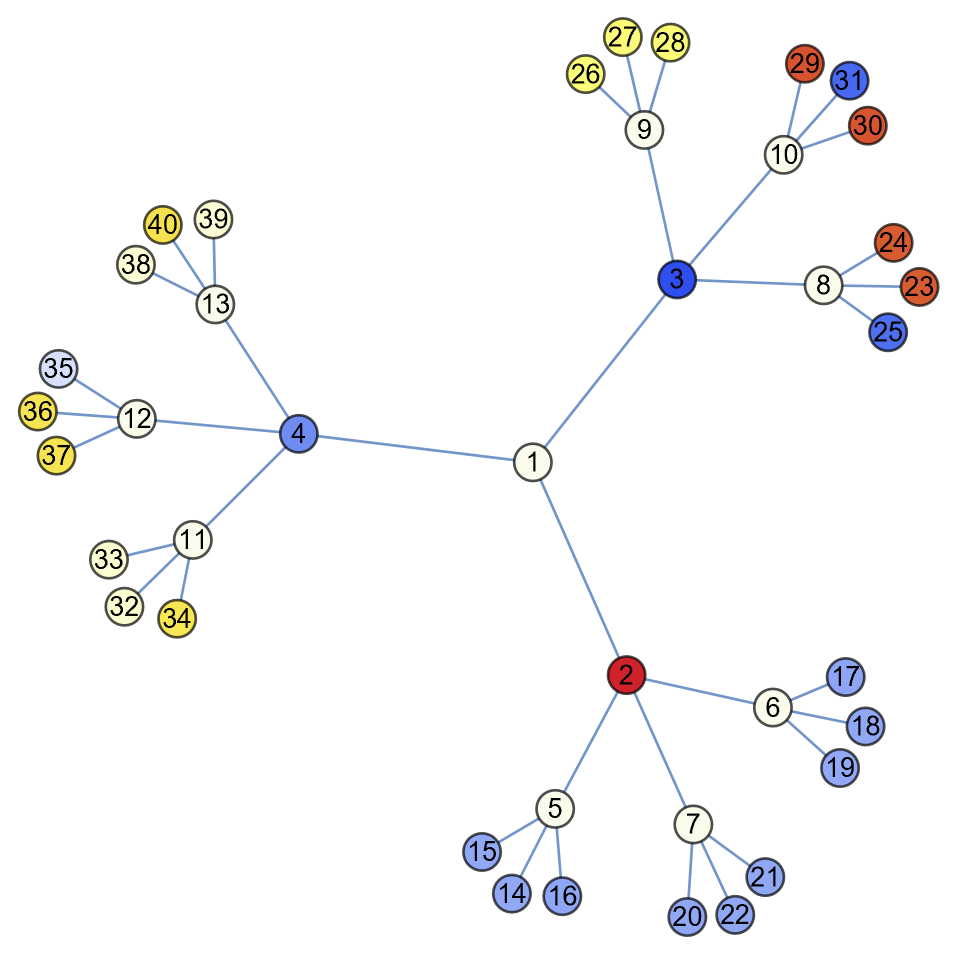}
\includegraphics[angle=0,width=.3\textwidth]{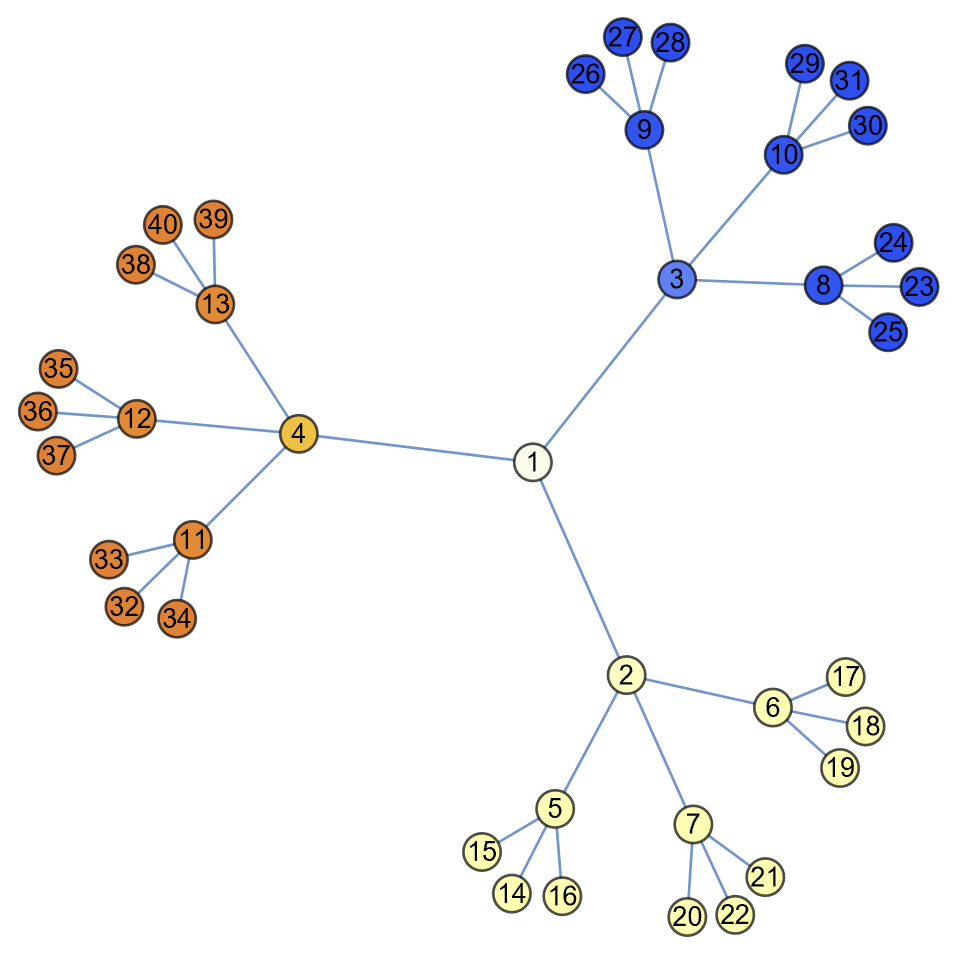}
(a)\;\;\;\;\;\;\;\;\;\;\;\;\;\;\;\;\;\;\;\;\;\;\;\;\;\;\;\;\;\;\;\;\;\;\;\;\;\;\;\;\;\;\; (b) \;\;\;\;\;\;\;\;\;\;\;\;\;\;\;\;\;\;\;\;\;\;\;\;\;\;\;\;(c)
\includegraphics[angle=0,width=.47\textwidth]{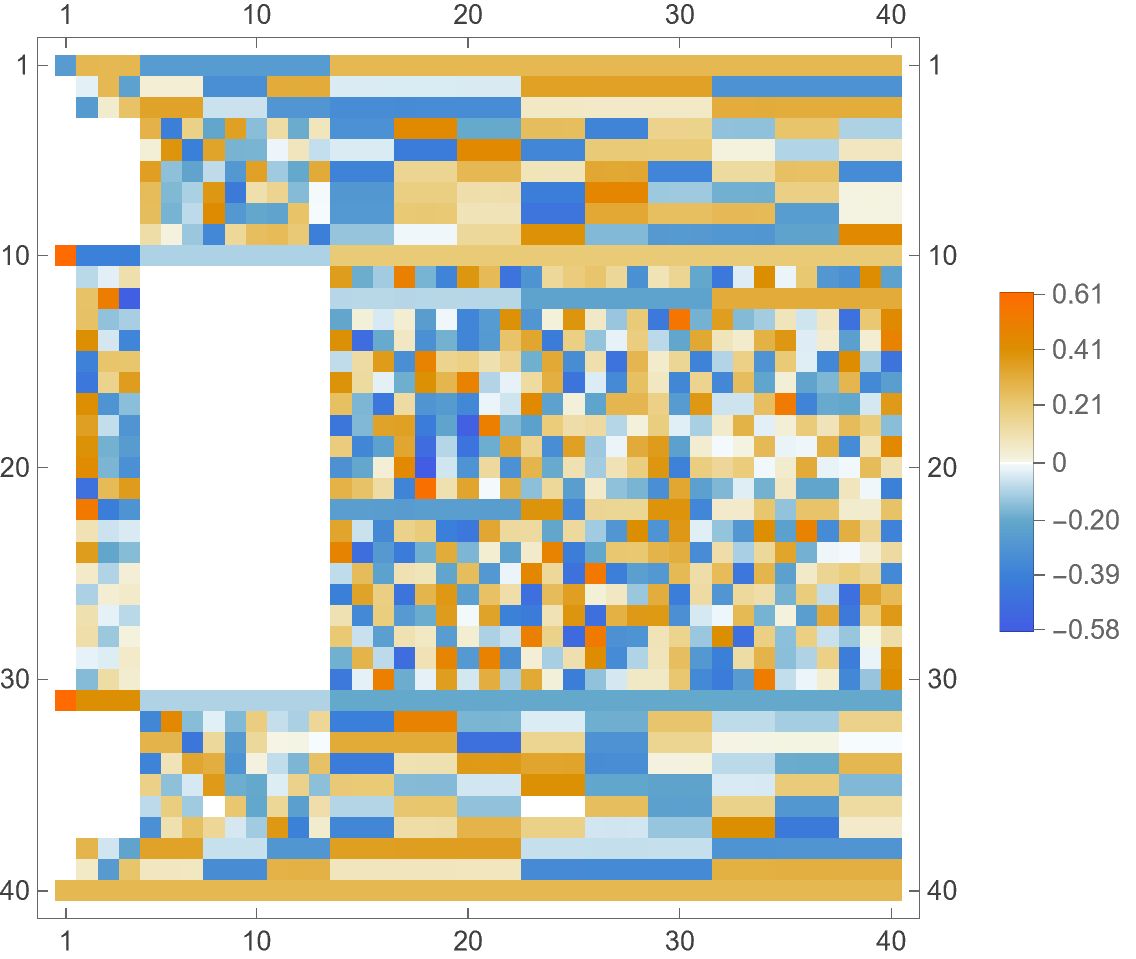} 
\includegraphics[angle=0,width=.5\textwidth]{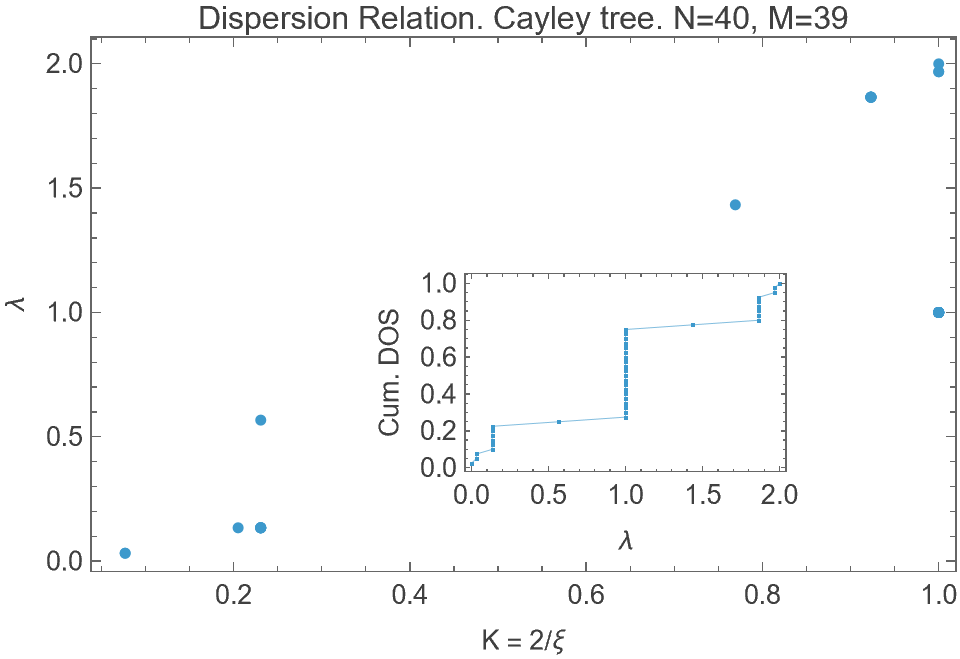} 
(d) \;\;\;\;\;\;\;\;\;\;\;\;\;\;\;\;\;\;\;\;\;\;\;\;\;\;\;\;\;\;\;\;\;\;\;\;\;\;\;\;\;\;\;\;\;\;\;\;\;\;\;\;\;\; (e)\\
\includegraphics[angle=0,width=.35\textwidth]{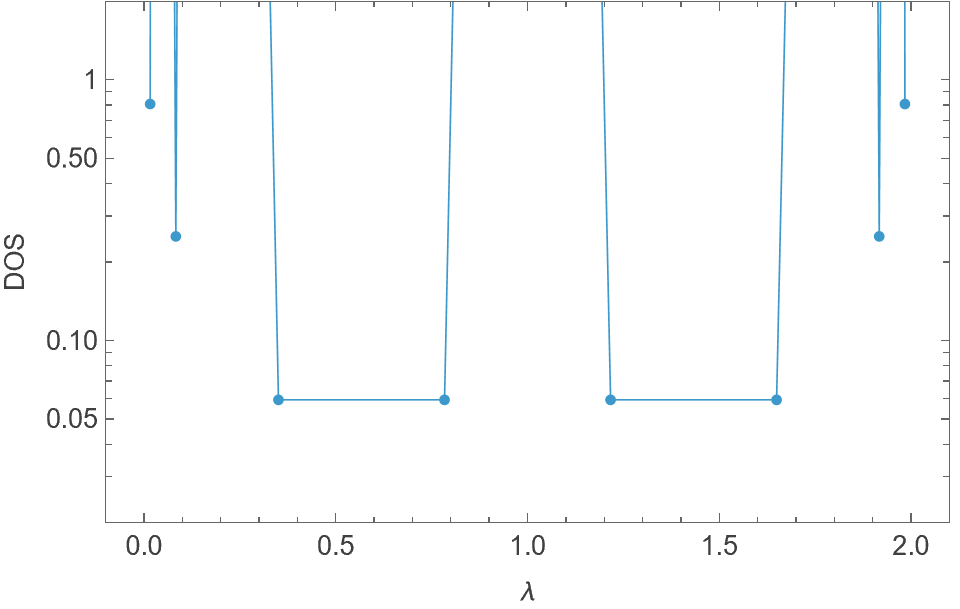} 
(f)
\end{center}
\vspace*{-0.5truecm}
\caption[]{
Cayley tree graph with a branching ratio of 3 and 4 layers, $N=40$ and $M=39$
\cite{TUNC15,TUNC20}.
(a) The bipartite eigenvector  $| \lambda_1 \rangle$ with 
$\lambda = 2$ and $K = 39/39 = 1$.
(b) The tripartite eigenvector  $| \lambda_{22} \rangle$, 
representative of the 20 degenerate eigenvectors with $\lambda = 1$ and $K = 39/39 = 1$.
(c) The Fiedler eigenvector  $| \lambda_{39} \rangle$, 
with $\lambda \approx 0.03175$ and $K = 3/39 \approx 0.07692$.
(d) Vertex values for all the $N=40$ eigenvectors.
The degenerate eigenvectors from $| \lambda_{11} \rangle$ through $| \lambda_{30} \rangle$ 
with eigenvalues $\lambda = 1$ have vertex values 0 on vertices 1 and 5-13 and varying 
positive and negative values on the rest. 
(e) Dispersion relation.
The inset shows the cumulative DOS.  
 (f) Discrete density of states (DOS). 
The divergences correspond to the degenerate eigenvectors. 
}
\label{fig:disprelsT40}
\end{figure}

Next, we consider a more complicated tree graph: 
a Cayley tree with a branching ratio of $b=3$ and $R=4$ layers, $N=40$ and $M=39$.
As it is defined in Refs.~\cite{TUNC15,TUNC20},
it consists of a ``root'' vertex with degree 3 and its three neighbors in layer 2, each with 
degree 4. These are connected to 3 neighbors in layer 3, etc. Thus, each layer 
contains three times the number of vertices as the previous one. Figures~\ref{fig:disprelsT40} 
(a)--(c) show this structure. Each subtree rooted at layer $r$ has three-fold 
permutational symmetry around its root \cite{TUNC20}. 
Its global parameters are $\langle d \rangle = 1.95$, $\rho = 0.05$, $C=0$, $D=6$, and 
$L \approx 4.3615$. 
We find the ratio, $L/L_{\rm rand} \approx 0.7896 < 1.5$, indicating the small-world 
property. This may at first seem surprising, as the Cayley tree has only local connections. 
However, it is consistent with the repeated, three-fold branching structure, which leads to 
$N \sim e^{D/2}$ or $D \sim 2 \ln(N)$. 

The bipartite eigenvector  $| \lambda_1 \rangle$ with $\lambda = 2$ and $K = 39/39 = 1$ 
is shown in Fig.~\ref{fig:disprelsT40}(a).
The eigenvector  $| \lambda_{22} \rangle$ is 
representative of the 20 degenerate eigenvectors with $\lambda = 1$ and $K = 39/39 = 1$. 
It is shown in Fig.~\ref{fig:disprelsT40}(b).
These eigenvectors are actually tripartite, as each edge connects a pair of vertices with different 
values among the three possibilities, $-$, 0, and $+$. 
The Fiedler eigenvector  $| \lambda_{39} \rangle$ 
with $\lambda \approx 0.03175$ and $K = 3/39 \approx 0.07692$ 
is shown in Fig.~\ref{fig:disprelsT40}(c). The 20 
degenerate eigenvectors from $| \lambda_{11} \rangle$ through $| \lambda_{30} \rangle$ 
with eigenvalues $\lambda = 1$ have vertex values 0 on vertices 1 (the root) 
and 5-13 (layer 3) and varying 
positive and negative values on layers 2 and 4. 
The vertex values for all the $N=40$ eigenvectors
are shown in Fig.~\ref{fig:disprelsT40}(d). 
We note that we have not orthogonalized the matrix of eigenvectors. Therefore, the 
matrix with elements $\langle \lambda_n | \lambda_m \rangle$ is not in general diagonal, 
but block-diagonal.
Groups of $r$ degenerate eigenvectors will form $r \times r$ square blocks 
along the diagonal. This does not present a problem for our considerations in this paper. 
Exact symbolic calculation with Mathematica~14 software shows that all the 
eigenvalues are symmetric about 1, denoted as $(1 \pm x) \times$ degeneracy: 
$(1 \pm 1) \times 1$, $(1 \pm \sqrt{15}/4) \times 2$, $(1 \pm \sqrt{3}/2) \times 6$, 
 $(1 \pm \sqrt{3}/4) \times 1$,  and $1 \times 20$. 
 
The dispersion relation, $\lambda$ vs $K$, is shown in Fig.~\ref{fig:disprelsT40}(e).
The inset shows the cumulative DOS for the eigenvalues.  
The discrete density of states (DOS) is shown in Fig.~\ref{fig:disprelsT40}(f).
The wide, divergent peaks correspond to the degenerate sets of eigenvectors. 
These contain a large number of exact zero vertex values, due to motif joinings and 
duplications in this highly symmetric graph \cite{BANE09}. 
\clearpage

\clearpage

\subsection{Cayley tree with random shortcuts}
\label{Sec:Cayley40L}
We next consider a network generated from the Cayley tree with a 
branching ratio of 3 and 4 layers of Fig.~\ref{fig:disprelsT40} 
by adding 14 random shortcuts \cite{WS98}, $N=40$ and $M=53$. 
Its global parameters are $\langle d \rangle = 2.65$, $\rho \approx 0.06795$, 
$C=0.09091$, $D=6$, and $L \approx 3.62564$. 
We find the ratio, $L/L_{\rm rand} \approx 0.9579 < 1.5$, indicating the small-world property.

\begin{figure}[h]
\begin{center}
\vspace*{-0.1truecm}
\includegraphics[angle=0,width=.3\textwidth]{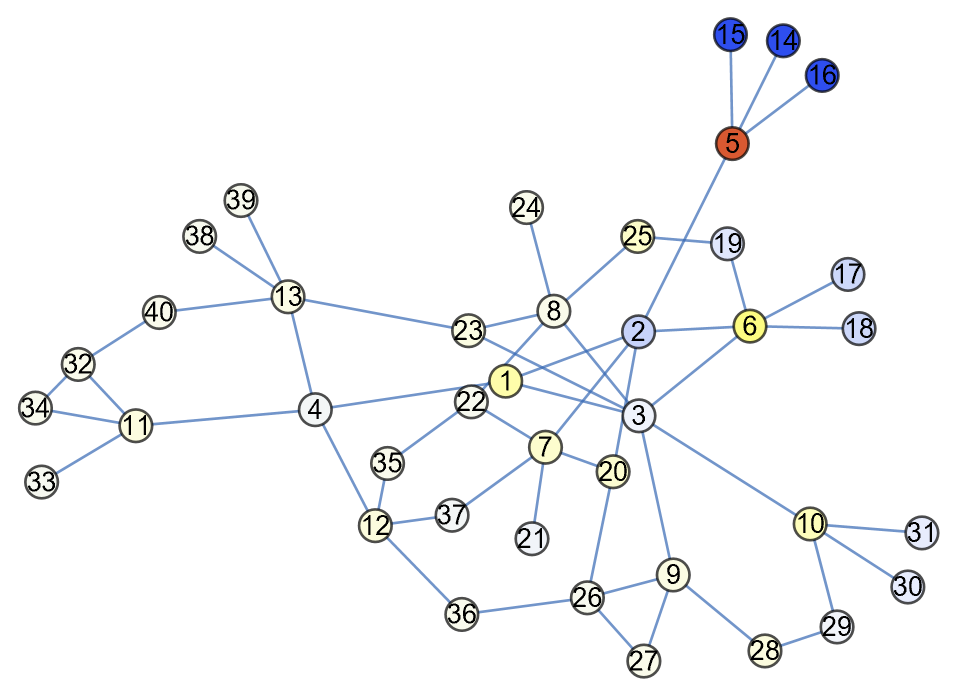}
\includegraphics[angle=0,width=.3\textwidth]{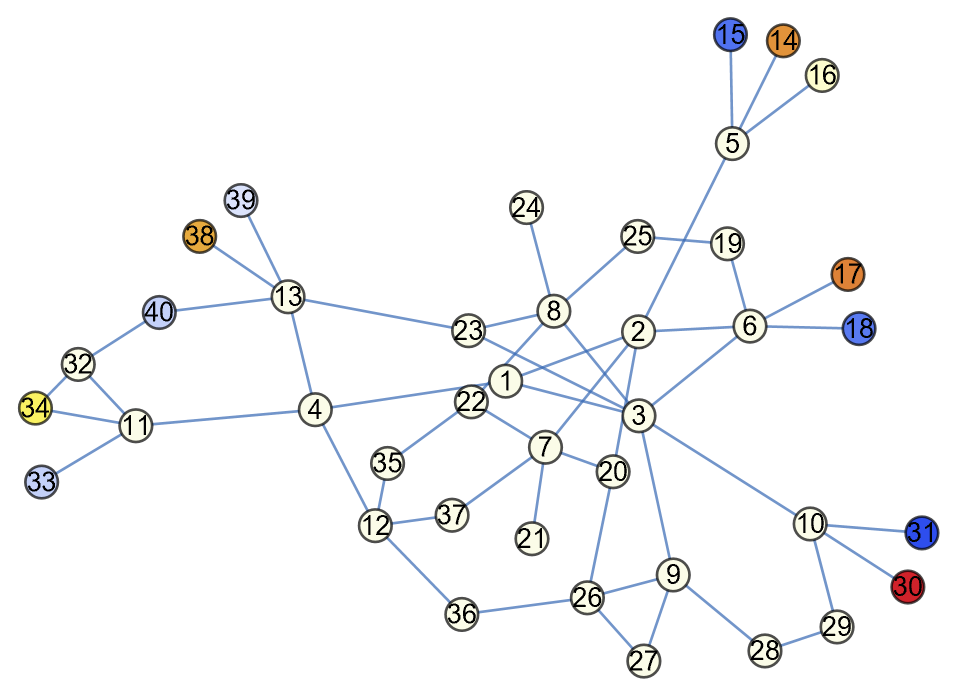}
\includegraphics[angle=0,width=.3\textwidth]{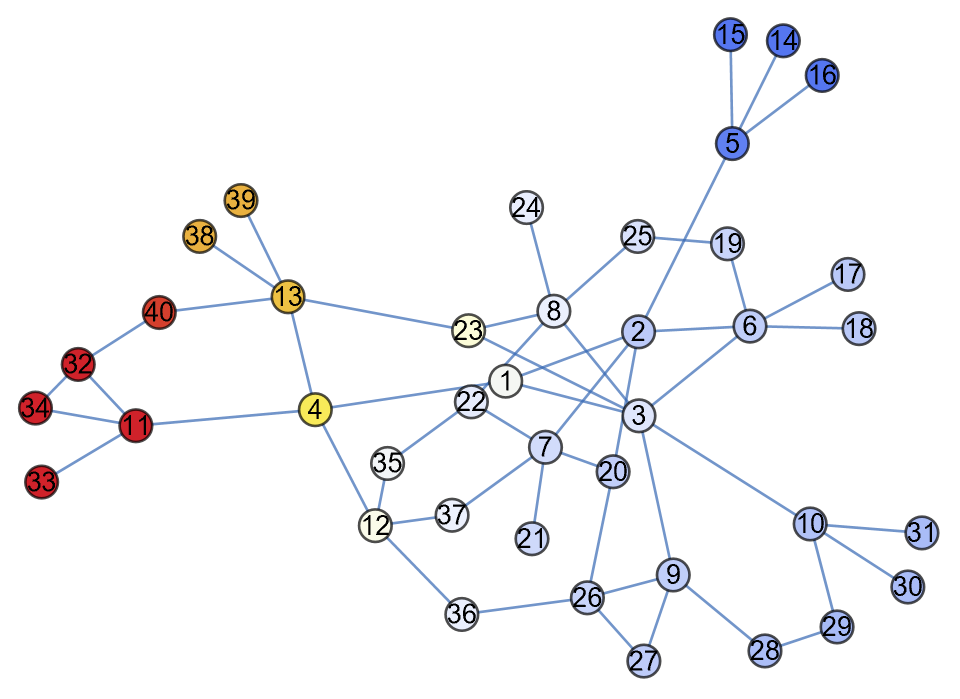}
(a)\;\;\;\;\;\;\;\;\;\;\;\;\;\;\;\;\;\;\;\;\;\;\;\;\;\;\;\;\;\;\;\;\;\;\;\;\;\;\;\;\;\;\; (b) \;\;\;\;\;\;\;\;\;\;\;\;\;\;\;\;\;\;\;\;\;\;\;\;\;\;\;\;(c)
\includegraphics[angle=0,width=.47\textwidth]{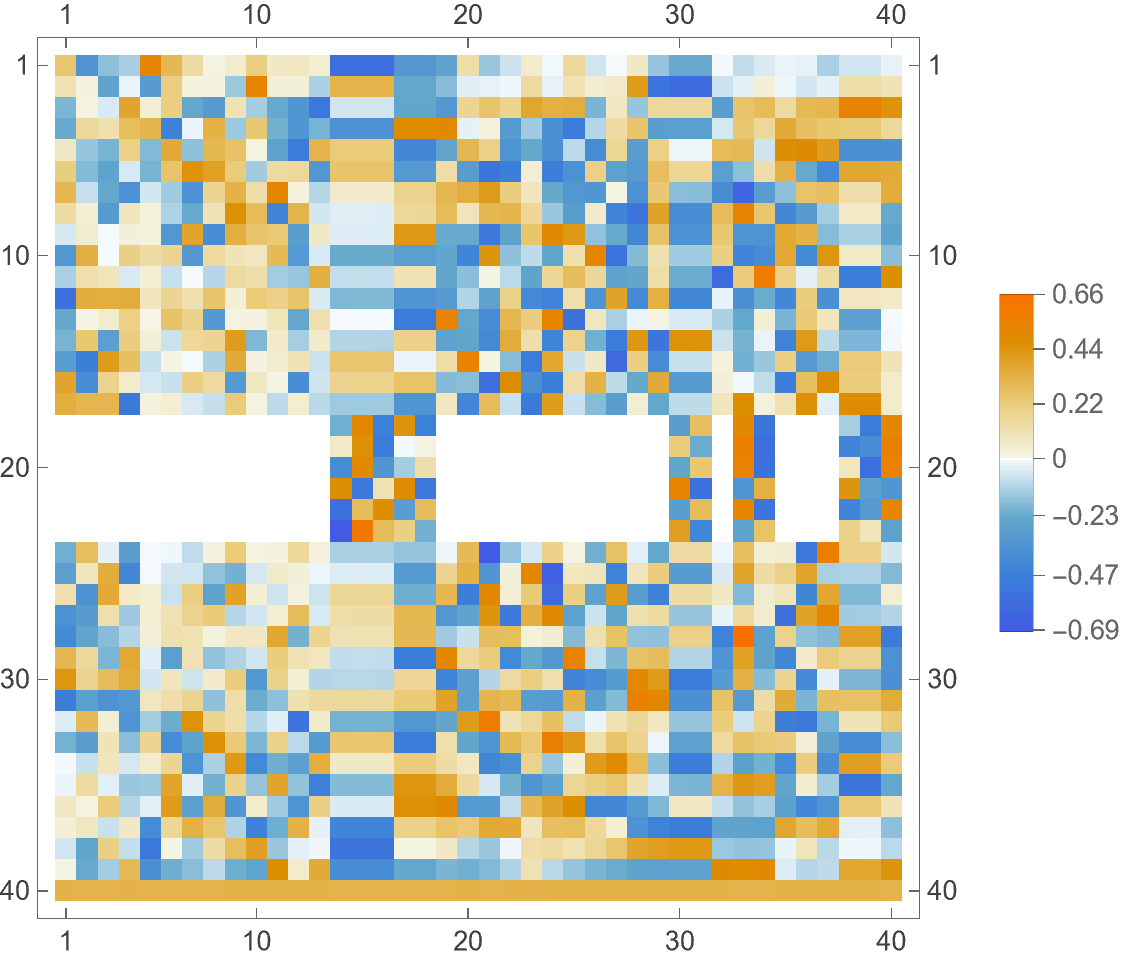} 
\includegraphics[angle=0,width=.5\textwidth]{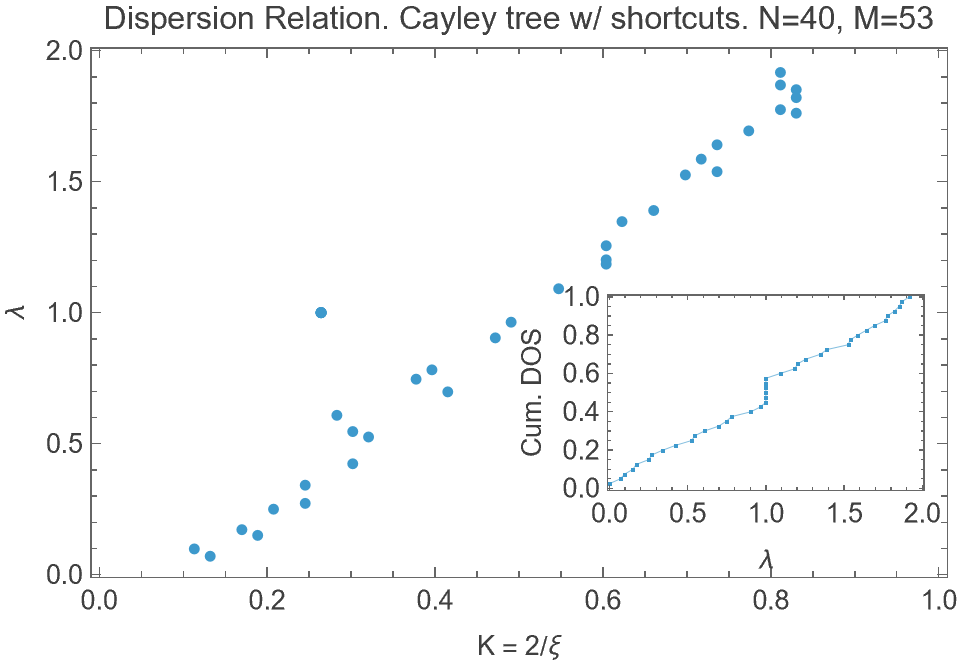} 
(d) \;\;\;\;\;\;\;\;\;\;\;\;\;\;\;\;\;\;\;\;\;\;\;\;\;\;\;\;\;\;\;\;\;\;\;\;\;\;\;\;\;\;\;\;\;\;\;\;\;\;\;\;\;\; (e)\\
\includegraphics[angle=0,width=.35\textwidth]{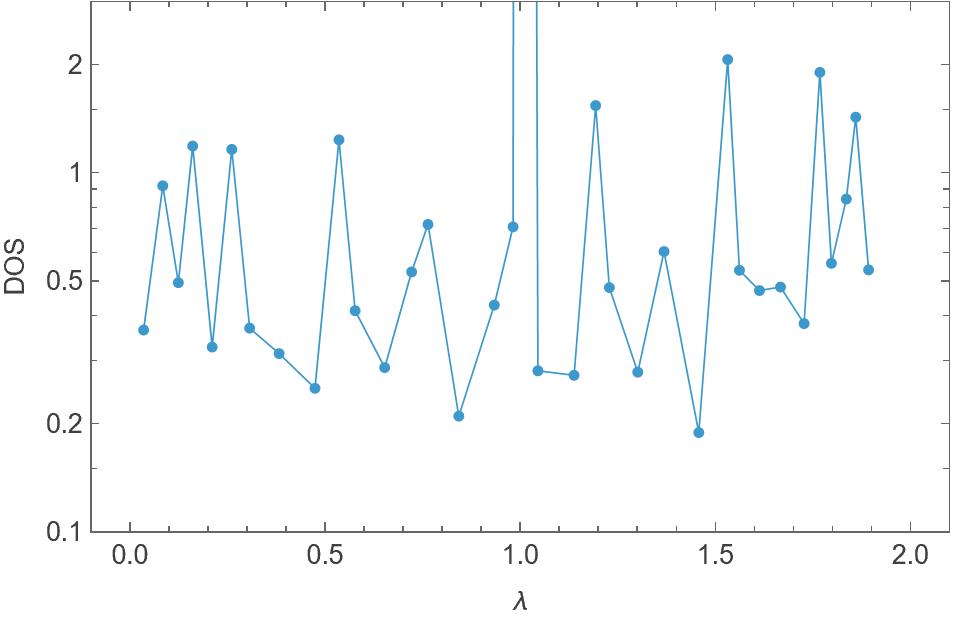} 
(f)
\end{center}
\vspace*{-0.5truecm}
\caption[]{
A small-world network \cite{WS98} generated from the Cayley tree graph with a 
branching ratio of 3 and 4 layers of Fig.~\ref{fig:disprelsT40} 
by adding 14 random shortcuts, $N=40$ and $M=53$. 
(a) $| \lambda_1 \rangle$ with $\lambda \approx 1.9168$ and 
$K = 43/53 \approx 0.8113$.
(b) $| \lambda_{21} \rangle$, 
representative of the 6 degenerate eigenvectors with $\lambda = 1$ and  
$K = 14/53 \approx 0.2642$.
(c) The Fiedler eigenvector  $| \lambda_{39} \rangle$, 
with $\lambda \approx 0.07028$ and $K = 7/53 \approx 0.1321$.
Only vertex 1 has exactly zero value in $| \lambda_{39} \rangle$. 
(d) Vertex values for the $N=40$ eigenvectors. 
The six degenerate eigenvectors from $| \lambda_{18} \rangle$ 
through $| \lambda_{23} \rangle$ 
with eigenvalues $\lambda = 1$ have vertex values 0 on vertices 1-13, 19-29, 32, 35-37, 
and varying positive and negative values on the rest. 
(e) Dispersion relation. 
Contiguous regions with $w_i = 0$ for $\lambda = 1$ leads to a depression of $K$. 
The inset shows the cumulative DOS for the eigenvalues.  
(f) Discrete density of states (DOS). 
The divergent peak corresponds to the six degenerate eigenvectors with $\lambda = 1$. 
}
\label{fig:disprelsT40L14}
\end{figure}

The configuration of $| \lambda_1 \rangle$ with $\lambda \approx 1.9168$ and 
$K = 43/53 \approx 0.8113$ is shown in Fig.~\ref{fig:disprelsT40L14}(a).
$| \lambda_{21} \rangle$, 
representative of the 6 degenerate eigenvectors with $\lambda = 1$ and  
$K = 14/53 \approx 0.2642$ is shown in Fig.~\ref{fig:disprelsT40L14}(b).
The Fiedler eigenvector  $| \lambda_{39} \rangle$, 
with $\lambda \approx 0.07028$ and $K = 7/53 \approx 0.1321$ is shown in 
Fig.~\ref{fig:disprelsT40L14}(c).
Only $v_1$ has exactly zero value in $| \lambda_{39} \rangle$. 
Vertex values for all the $N=40$ eigenvectors are shown in Fig.~\ref{fig:disprelsT40L14}(d). 
The six degenerate eigenvectors from $| \lambda_{18} \rangle$ 
through $| \lambda_{23} \rangle$ 
with eigenvalues $\lambda = 1$ have values 0 on the original root $v_1$, 
vertices 5-13 (the original layer 3), and 19-29, 32, 35-37 (15 of the 27 vertices 
in the original layer 4), 
and varying positive and negative values on the rest. A few scattered zeroes are also 
found on the other eigenvectors.
The dispersion relation is shown in Fig.~\ref{fig:disprelsT40L14}(e).
The inset shows the cumulative DOS for the eigenvalues.  
Except for the degenerate states, the data points are scattered along an approximately 
straight line, strikingly different from the unmodified Cayley tree. 
The discrete density of states (DOS)  is shown in Fig.~\ref{fig:disprelsT40L14}(f). 
The divergent peak corresponds to the degenerate eigenvectors with $\lambda = 1$. 
The rest of the points describe a succession of local maxima and minima. 

\clearpage

\clearpage


\subsection{{C.~elegans nervous system}}
\label{Sec:CE277}

Next we consider an unweighted, undirected graph representing the $N=277$ 
neurons of the roundworm {\it Caenorhabditis elegans} as its vertices and their 
$M=1918$ connections as its edges. 
The data are extracted from the data sets contained in \cite{CE277} and 
discussed in \cite{KAIS06,BANE08,BANE09,ARNA18}. 
This is, to our knowledge, the first fully mapped neuronal network of any 
organism \cite{WHIT86}. 

\begin{figure}[h]
\begin{center}
\vspace*{-0.1truecm}
\includegraphics[angle=0,width=.22\textwidth]{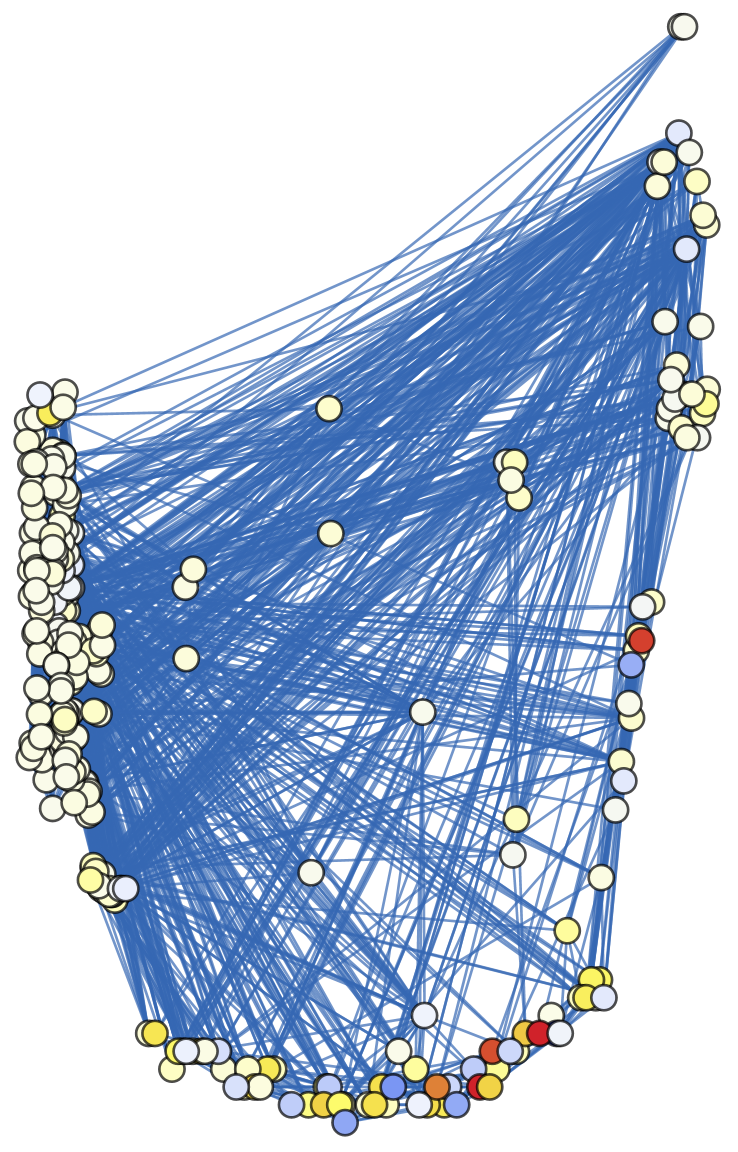}
\hspace*{1.5truecm}
\includegraphics[angle=0,width=.22\textwidth]{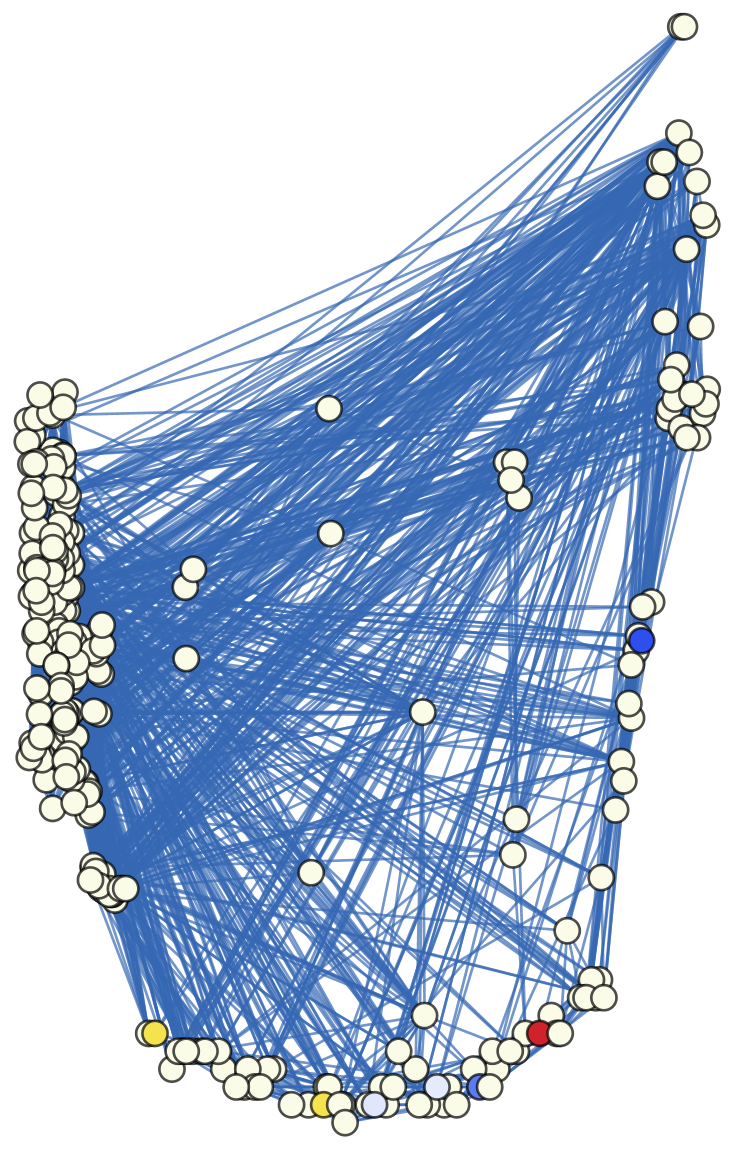}
\hspace{2.5truecm}
\includegraphics[angle=0,width=.22\textwidth]{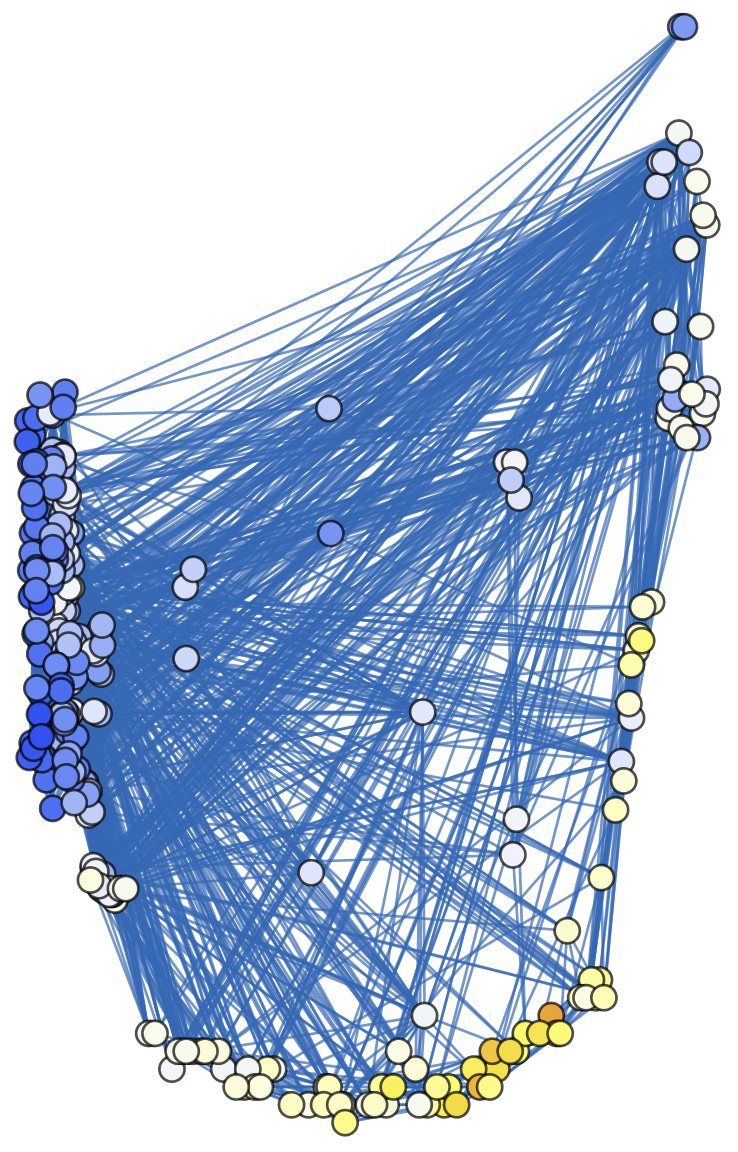}
\hspace*{-0.5truecm}
(a)\;\;\;\;\;\;\;\;\;\;\;\;\;\;\;\;\;\;\;\;\;\;\;\;\;\;\;\;\;\;\;\;\;\;\;\;\;\;\;\;\;\;\; (b) \;\;\;\;\;\;\;\;\;\;\;\;\;\;\;\;\;\;\;\;\;\;\;\;\;\;\;\;(c)
\hspace*{1.0truecm}
\includegraphics[angle=0,width=.4\textwidth]{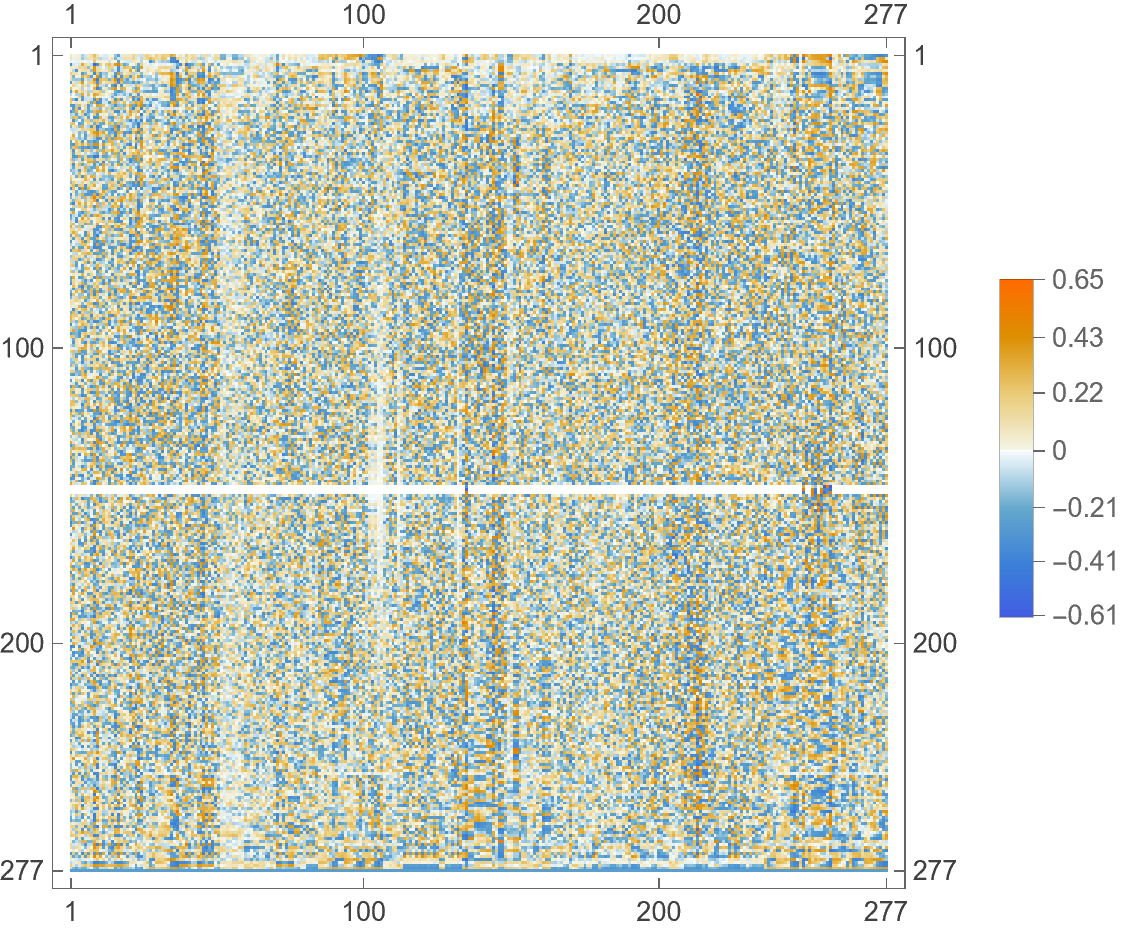} 
\includegraphics[angle=0,width=.45\textwidth]{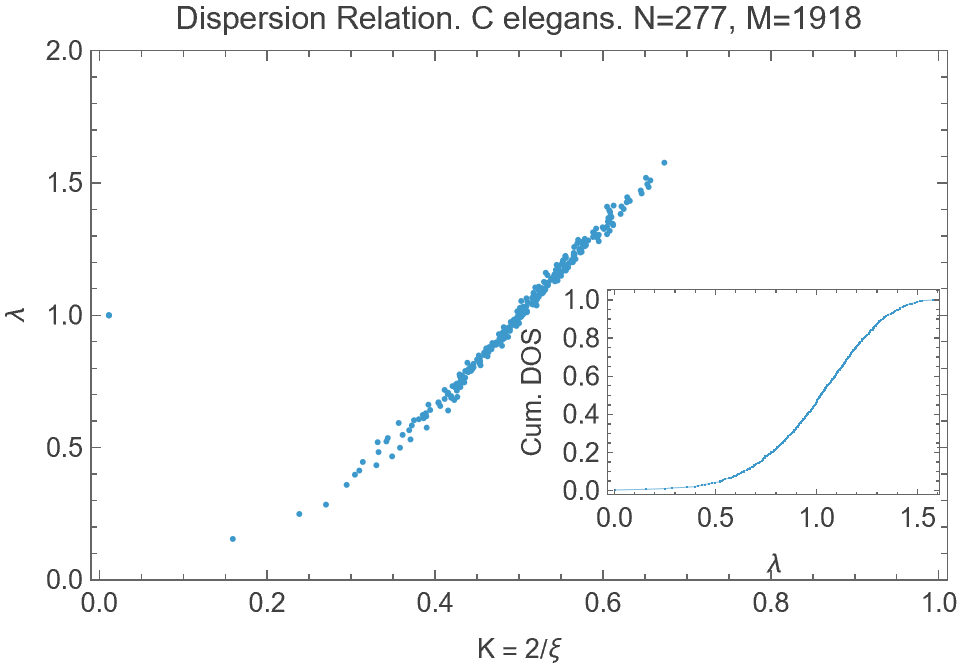} 
(d)\;\;\;\;\;\;\;\;\;\;\;\;\;\;\;\;\;\;\;\;\;\;\;\;\;\;\;\;\;\;\;\;\;\;\;\;\;\;\;\;\;\;\; (e)\\
\includegraphics[angle=0,width=.35\textwidth]{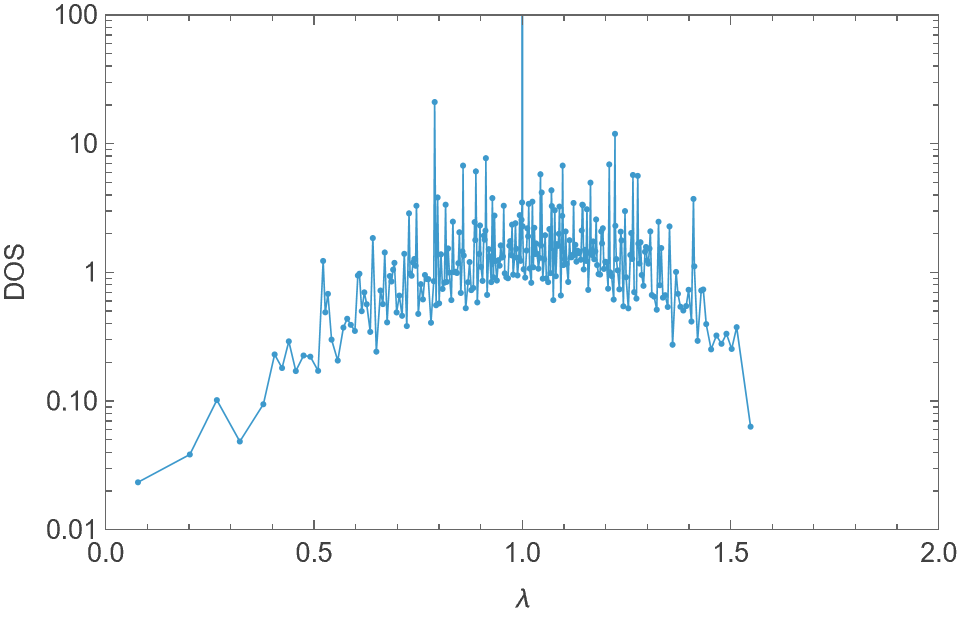} 
(f)
\end{center}
\vspace*{-0.5truecm}
\caption[]{
Neuronal network of the roundworm {\it C.~elegans}. Neurons are 
shown as vertices and connections as edges. $N=277$ and $M=1918$ 
\cite{KAIS06,BANE08,BANE09,ARNA18}.
Parts (a-c) are oriented as in Fig.~3(c) of \cite{KAIS06}, with the head to the left. 
The vertical coordinate is magnified by 20 for improved visibility. 
(a) The eigenvector  $| \lambda_{1} \rangle$ with 
$\lambda \approx 1.5768$ and $K = 1291/1918 \approx 0.6731$.
(b) $| \lambda_{148} \rangle$, representative of the 
three degenerate eigenvectors with $\lambda = 1$ and $K = 22/1918 \approx 0.01147$ that 
are characterized by large contiguous regions of vertices with 0 values. 
(c)  The Fiedler eigenvector  $| \lambda_{276} \rangle$ with 
$\lambda \approx 0.1548$  and $K = 305/1918 \approx 0.1590$.
(d) Vertex values for the $N=277$ eigenvectors. 
(e) Dispersion relation. 
The inset shows the cumulative DOS.  
(f) Density of states (DOS). 
The divergent peak corresponds to the three degenerate eigenvectors with $\lambda = 1$. 
}
\label{fig:disprelCelegans277}
\end{figure}

Projections of the network onto a vertical plane, oriented longitudinally with the body 
axis are shown in Fig.~\ref{fig:disprelCelegans277} (a--c). They are 
oriented as in Fig.~3(c) of Ref.~\cite{KAIS06}, 
with the head to the left and the abdomen down. 
The vertical coordinate is magnified by 20 for improved visibility of the long-distance connections. 
The global parameters of this graph are 
$\langle d \rangle \approx 13.8484$, $\rho = 0.05018$, $C=0.1981$, $D=6$, 
and $L \approx 2.6389$. 
The small values of  $D$ and $L$, with $L/L_{\rm rand} \approx 1.2332 < 1.5$,
qualify it as a small-world network. 

The configuration of  $| \lambda_{1} \rangle$ with 
$\lambda \approx 1.5768$ and $K = 1291/1918 \approx 0.6731$ 
is shown in Fig.~\ref{fig:disprelCelegans277}(a).
The eigenvector $| \lambda_{148} \rangle$, representative of the 
three degenerate eigenvectors with $\lambda = 1$ and $K = 22/1918 \approx 0.01147$ that 
are characterized by large, contiguous regions of vertices with value exactly 0, 
and consequently by a 
very small value of $K$,
is shown in Fig.~\ref{fig:disprelCelegans277}(b). 
The degenerate eigenvectors are strongly localized, containing only 8 vertices with nonzero value. 
These vertices have small degrees between 1 and 4, much smaller than the 
global $\langle d \rangle \approx 14$. This supports the notion that $\lambda = 1$ 
is related to duplication or addition of small motifs. 
The Fiedler eigenvector  $| \lambda_{276} \rangle$ with 
$\lambda \approx 0.1548$  and $K = 305/1918 \approx 0.1590$
is shown in Fig.~\ref{fig:disprelCelegans277}(c).
Vertex values for all the $N=277$ eigenvectors are shown 
in Fig.~\ref{fig:disprelCelegans277}(d).  
The dispersion relation is shown in Fig.~\ref{fig:disprelCelegans277}(e).
The inset shows the cumulative DOS for the eigenvalues. We notice that $\lambda_1$ 
is unusually low and $\lambda_{N-1}$ is unusually high. Also, the gap between 
$\lambda_{N-1}$ and $\lambda_{N-2}$ is quite large. 
The discrete density of states (DOS), is shown in Fig.~\ref{fig:disprelCelegans277}(f).
The divergent peak corresponds to the degenerate eigenvectors with $\lambda = 1$. 
We also note that the asymmetric shape of the DOS resembles that of a smoothed version of
analogous data, shown in Fig.~2(d) of Ref.~\cite{BANE09}. 

\clearpage

\clearpage

\subsection{St.~Mark's food web}
\label{Sec:SM48}
\begin{figure}[h]

\begin{center}
\vspace*{-0.1truecm}
\includegraphics[angle=0,width=.30\textwidth]{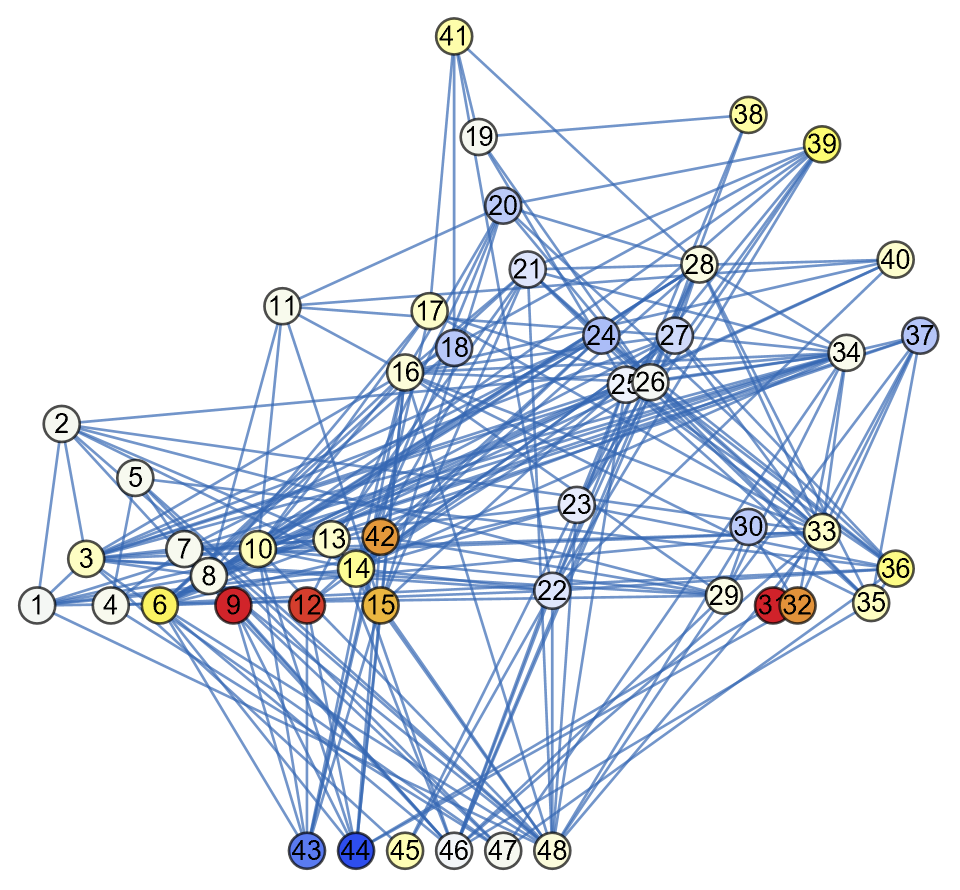}
\includegraphics[angle=0,width=.30\textwidth]{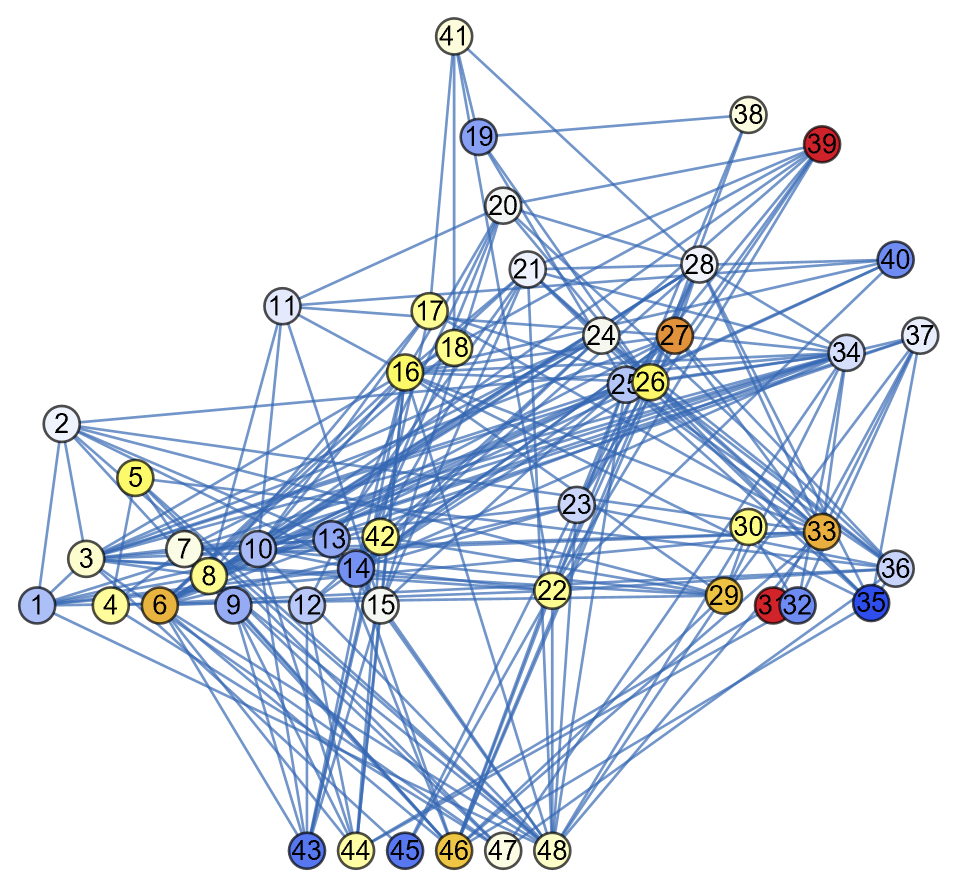}
\includegraphics[angle=0,width=.30\textwidth]{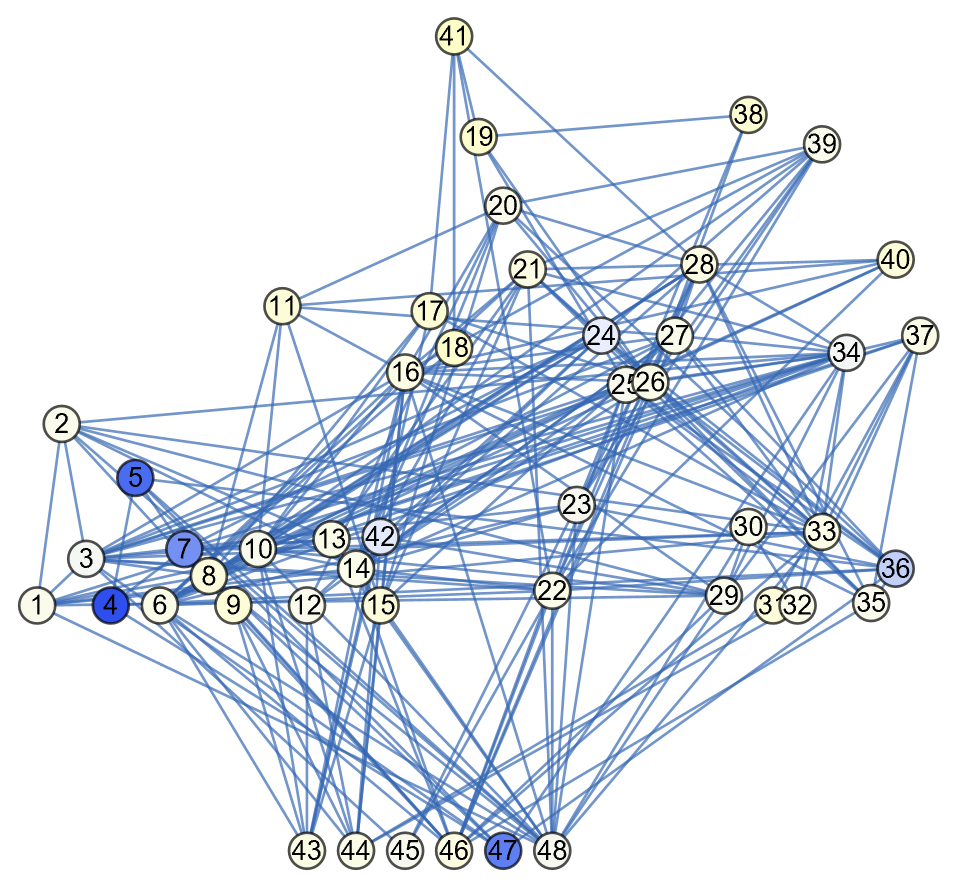}
(a)\;\;\;\;\;\;\;\;\;\;\;\;\;\;\;\;\;\;\;\;\;\;\;\;\;\;\;\;\;\;\;\;\;\;\;\;\;\;\;\;\;\;\; (b) \;\;\;\;\;\;\;\;\;\;\;\;\;\;\;\;\;\;\;\;\;\;\;\;\;\;\;\;(c)
\hspace*{1.0truecm}
\includegraphics[angle=0,width=.4\textwidth]{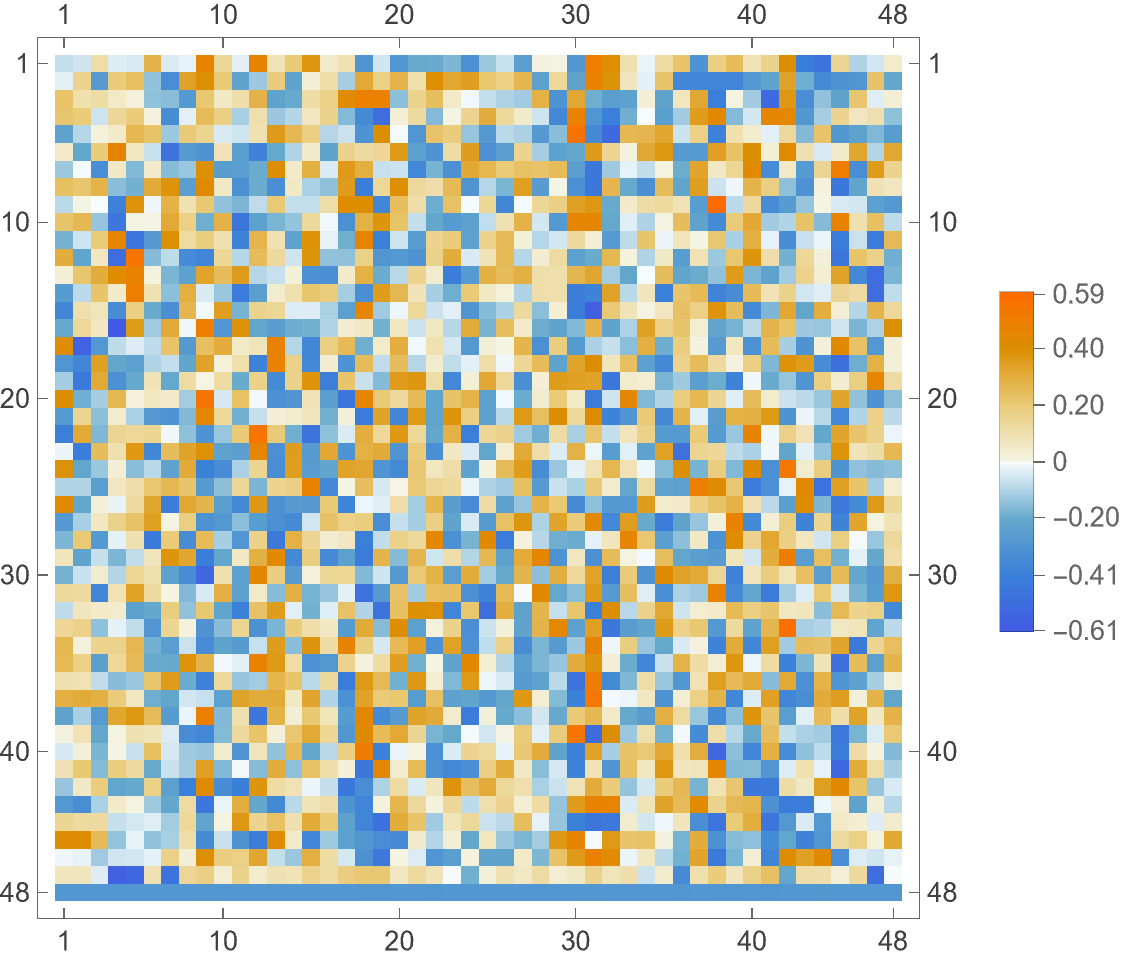} 
\includegraphics[angle=0,width=.45\textwidth]{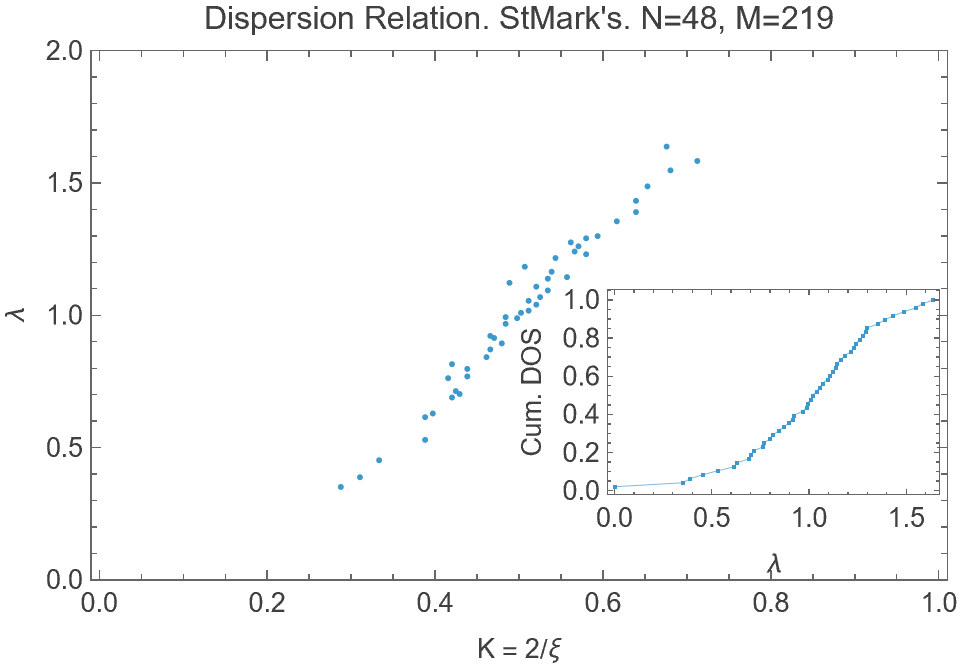} 
(d)\;\;\;\;\;\;\;\;\;\;\;\;\;\;\;\;\;\;\;\;\;\;\;\;\;\;\;\;\;\;\;\;\;\;\;\;\;\;\;\;\;\;\; (e)\\
\includegraphics[angle=0,width=.35\textwidth]{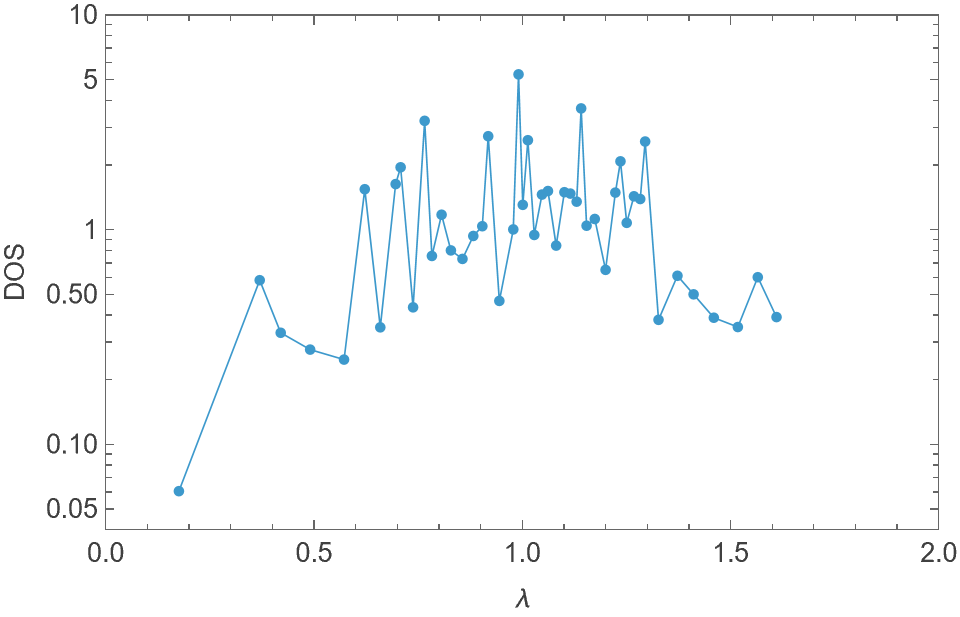} 
(f)
\end{center}
\vspace*{-0.5truecm}
\caption[]{
Undirected, unweighted graph generated from the directed, weighted graph representing a 
food web in the St.~Mark's wetlands on the Gulf coast of Florida, USA.
Trophic species are shown as vertices and prey-predator connections as edges. 
$N=48$ and $M=219$.
The vertical axes in (a--c) represent effective trophic levels \cite{CHRI99}. 
(a) The eigenvector  $| \lambda_{1} \rangle$ with 
$\lambda \approx 1.6375$ and $K = 148/219 = 0.6758$.
(b) The eigenvector $| \lambda_{27} \rangle$ with $\lambda = 0.993058$, 
which also forms the 
most closely degenerate pair with  $| \lambda_{28} \rangle$ with 
$\lambda = 0.989033$. 
(c)  The Fiedler eigenvector  $| \lambda_{47} \rangle$ with 
$\lambda \approx 0.3514$ and $K = 63/219 = 0.2877$.
(d) Vertex values for the $N=48$ eigenvectors. 
(e) Dispersion relation.
The inset shows the cumulative DOS for the eigenvalues.  
(f) Discrete density of states (DOS). 
The maximum near $\lambda = 1$ corresponds to the nearly degenerate pair, 
$| \lambda_{27} \rangle$ and  $| \lambda_{28} \rangle$.
}
\label{fig:disprelStMarks}
\end{figure}

Next we consider an 
undirected, unweighted graph generated from the directed, weighted graph 
representing a food web in the St.~Mark's wetlands on the Gulf coast of Florida, USA. 
The original data are from Table~2 of Ref.~\cite{CHRI99}.
Trophic species, which are groups of species that share all their prey and predators 
\cite{WILL02,PAR07}, are shown as vertices and prey-predator connections as edges. 
$N=48$ and $M=219$.
The vertical axes in Fig.~\ref{fig:disprelStMarks} (a--c) represent effective trophic levels, 
from 1 for primary producers and detritus to 4.5 for the top predator \cite{CHRI99}. 
The global parameters of this food web are 
$\langle d \rangle = 9.125$, $\rho = 0.194149$, $C \approx0.2946$, $D=4$,
 and $L \approx 2.0860$, 
 with $L/L_{\rm rand} \approx 1.1914 < 1.5$, 
which clearly qualify it as a small-world network. This value of $L$ also lies within 
one standard deviation of the mean for the seven food webs 
considered in Ref.~\cite{WILL02}. 

The eigenvector  $| \lambda_{1} \rangle$ with 
$\lambda \approx 1.6375$ and $K = 148/219 \approx 0.6758$ is shown in 
Fig.~\ref{fig:disprelStMarks}(a).
The eigenvector $| \lambda_{27} \rangle$ with $\lambda \approx 0.9931$ 
and $K = 106/219 \approx 0.4840$, which also forms the 
most closely degenerate pair with  $| \lambda_{28} \rangle$ with 
$\lambda \approx 0.9890$ and $K = 109/219 \approx 0.4977$ 
is shown in Fig.~\ref{fig:disprelStMarks}(b). 
The Fiedler eigenvector  $| \lambda_{47} \rangle$ with 
$\lambda \approx 0.3514$ and $K = 63/219 \approx 0.2877$
is shown in Fig.~\ref{fig:disprelStMarks}(c).
Vertex values for the $N=47$ eigenvectors 
are  shown in Fig.~\ref{fig:disprelStMarks}(d). 
The dispersion relation is shown in Fig.~\ref{fig:disprelStMarks}(e).
The inset shows the cumulative DOS for the eigenvalues.  
The ranges occupied by the nonzero values of $\lambda$ and $K$ 
are even smaller than those seen in Fig.~\ref{fig:disprelCelegans277}(e).
The discrete density of states (DOS)  is shown in Fig.~\ref{fig:disprelStMarks}(f). 
The global maximum near $\lambda = 1$ corresponds to the nearly degenerate pair, 
$| \lambda_{27} \rangle$ and $| \lambda_{28} \rangle$. 
This graph has no vertices with value exactly zero. 
\clearpage

\clearpage

\subsection{Dolphin community}
\label{Sec:Dol62}
\begin{figure}[h]
\begin{center}
\vspace*{-0.1truecm}
\includegraphics[angle=0,width=.32\textwidth]{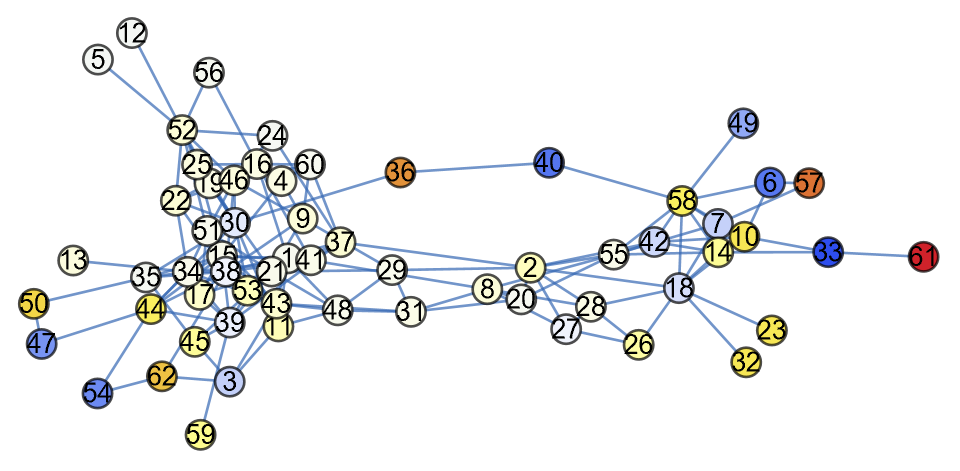}
\includegraphics[angle=0,width=.32\textwidth]{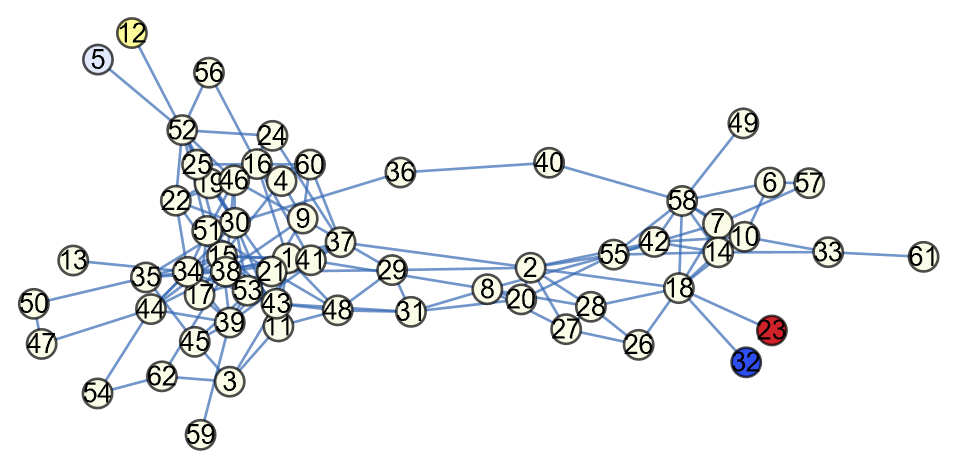}
\includegraphics[angle=0,width=.32\textwidth]{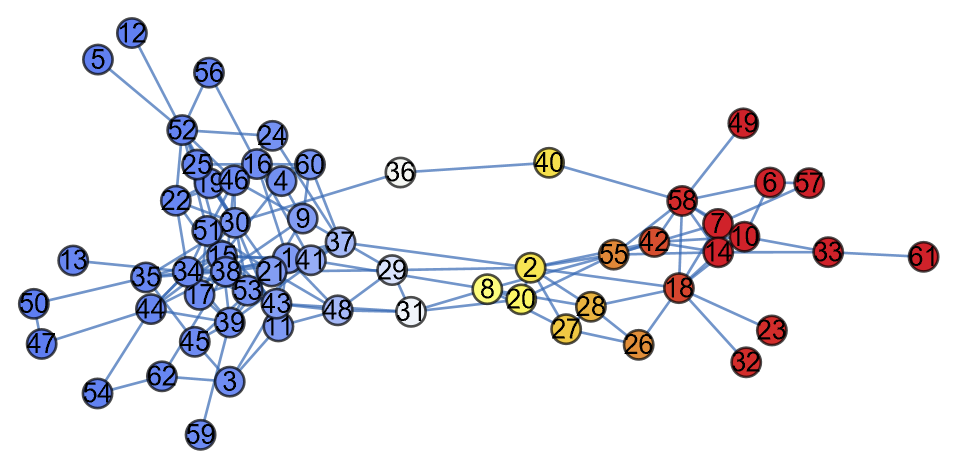}
(a)\;\;\;\;\;\;\;\;\;\;\;\;\;\;\;\;\;\;\;\;\;\;\;\;\;\;\;\;\;\;\;\;\;\;\;\;\;\;\;\;\;\;\; (b) \;\;\;\;\;\;\;\;\;\;\;\;\;\;\;\;\;\;\;\;\;\;\;\;\;\;\;\;(c)
\hspace*{1.0truecm}
\includegraphics[angle=0,width=.4\textwidth]{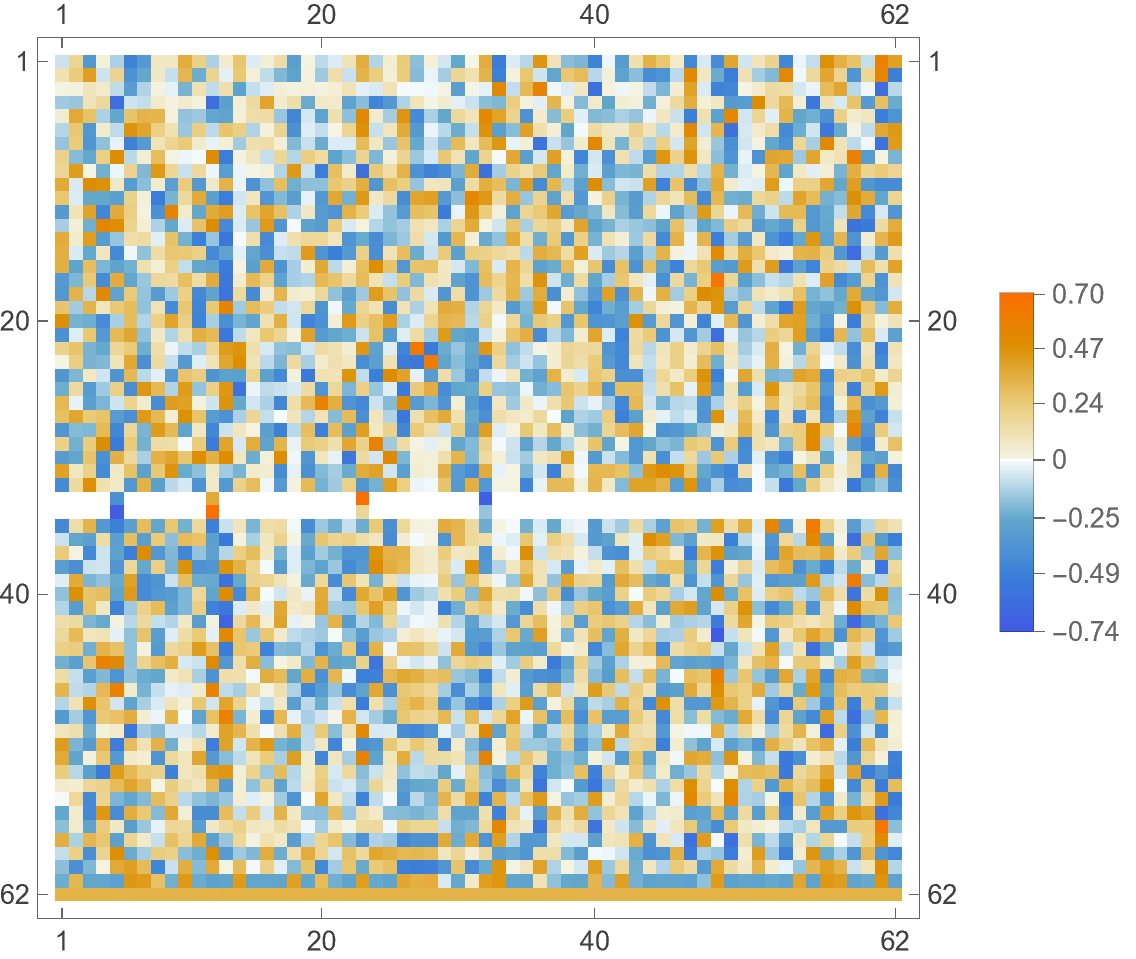} 
\includegraphics[angle=0,width=.45\textwidth]{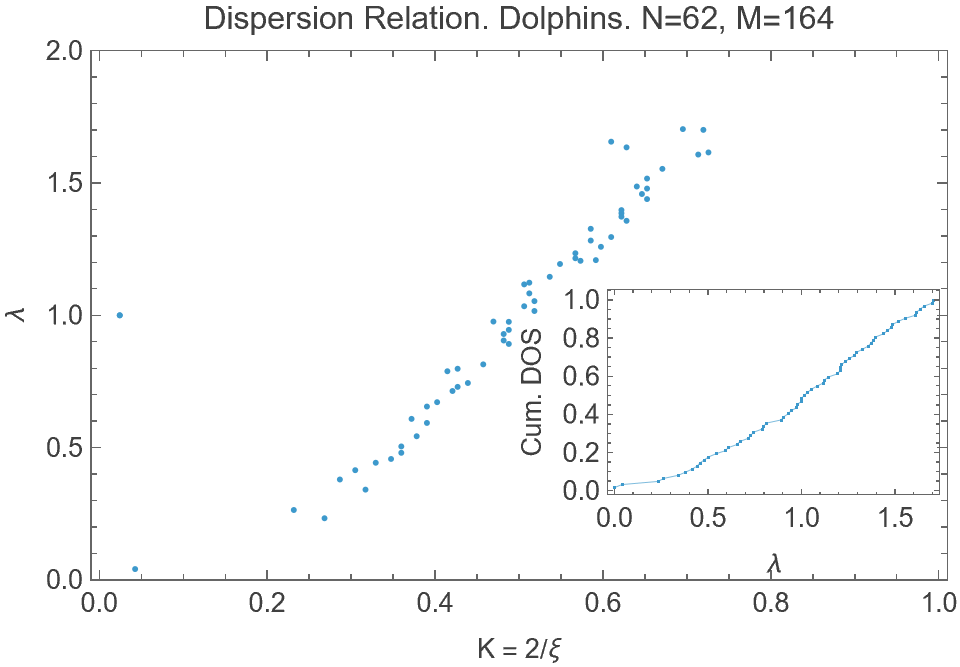} 
(d)\;\;\;\;\;\;\;\;\;\;\;\;\;\;\;\;\;\;\;\;\;\;\;\;\;\;\;\;\;\;\;\;\;\;\;\;\;\;\;\;\;\;\; (e)\\
\includegraphics[angle=0,width=.35\textwidth]{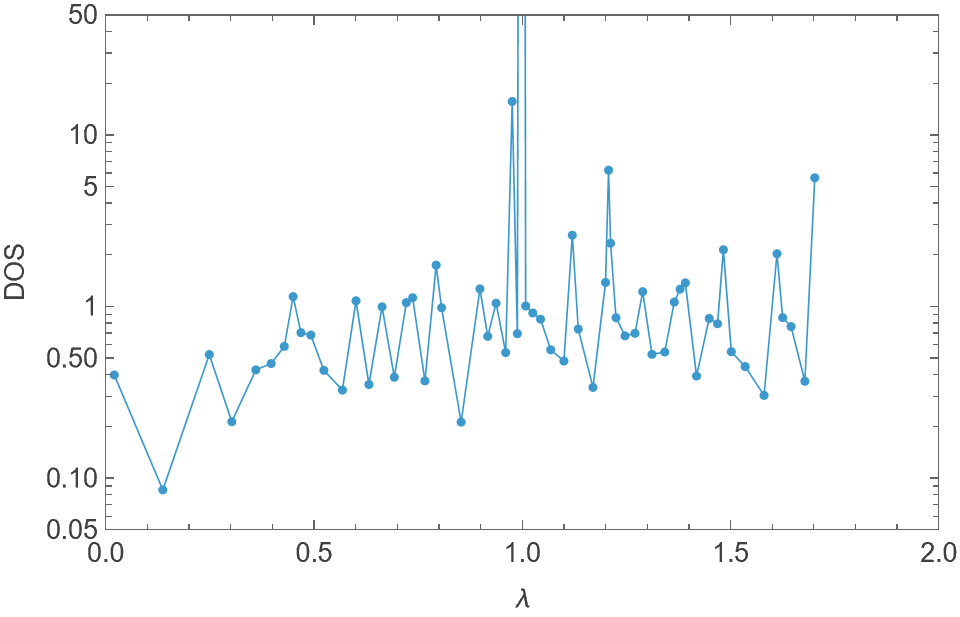} 
(f)
\end{center}
\vspace*{-0.5truecm}
\caption[]{
Undirected, unweighted graph representing long-lasting social relationships in a 
community of dolphins. 
$N=62$ and $M=164$ \cite{LUSS03B,LUSS03,DNOTE}. 
(a) The eigenvector  $| \lambda_{1} \rangle$ with 
$\lambda \approx 1.7037$ and $K = 114/164 \approx 0.6951$.
(b) The eigenvector $| \lambda_{33} \rangle$ with $\lambda = 1$
 and $K = 4/164 \approx 0.02439$, which forms a 
degenerate pair with  $| \lambda_{34} \rangle$ with the same $K$. 
(c)  The Fiedler eigenvector  $| \lambda_{61} \rangle$ with 
$\lambda \approx 0.04109$ and $K = 7/164 \approx 0.04268$.
(d) Vertex values for the $N=62$ eigenvectors. 
(e) Dispersion relation.
The inset shows the cumulative DOS for the eigenvalues.  
(f) Density of states (DOS). 
The divergence at $\lambda = 1$ corresponds to the degenerate pair.
}
\label{fig:disprelDolphins}
\end{figure}

As an example of a simple social network, we next consider 
an undirected, unweighted graph representing long-lasting social relationships in a 
community of bottlenose dolphins living off Doubtful Sound, New Zealand. 
$N=62$ and $M=164$ \cite{LUSS03B,LUSS03,DNOTE}. 
The global parameters of this social network are 
$\langle d \rangle = 5.2903$, $\rho = 0.08673$, $C \approx0.3036$, $D=8$, 
and $L \approx 3.2893$, with $L/L_{\rm rand} \approx 1.3277 < 1.5$, 
which qualify it as a small-world network. 
The organization of this network is shown in a spring-electric layout in 
Fig.~\ref{fig:disprelDolphins} (a-c). Visual inspection suggests that this may show 
an early stage of the merging of two groups of similar size. A thorough discussion of the 
group structure of this community is given in Ref.~\cite{LUSS03}. 

The eigenvector  $| \lambda_{1} \rangle$ with 
$\lambda \approx 1.7037$ and $K = 114/164 \approx 0.6951$
is shown in Fig.~\ref{fig:disprelDolphins}(a).
The eigenvector $| \lambda_{33} \rangle$ with $\lambda = 1$
 and $K = 4/164 \approx 0.02439$ is shown in Fig.~\ref{fig:disprelDolphins}(b). 
 As the figure shows, it is strongly localized on vertices 23 and 32. This eigenvector forms a 
degenerate pair with  $| \lambda_{34} \rangle$ with the same $K$, localized on 
vertices 5 and 12. All the other vertex values in these two eigenvectors are 
exactly zero. This graph shows a relatively simple example of the ways eigenvectors with 
$\lambda =1$ can be generated by addition or duplication of small motifs. 
Many additional examples are included as theorems in Ref.~\cite{BANE08B}. 
The Fiedler eigenvector  $| \lambda_{61} \rangle$ with 
$\lambda \approx 0.04109$ and $K = 7/164 \approx 0.04268$ 
is shown in Fig.~\ref{fig:disprelDolphins}(c). The two above-mentioned groups are connected by just seven edges.
(d) Vertex values for the $N=62$ eigenvectors are shown 
in Fig.~\ref{fig:disprelDolphins}(d).
The dispersion relation is shown in Fig.~\ref{fig:disprelDolphins}(e).
The inset shows the cumulative DOS for the eigenvalues.  
Density of states (DOS) is shown 
on a logarithmic scale in Fig.~\ref{fig:disprelDolphins}(f). 
The divergence at $\lambda = 1$ corresponds to the degenerate pair.
The gap, $\lambda_{N-1} - \lambda _{N-2}$, is quite large and corresponds to the 
global minimum in the DOS. 
\clearpage

\clearpage

\subsection{Florida power grid}
\label{Sec:FLpow}
As an example of a geographically constrained, technological transport system, 
we next consider an 
undirected, unweighted graph generated from the undirected, weighted graph representing the 
high-voltage power grid in the state of Florida, USA.
$N=84$ and $M=137$. The coordinates of the graph figures give the geographical locations of 
vertices representing power plants and distribution stations. The raw data, 
describing the network around the year 2008, were obtained from 
Ref.~\cite{FLmap}.  
This network has previously been studied by spectral and simulational methods 
\cite{ABOU10B,ABOU11,Rikvold,XU14}. 
The global parameters are 
$\langle d \rangle = 3.2619$, $\rho = 0.0393$, $C \approx 0.2112$, $D=13$, and 
$L \approx 5.1394$, with $L/L_{\rm rand} \approx 1.3714 < 1.5$, which qualify it as a 
marginal small-world network by our definition. 

\begin{figure}[h]
\begin{center}
\vspace*{-0.1truecm}
\includegraphics[angle=0,width=.25\textwidth]{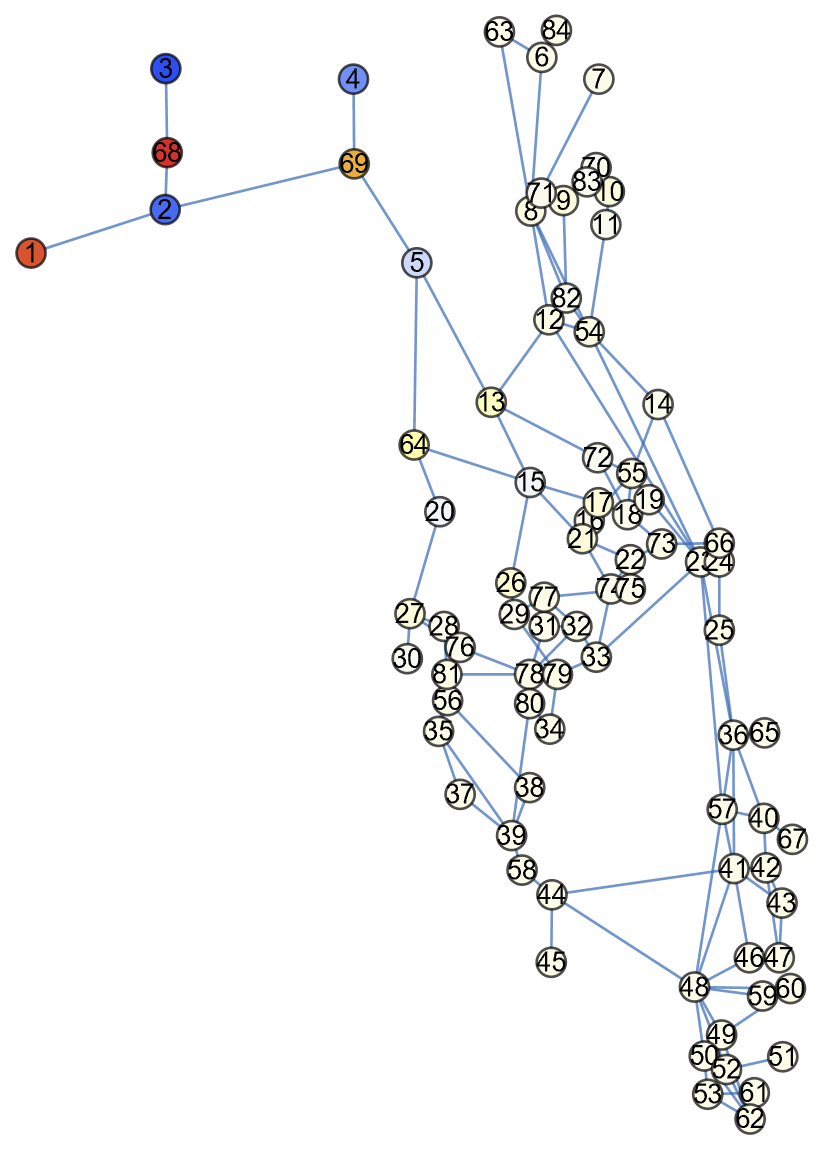}
\includegraphics[angle=0,width=.25\textwidth]{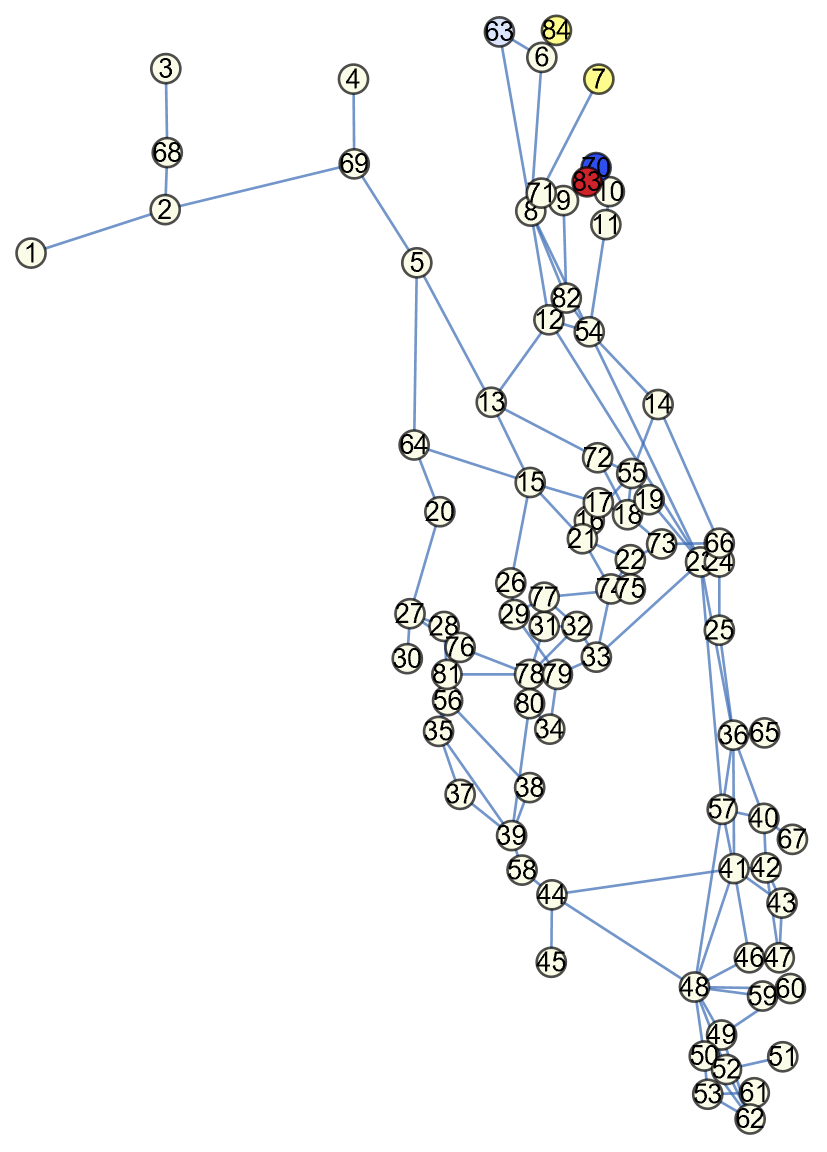}
\includegraphics[angle=0,width=.25\textwidth]{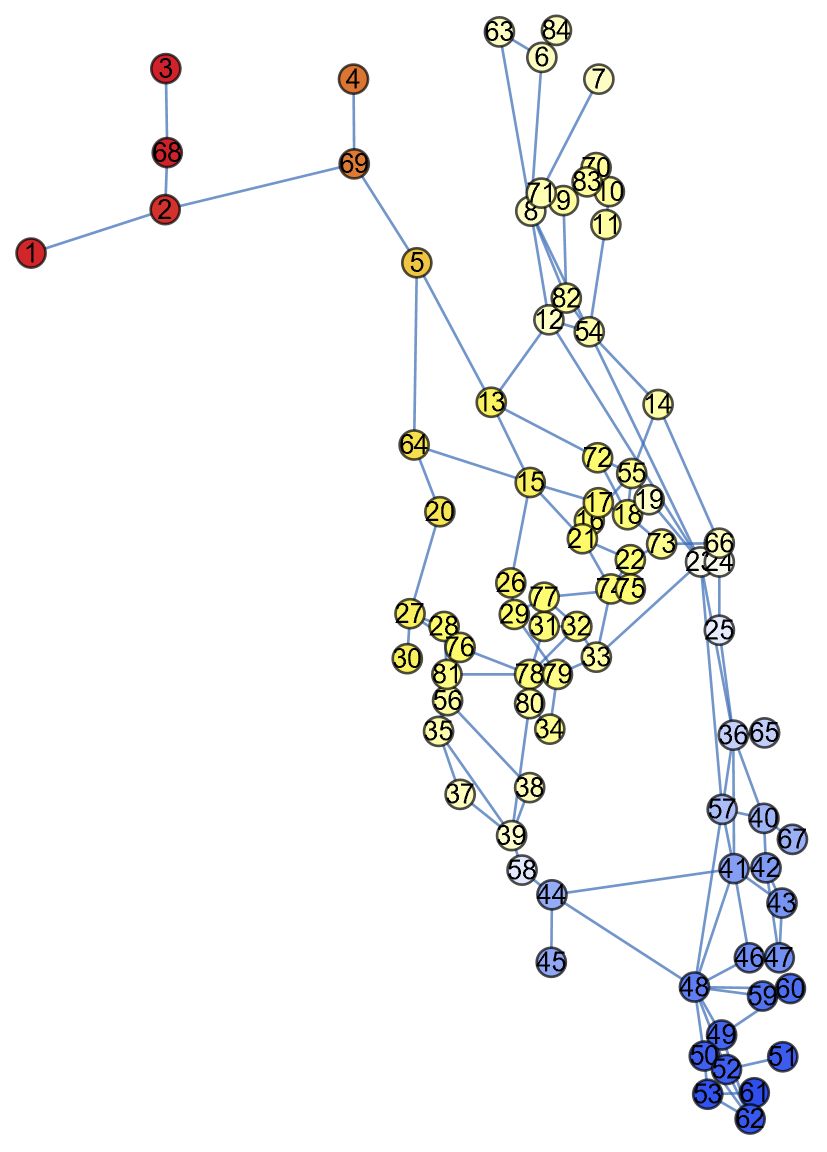}
(a)\;\;\;\;\;\;\;\;\;\;\;\;\;\;\;\;\;\;\;\;\;\;\;\;\;\;\;\;\;\;\;\;\;\;\;\;\;\;\;\;\;\;\; (b) \;\;\;\;\;\;\;\;\;\;\;\;\;\;\;\;\;\;\;\;\;\;\;\;\;\;\;\;(c)
\hspace*{1.0truecm}
\includegraphics[angle=0,width=.4\textwidth]{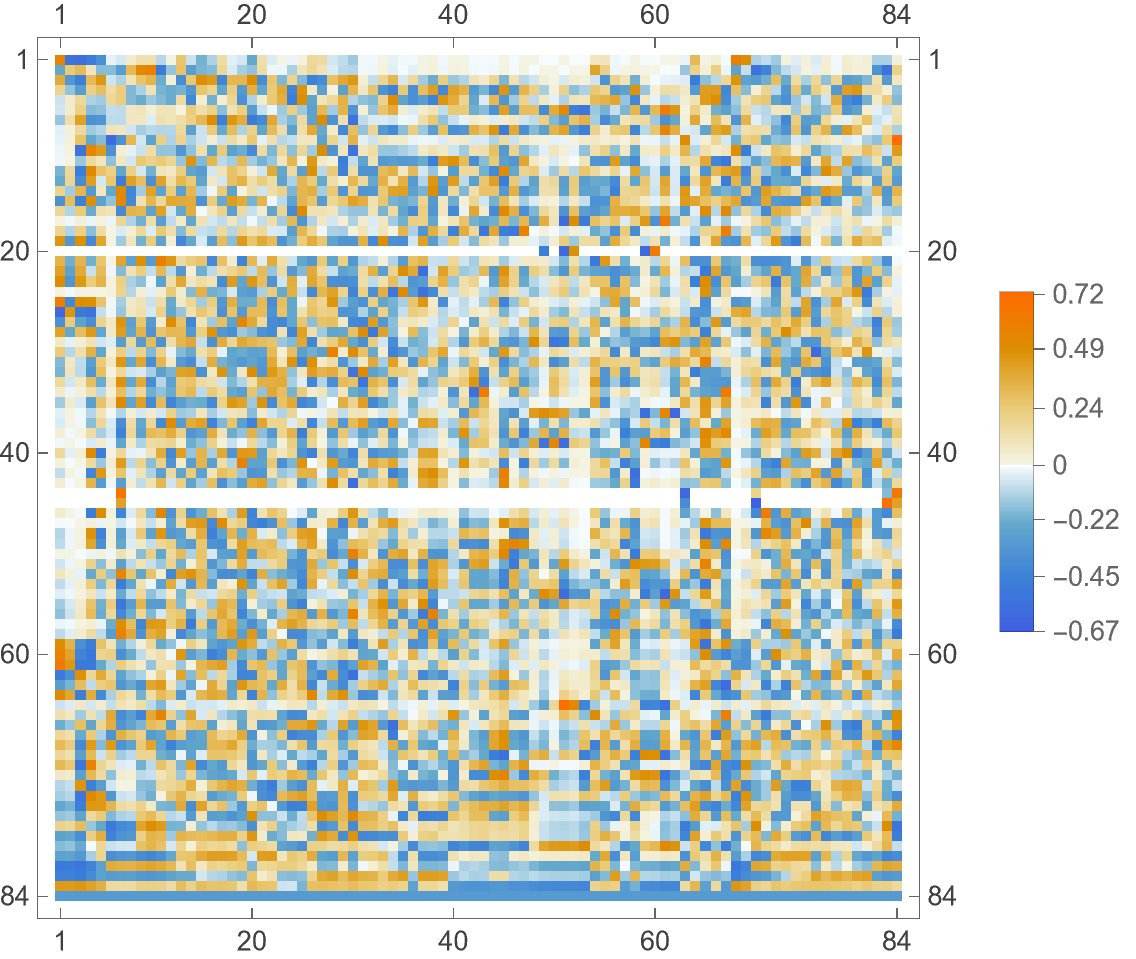} 
\includegraphics[angle=0,width=.45\textwidth]{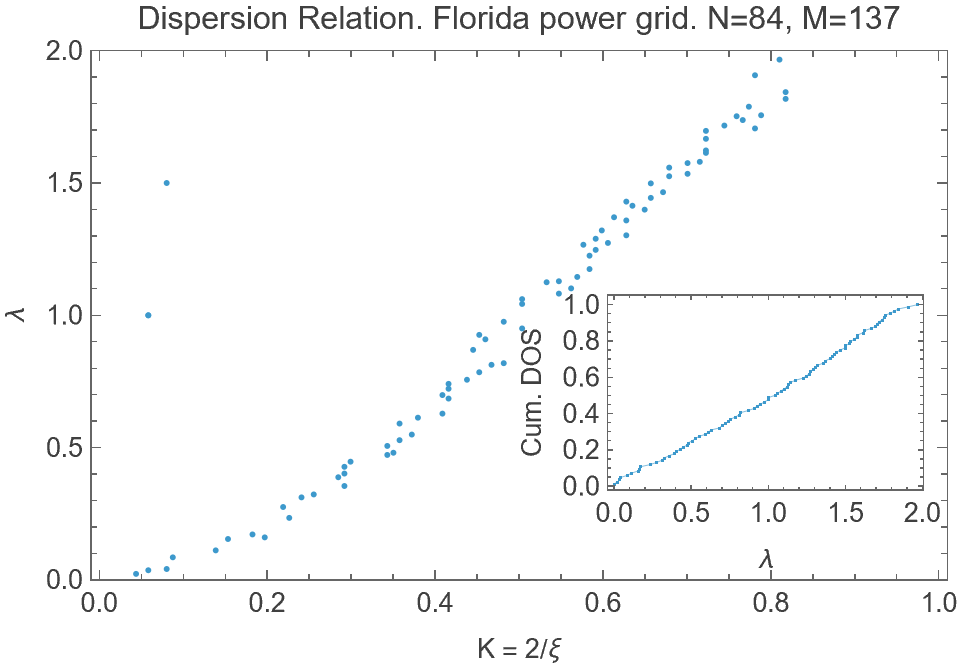} 
(d)\;\;\;\;\;\;\;\;\;\;\;\;\;\;\;\;\;\;\;\;\;\;\;\;\;\;\;\;\;\;\;\;\;\;\;\;\;\;\;\;\;\;\; (e)\\
\includegraphics[angle=0,width=.35\textwidth]{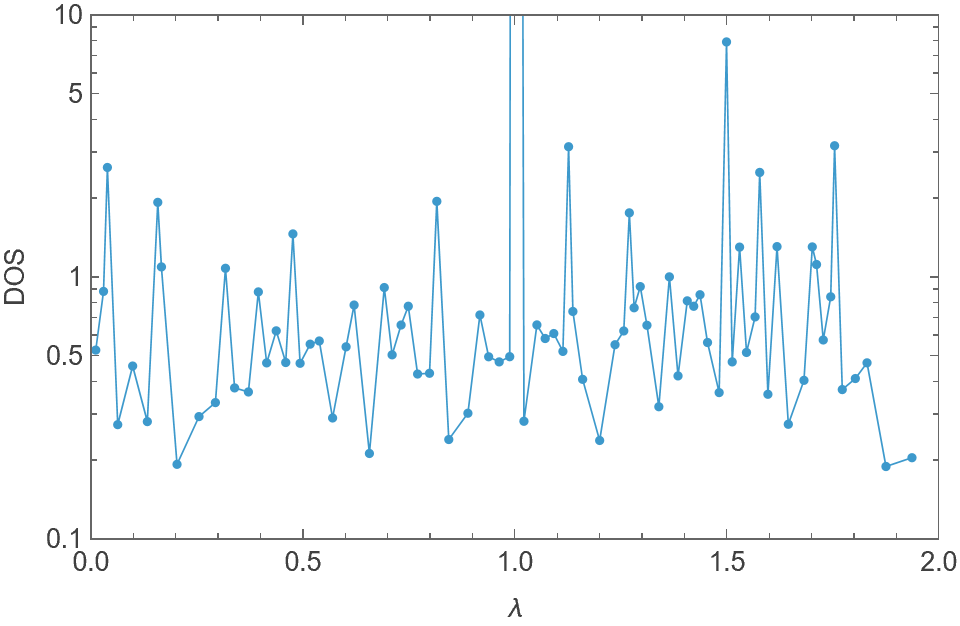} 
(f)
\end{center}
\vspace*{-0.5truecm}
\caption[]{
Undirected, unweighted graph generated from the undirected, weighted graph representing the 
high-voltage power grid in the state of Florida, USA.
$N=84$ and $M=137$. The coordinates of the graph figures give the geographical locations of 
vertices representing power plants and distribution stations 
\cite{ABOU10B,ABOU11,Rikvold,XU14}. 
(a) The eigenvector  $| \lambda_{1} \rangle$ with 
$\lambda \approx 1.9662$ and $K = 111/137 \approx 0.8102$.
(b) The eigenvector $| \lambda_{45} \rangle$ with $\lambda = 1$
 and $K = 8/137 \approx 0.05839$, which forms a 
degenerate pair with  $| \lambda_{44} \rangle$ that also has the same $K$. 
(c)  The Fiedler eigenvector  $| \lambda_{83} \rangle$ with 
$\lambda \approx 0.02292$ and $K = 6/137 \approx 0.04380$.
(d) Vertex values for the $N=84$ eigenvectors. 
Eigenvector $| \lambda_{20} \rangle$ with five nonzero vertices 
has $\lambda = 3/2$ and $K = 11/137 \approx 0.0829$.  
(e) Dispersion relation. The inset shows the cumulative DOS.   
(f) Density of states (DOS). 
The divergence at $\lambda = 1$ corresponds to the degenerate pair.
}
\label{fig:disprelFLgrid}
\end{figure}

The eigenvector  $| \lambda_{1} \rangle$ with 
$\lambda \approx 1.9662$ and $K = 111/137 \approx 0.8102$ 
is shown in Fig.~\ref{fig:disprelFLgrid}(a).
As it may not be obvious from the color-coded vertices, this eigenvector has no vertices with 
value zero. While dominated by vertices in the north-western corner, the rest of the 
vertices bear small, nonzero values that are arranged in an approximately 
bipartite configuration with $K \approx 81 \%$ broken bonds. 
In Fig.~\ref{fig:disprelFLgrid}(b) we show $| \lambda_{45} \rangle$ with $\lambda = 1$
 and $K = 8/137 \approx 0.05839$. It forms a 
degenerate pair with  $| \lambda_{44} \rangle$ that also has the same $K$.
Both these eigenvectors contain only the same five nonzero vertices: 7, 63, 70, 83, and 84 
in the north-eastern corner of the grid.   
However, $| \lambda_{45} \rangle$ is dominated by vertices 70 and 83, while 
$| \lambda_{44} \rangle$  is dominated by vertices 7, 63, and 84. 
Such eigenfunctions with $\lambda = 1$ and large contiguous regions of vertices 
with value zero are often formed by duplication or addition 
of small motifs \cite{BANE08B,BANE09}. 
Indeed, several of these vertices in the northern part of the grid have connections to the 
the grids of other states, which are not included in this graph. 
It is also worth noting one more eigenvector with an exact eigenvalue: $| \lambda_{20} \rangle$
with $\lambda = 3/2$ and $K = 11/137 \approx 0.08029$. 
Its only nonzero vertices are 49, 51, 52, 59, and 60, all 
near the southern tip of the grid. 
It is almost degenerate with the more strongly extended 
$| \lambda_{21} \rangle$with $\lambda \approx 1.4985$.

The Fiedler eigenvector  $| \lambda_{83} \rangle$ with 
$\lambda \approx 0.02292$ and $K = 6/137 \approx 0.04380$ 
is shown in Fig.~\ref{fig:disprelFLgrid}(c). It bisects the network into a northern and a southern 
part of similar size. 
Vertex values for the $N=84$ eigenvectors are shown 
in Fig.~\ref{fig:disprelFLgrid}(d). 
The dispersion relation is shown in Fig.~\ref{fig:disprelFLgrid}(e). 
The inset shows the cumulative DOS.  The density of states (DOS) is shown 
in Fig.~\ref{fig:disprelFLgrid}(f). 
The divergence at $\lambda = 1$ corresponds to the degenerate pair, while the highest of 
the nondivergent peaks corresponds to the almost degenerate pair. 
\clearpage
\clearpage

\subsection{Pore network from random sphere pack}
\label{sec:IV}

\begin{figure}[h]
\begin{center}
\vspace*{-0.1truecm}
\includegraphics[angle=0,width=.29\textwidth]{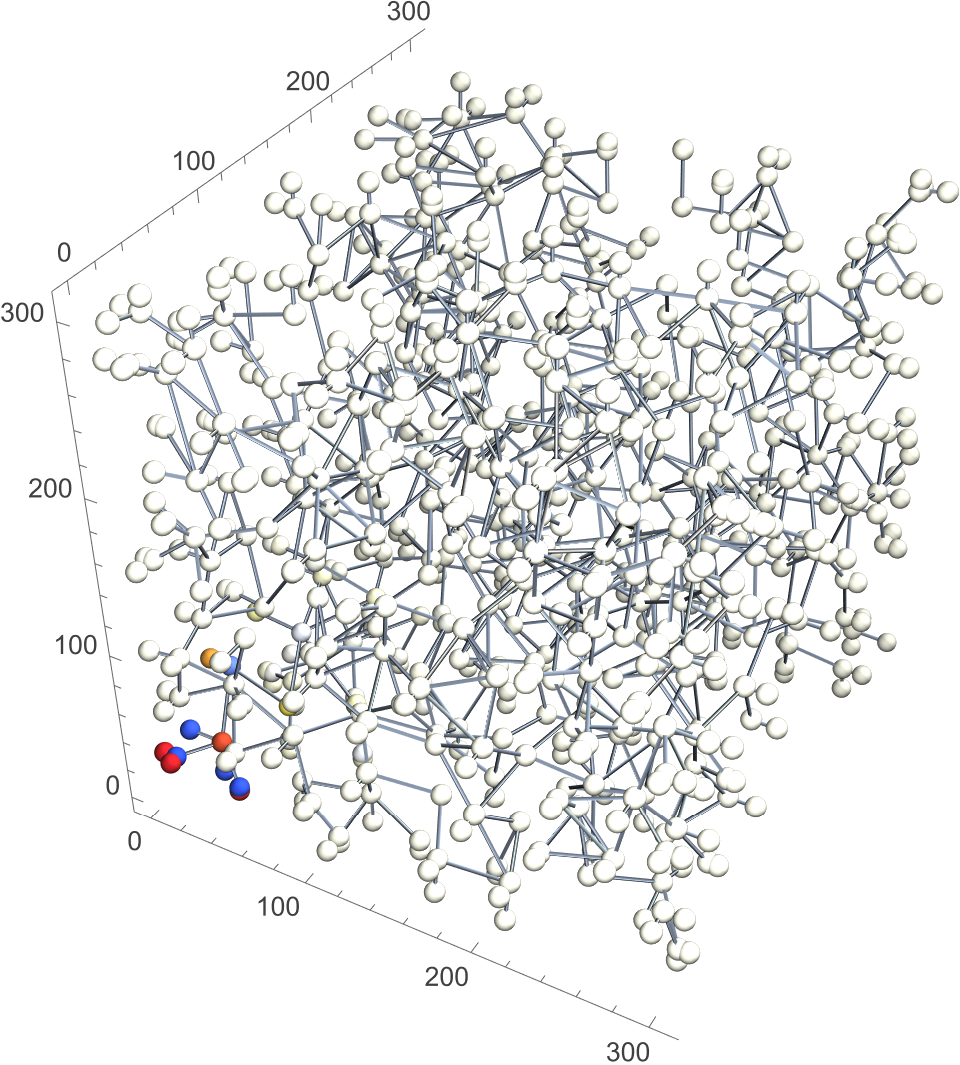}
\includegraphics[angle=0,width=.29\textwidth]{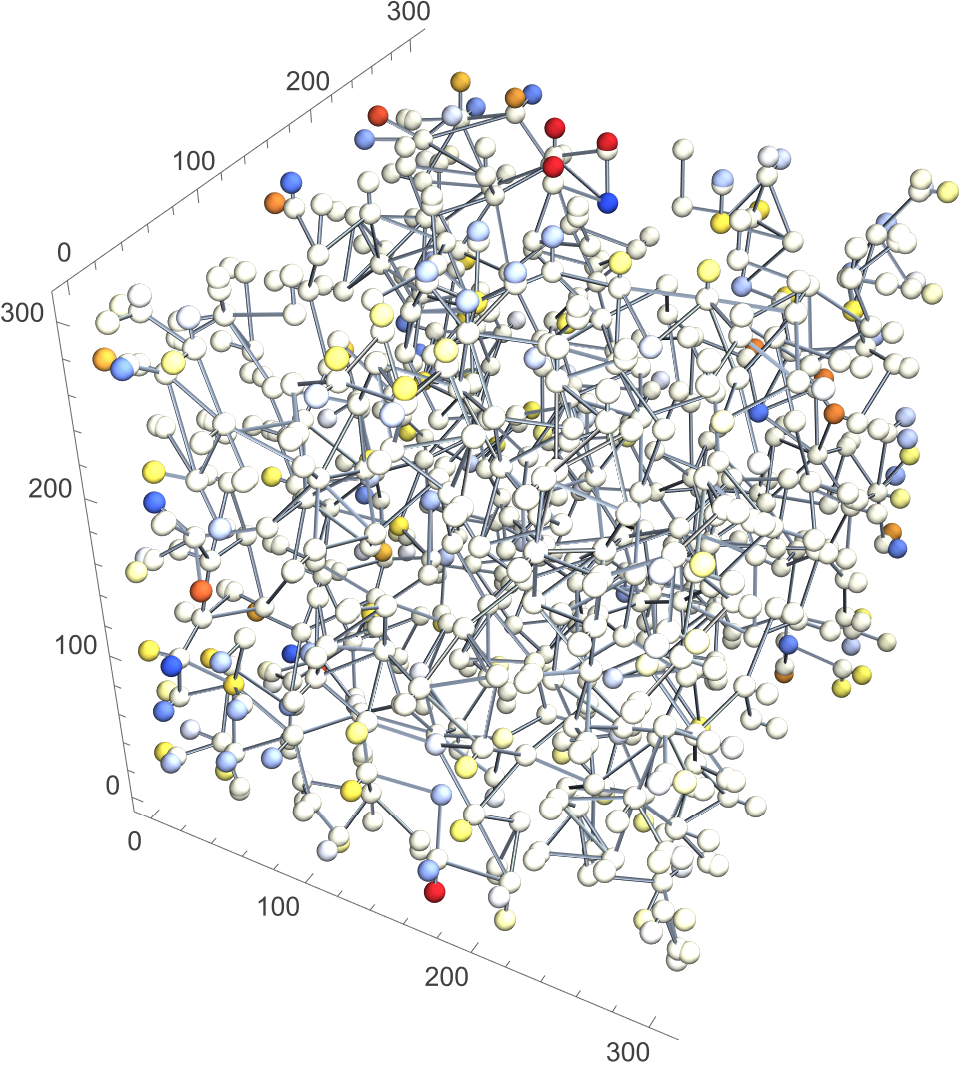}
\includegraphics[angle=0,width=.29\textwidth]{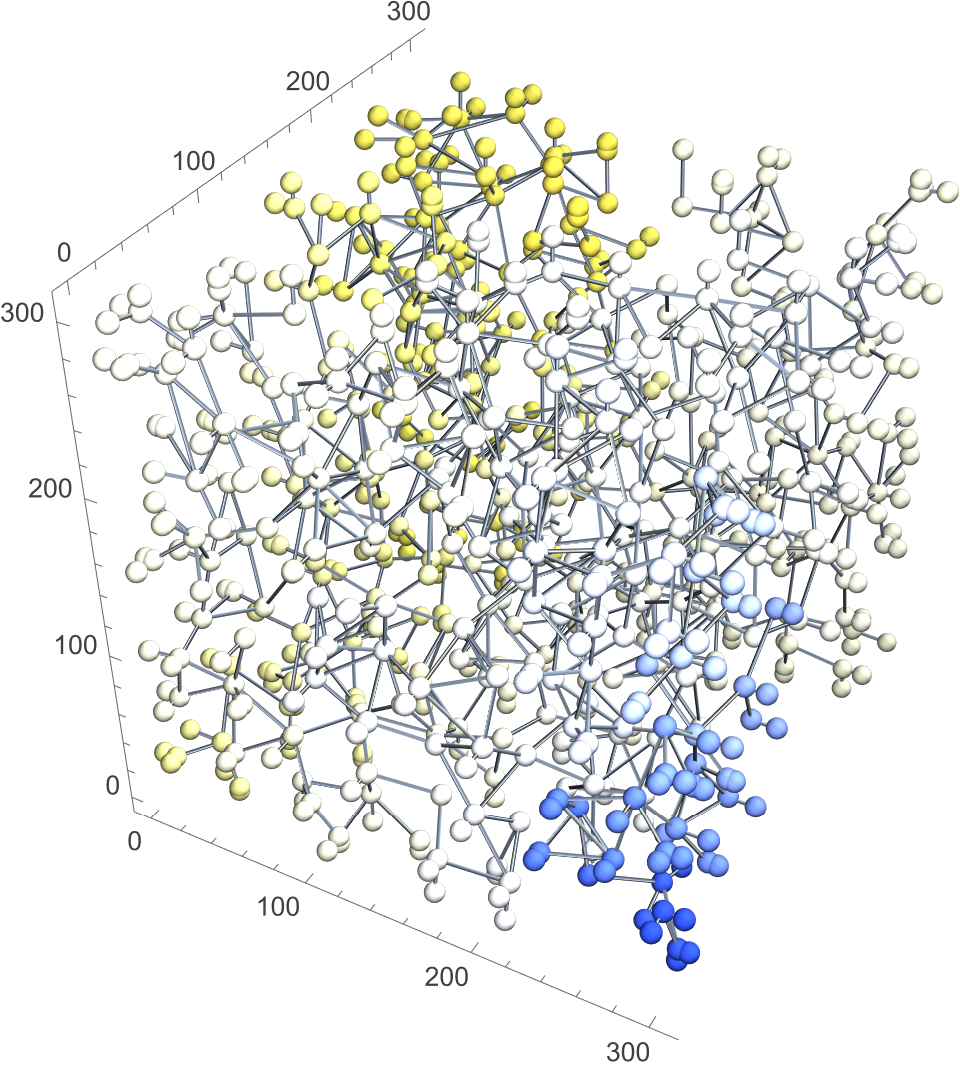}
(a)\;\;\;\;\;\;\;\;\;\;\;\;\;\;\;\;\;\;\;\;\;\;\;\;\;\;\;\;\;\;\;\;\;\;\;\;\;\;\;\;\;\;\; (b) \;\;\;\;\;\;\;\;\;\;\;\;\;\;\;\;\;\;\;\;\;\;\;\;\;\;\;\;(c)
\includegraphics[angle=0,width=.47\textwidth]{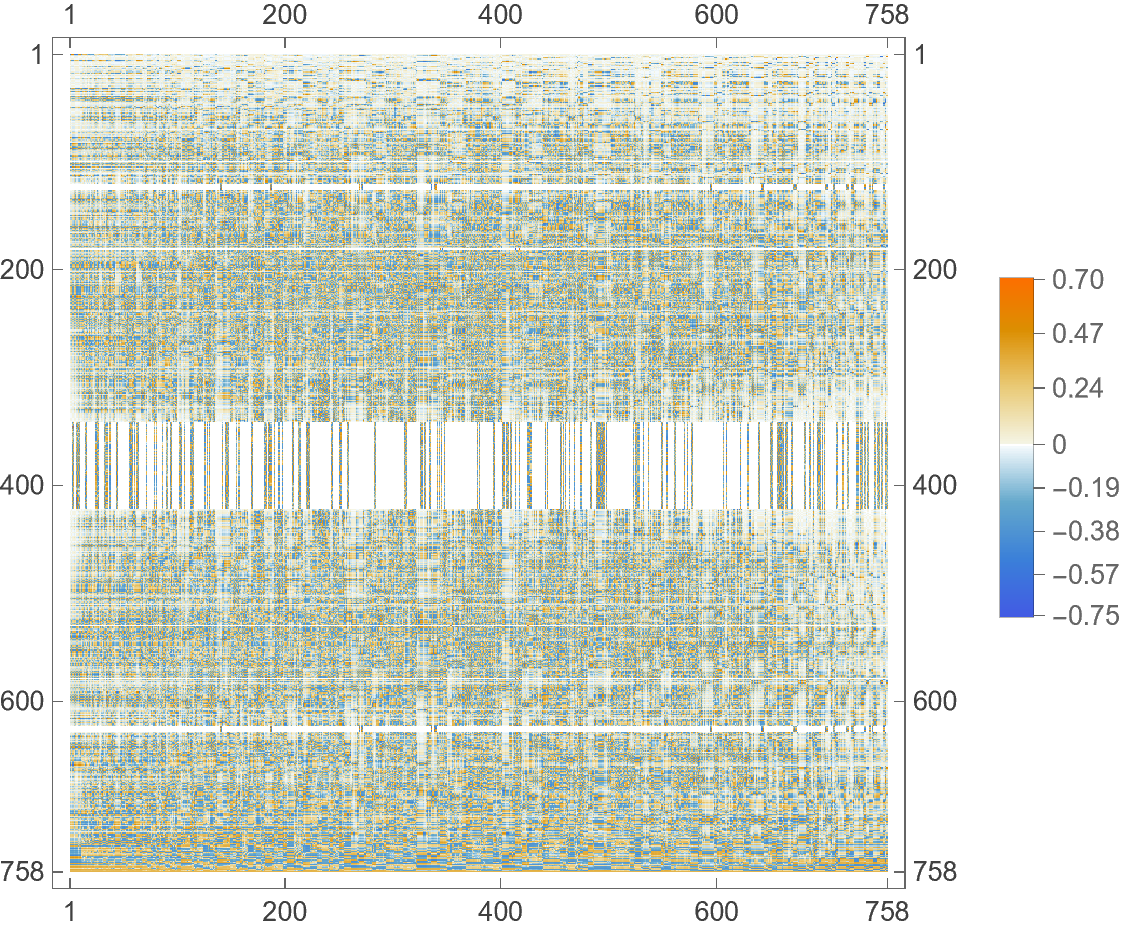} 
\includegraphics[angle=0,width=.5\textwidth]{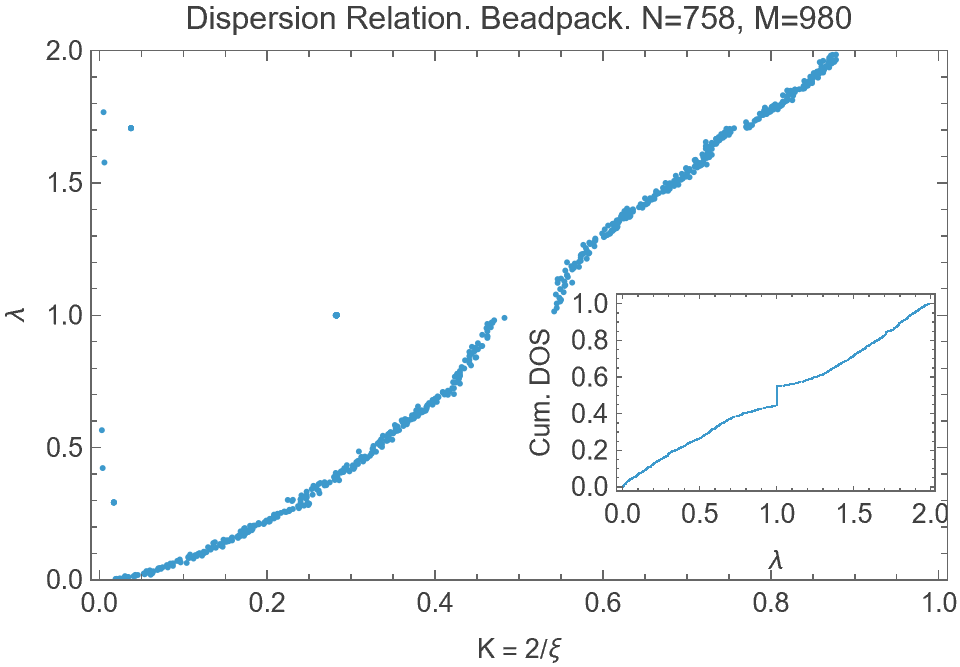} 
(d) \;\;\;\;\;\;\;\;\;\;\;\;\;\;\;\;\;\;\;\;\;\;\;\;\;\;\;\;\;\;\;\;\;\;\;\;\;\;\;\;\;\;\;\;\;\;\;\;\;\;\;\;\;\; (e)\\
\includegraphics[angle=0,width=.35\textwidth]{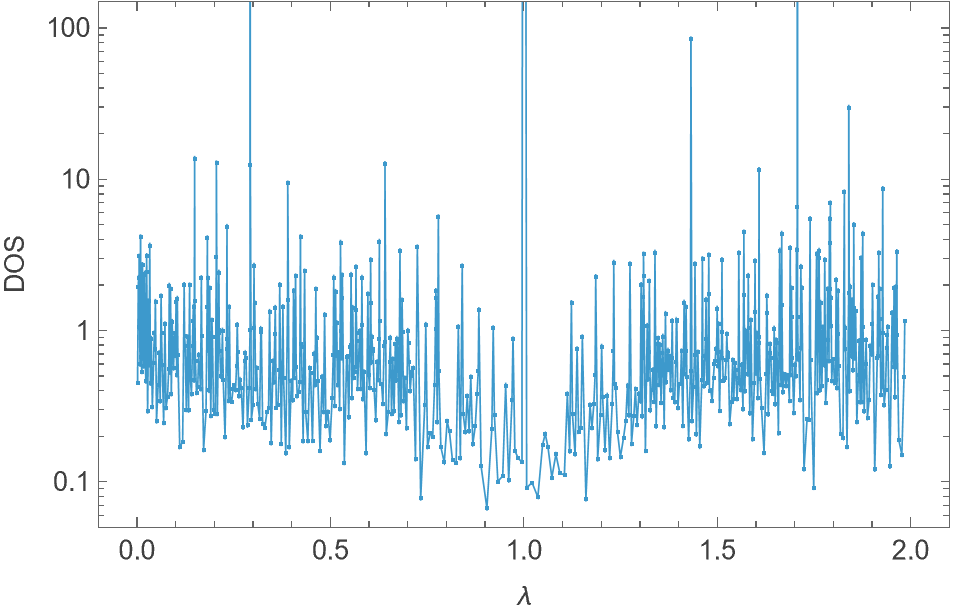} 
(f)
\end{center}
\vspace*{-0.5truecm}
\caption[]{
Pore network extracted from a random packing of 3-mm diameter glass beads. Pores are 
shown as vertices and pore throats as edges. $N=758$ and $M=980$ \cite{BROD25}.
(a) $| \lambda_{1} \rangle$ with $\lambda \approx 1.9853$ 
and $K = 860/980 \approx 0.8776$. 
(b) $| \lambda_{390} \rangle$, representative of the 80
degenerate eigenvectors with $\lambda = 1$ and $K = 277/980 \approx 0.2827$,
which contain many vertices with value 0. 
(c) The Fiedler eigenvector  $| \lambda_{757} \rangle$ with 
$\lambda \approx 0.002929$ and $K = 21/980 \approx 0.02143$.
(d) Vertex values for the $N=758$ eigenvectors. 
(e) Dispersion relation.
The inset shows the cumulative DOS.  
(f) Discrete density of states (DOS). The degenerate 
eigenstates with $\lambda = 1$ are surrounded by two six-fold degenerate states with 
$\lambda = 1 \pm 1/\sqrt 2$, one pair with $\lambda = 1 \pm 1/\sqrt 3$,
and one pair with $\lambda = (7 \pm \sqrt{13})/6$ which together produce six points 
with $K$ near zero in the dispersion relation.  
}
\label{fig:disprelGlassBeads}
\end{figure}

As an example of a three-dimensional porous medium, we consider 
a pore network extracted from a random packing of 3-mm diameter glass beads. 
A segmentation map of the bead pack, saturated with an index-matched, fluorescent liquid, 
was produced by tomographic scanning with a laser sheet \cite{BROD25}. 
The map was converted into a network representation of the pore space using the SNOW algorithm \cite{GOST17} as implemented in the PoreSpy software \cite{GOST19}. 
The resulting graph is shown in Fig.~\ref{fig:disprelGlassBeads}(a-c)
with pores as vertices and pore throats as edges. $N=758$ and $M=980$.
The global parameters of this pore network are 
$\langle d \rangle = 2.5858$, $\rho \approx 0.003416$, $C \approx 0.1329$, $D=31$, 
and $L \approx 12.6345$. The ratio, $L/L_{\rm rand} \approx 1.8103 > 1.5$, 
indicates that this is not a small-world network, consistent with the absence of shortcut 
connections.  
The low $\langle d \rangle $ and $\rho$ are 
due to the fact that 
303 of the 758 vertices have degree one. Most of these lie near the boundaries, where the 
sample has been cut away from a larger network. 

The configuration of $| \lambda_{1} \rangle$ with $\lambda \approx 1.9853$ 
and $K = 860/980 \approx 0.8776$ is shown in Fig.~\ref{fig:disprelGlassBeads}(a). 
As it may not be obvious from the color-coded vertices, this eigenvector has no vertices with 
value zero. While dominated by the ten strongly colored vertices in the lower left corner, 
the rest of the vertices bear small values that are arranged in an approximately 
bipartite configuration with $K \approx 88 \%$ broken bonds. 
The configuration of $| \lambda_{390} \rangle$, representative of the 80
degenerate eigenvectors with $\lambda = 1$ and $K = 277/980 \approx 0.2827$ 
is shown in Fig.~\ref{fig:disprelGlassBeads}(b).
They are characterized by large, contiguous clusters of vertices with exact zero value. 
The Fiedler eigenvector  $| \lambda_{757} \rangle$ with 
$\lambda \approx 0.002929$ and $K = 21/980 \approx 0.02143$ 
is shown in Fig.~\ref{fig:disprelGlassBeads}(c). It approximately bisects the network 
into one positive and one negative part. 
Vertex values for the $N=758$ eigenvectors are shown 
in Fig.~\ref{fig:disprelGlassBeads}(d).
The degenerate 
eigenstates with $\lambda = 1$ are surrounded by two six-fold degenerate states with 
$\lambda = 1 \pm 1/\sqrt 2$ and $K=37/980 \approx 0.03776$ and 
$K=17/980 \approx 0.01735$, respectively, 
and one pair of states localized on four vertices 
with nonzero values and $\lambda = 1 \pm 1/\sqrt 3$ and $K = 6/980 \approx 0.006122$ 
and $K = 4/980 \approx 0.004082$, respectively. 
The former are caused by joined pairs of vertices, one of which has degree 1 and the 
other degree 2. The latter are also caused by joined pairs of vertices with degrees 
1 and 3, respectively \cite{BANE08B,BANE08}. 
In addition to these groups of states with $\lambda$ symmetric about 1, 
we have also detected one localized pair of states with 
$\lambda = (7 \pm \sqrt {13})/6$, symmetric about $7/6$, also with just four vertices 
with nonzero values  and $K = 5/980 \approx 0.005102$ and $K = 3/980 \approx 0.003061$, 
respectively. 
The dispersion relation is shown in Fig.~\ref{fig:disprelGlassBeads}(e).
The inset shows the cumulative DOS.  
We note that the strongly localized eigenvectors with very large, contiguous regions of vertices 
bearing the exact value zero have much lower values of $K$ than the extended majority of states. 
The discrete density of states (DOS), is shown in 
Fig.~\ref{fig:disprelGlassBeads}(f). The divergences correspond to the degenerate 
eigenvalues.
\clearpage

\clearpage

\section{Summary and Conclusions}
\label{sec:V}

With the purpose of estimating effective length scales for individual eigenvectors 
of a Laplacian matrix applied to graph representations of complex networks, we have 
adapted a method previously introduced in condensed-matter physics \cite{DEBY57}. 
This method is based on counting the relative 
number of edges in the graph that connect vertices with values of different sign 
(``broken bonds''). 
After verifying that the method correctly reproduces the dispersion relation for a simple 
line graph, we have applied it to several artificial and natural networks, identifying both 
distributed and localized eigenvectors. Dispersion relations, as well as discrete densities of states 
(DOS) and images of selected eigenvectors have been 
represented in graphical and tabular forms. 
With the exception of the unmodified Cayley tree in Sec.~\ref{Sec:Cayley40}, 
which has mostly localized eigenvectors, 
most of the dispersion relations are separated into a majority of extended eigenvectors 
and a minority of strongly localized ones. 

The extended eigenvectors are represented in the dispersion relations by collections of points, 
$(K , \lambda)$, that are scattered around a monotonically increasing curve, extending from 
a joint minimum corresponding to the Fiedler vector, $| \lambda_{N-1} \rangle$, to the 
eigenvector with the largest eigenvalue, $| \lambda_{1} \rangle$. 
For the calibration case of the line graph discussed in Sec.~\ref{Sec:Lin64}, the relation between 
the ``wavenumber'' $K$ and the Laplacian eigenvalue $\lambda$, given by the ``broken bond'' 
estimate of Eq.~(\ref{eq:DebyeK}), was shown to agree almost exactly 
with the analytical dispersion relation for an infinitely long line, Eq.~(\ref{eq:lin}). 
The shape and range of the part of the dispersion relation representing the extended 
eigenvectors depend on the specific network. The smallest range, corresponding to the largest 
edge density, is found for St.~Mark's food web, Sec.~\ref{Sec:SM48}. 
The largest range, corresponding to the smallest edge density, is found for the line graph. 
(See Table \ref{tab:table2}.) 
Both these graphs have only extended states. 
In the discrete DOS, the fluctuations of distributed 
eigenvalues around a monotonic curve causes a 
succession of relative maxima and minima, while localized states with degenerate 
eigenvalues cause divergent peaks. 
Even though our numerical data are sparse in the small-$K$ region, the 
dispersion relations in this regime are convex, as observed in Sec.~\ref{sec:IId}. 
In the large-$K$, large-$\lambda$ region, 
most of the dispersion relations are scattered around an almost linear curve. 

The localized eigenvectors contain a small number of vertices with nonzero values, while the 
remaining majority of vertices bear the value zero. 
These ``defect'' configurations typically 
represent duplication or addition of small {\it motifs} \cite{BANE08B,BANE08,BANE09}
or {\it singular cores} \cite{SCIR07}, connected to  
other vertices of degree at least two. They are often related to degenerate 
eigenvalues, or to pairs of eigenvalues located symmetrically around 1 or some other rational 
number. Often, but not always, we see such motifs near ``boundaries'' of the 
graph, where it has been disconnected from a larger network. Two examples in this work, 
where this is known to be the case, are the Florida power grid in Sec.~\ref{Sec:FLpow} and 
the model porous medium in Sec.~\ref{sec:IV}.
Such large, contiguous regions of zero-value vertices in the graph lead to a reduced number 
of broken bonds, which causes anomalously small $K$ values, relative to the corresponding 
eigenvalue. For examples, see 
Figs.~{\ref{fig:disprelsT40}--{\ref{fig:disprelCelegans277}(e) and 
{\ref{fig:disprelDolphins}--{\ref{fig:disprelGlassBeads}(e). 

This work presents several opportunities for future study. Among these are the possibilities for  
extending the ``broken-bond'' approach to weighted and/or directed graphs, and to the 
unnormalized graph Laplacian, $\bf L$, appropriate for the study of oscillations on networks.  
Another direction would be an extension to very large networks, which might make visualization 
difficult but on the other hand could enable finite-size scaling studies. 
We hope the results of this study 
can encourage further research on the relations between structure and 
function of complex networks in various branches of science.

\section*{Acknowledgments}

We thank K.~Pierce and J.~F. Brodin for producing and providing the network data 
analyzed in Sec.~\ref{sec:IV}, and
G.~M.\ Buend{\'i}a, M.~Moura, P.~Reis, and M.~A.\ Novotny 
for helpful comments on the manuscript.  

Supported by the Research Council of Norway through the Center of Excellence
funding scheme, Project No. 262644.

\newpage

\appendix*
\section{
Mathematical details}
\label{sec:APP}

\subsection{Eigenvalue spectrum of normalized Laplacians}
\label{sec:spec}

The walk-normalized Laplacian, ${\bf W} = {\bf D}^{-1} {\bf L}$ is related to the 
symmetrically normalized Laplacian, ${\mathcal L} = {\bf D}^{-1/2} {\bf L}{\bf D}^{-1/2}$, 
by the similarity transformation, ${\mathcal L} = {\bf D}^{1/2} {\bf W}{\bf D}^{-1/2}$, 
and the two normalized forms therefore have the same set of eigenvalues, $\{ \lambda_i \}$, 
which is confined to the interval, $ [0,2]$. Further basic properties of the 
normalized spectrum for a graph $G$ with $N$ vertices are listed 
in Lemma~1.7 of Ref.~\cite{CHUN96}. We give those here, with the only difference that we 
list the eigenvalues in descending order, from $\lambda_1$ for the maximum to 
$\lambda_N = 0$ for the minimum. 
Since all the row sums of ${\bf W}$ (and also of ${\bf L}$) vanish, the 
right eigenvector corresponding 
to $\lambda_N = 0$ is the Perron-Frobenius vector, which has all elements equal. 
\begin{itemize}
\item[\bf (i)]
$\sum_i \lambda_i \le N$, with equality if and only if $G$ has no isolated vertices. 
Assuming that the latter is the case, we know that ${\bf W}$ has exactly one eigenvalue, 
$\lambda_N$, that equals zero. Thus, the average of the $N-1$ positive eigenvalues is 
\beq
\langle \lambda_{\rm pos} \rangle = N/(N-1) \;.
\label{eq:avepos}
\eeq
This result is important for the understanding of the following boundary theorems. 
\item[\bf (ii)]
For $N \ge 2$, $\lambda_{N-1} \le N/(N-1)$, with equality if and only if $G$ is the complete graph 
on $N$ vertices.\\
Also, if $G$ has no isolated vertices,  $\lambda_1 \ge N/(N-1)$. 
\item[\bf (iii)]
If $G$ is not a complete graph, then $\lambda_{N-1} \le 1$.
\item[\bf (iv)]
If $G$ is connected, then $\lambda_{N-1} > 0$. \\
If $\lambda_{N-j}=0$ and $\lambda_{N-j+1} > 0$, then $G$ has exactly $j+1$ 
connected components. 
\item[\bf (v)]
For all $i  \ge 1$, $\lambda_i \le 2$.\\
$\lambda_1 = 2$ if and only if $G$ is bipartite and contains at least one edge. 
\item[\bf (vi)]
The spectrum of a graph is the union of the spectra of its connected subgraphs. 
\end{itemize}

\subsection{Verification of numerical criterion for exact zero vertex values}
\label{sec:zzero}

Vertex values in some eigenvectors that are known to be exactly zero due to 
high symmetries 
in the graph or duplication or addition 
of motifs \cite{BANE08B,BANE09} are reported in the numerical calculations used here 
as very small, real numbers. As these will have randomly varying signs, this will 
increase the number of broken bonds in the affected eigenvectors. 
However, in all the systems we have considered here, 
the genuinely nonzero values are separated from those very small numbers by 
a large gap of 
several orders of magnitude. An example for the case of the dolphin community 
discussed in Subsec.~\ref{Sec:Dol62} is shown in Fig.~\ref{fig:App}. 
We therefore identify the exact zeros as the numbers less than a cutoff value chosen 
within this gap,
in the range $2 \times 10^{-15}$ to $2 \times 10^{-12}$.

To ensure that the exact zeroes are not an artifact of the integer vertex degrees, we have 
recalculated the dolphin data after adding to each of the unit elements of the 
symmetric adjacency matrix a 
random number of order $10^{-4}$. This did not change the results reported here in any way. 
\begin{figure}[h]
\begin{center}
\hspace*{-1.5truecm}
\includegraphics[angle=0,width=.35\textwidth]{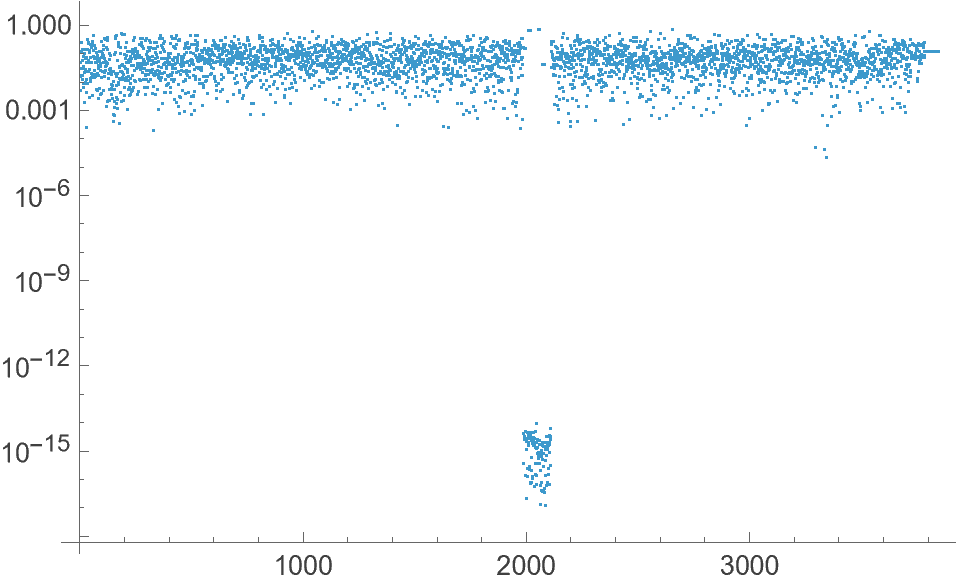}
\end{center}
\vspace*{-0.5truecm}
\caption[]{
This plot shows together on a logarithmic scale the magnitudes of all the 
$N^2$ vertex values 
as they were directly obtained from numerical calculation of the eigenvectors for the 
dolphin community discussed in Subsec.~\ref{Sec:Dol62}. This typical example shows 
the very large gap between the genuinely nonzero vertex values and the very small
noise values representing the exactly zero values. 
For Fig.~\ref{fig:disprelDolphins}(d), all numbers less than 
$10^{-13}$ were replaced by exact zeros. 
This procedure was applied for all the systems considered here, using cutoffs 
in the range $2 \times 10^{-15}$ to $2 \times 10^{-12}$.
}
\label{fig:App}
\end{figure}

\section*{Tables}
\begin{sidewaystable}[h]
\caption[]{
Table of important numerical data for the nine different graphs considered. 
$N$: Number of vertices. 
$M$: Number of edges.
$\langle d \rangle$: Mean degree. 
$C$: Global clustering coefficient. 
$D$: Graph diameter. 
$L$: Mean pair separation. 
$L_{\rm rand} = \ln N /\ln \langle d \rangle$: Mean $L$ for Erd{\"o}sz-R{\'e}nyi random graph 
with same $N$ and $\langle d \rangle$. 
$\rho$: Graph density. 
$\lambda_1$: Eigenvalue number 1. (Maximum eigenvalue.) 
$K_{1}$: Wavenumber corresponding to $\lambda_{1}$.
$\xi_{1}$: Correlation length corresponding to $\lambda_{1}$.
$\lambda_{N-1}$: Eigenvalue number $(N-1)$. (Smallest nonzero eigenvalue, a.k.a. the Fiedler 
value.) 
$K_{N-1}$: Wavenumber corresponding to $\lambda_{N-1}$.
$\xi_{N-1}$: Correlation length corresponding to $\lambda_{N-1}$.
$\lambda({\rm DOS_{max}})$: Eigenvalue corresponding to the maximum of the discrete 
density of states (DOS), indicating the closest degeneracy between eigenvalues.  
$n$-fold degeneracies are denoted as $\lambda \times n$. 
}
\begin{small}
\begin{tabular}{|l|c|c|c|c|c|c|c|c|c|c|c|c|c|c|c|}    
\hline
Graph type & $N$ & $M$ & $\langle d \rangle$ & $\rho$ & $C$ & $D$ & $L$ & $L/L_{\rm rand}$ & $\lambda_1$ & $K_1$ & $\xi_1$ & $\lambda_{N-1}$ & $K_{N-1}$ & $\xi_{N-1}$ & $\lambda$(DOS$_{\rm max}$)\\
\hline
\hline
Line graph & 64 & 63 & 1.9688 & 0.03125& 0 & 63 & 21.6667 & 3.5292
 &  2 & 1 & 2 & 0.0012 & 0.1587 & 126 &   0 and 2\\
\hline
Line w/ shortcuts & 64 & 85 & 2.6563 & 0.04216 & 0 & 10 & 4.5987 & 1.0803
 &  1.9545 & 0.9059 & 2.2078 & 0.0691 & 0.1059 & 18.8889  & 1.70\\
\hline
Cayley tree & 40\ & 39 & 1.95 & 0.05 & 0 & 6 & 4.3615 & 0.7896
 &  2 & 1 & 2 & 0.0318 & 0.0769 & 26 & $1 \times 20$\\
\hline
Cayley w/ shortcuts & 40 & 53 &  2.65 &  0.06795&  0.0909 & 6 & 3.6256 & 0.9579
 &  1.9168 & 0.8113 & 2.4651 & 0.0703 & 0.1321 & 15.1429 &  $1 \times 6$\\
\hline
C. elegans neurome & 277 & 1918 & 13.8484 & 0.0502& 0.1981 & 6 & 2.6389 & 1.2332
 &  1.5768 & 0.6731 & 2.9713 & 0.1548 & 0.1590 & 12.5770 & $1 \times 3$\\
\hline
St. Mark's food web & 48 & 219 & 9.125 & 0.1941& 0.2946 & 4 & 2.0860 & 1.1914
 &  1.6375 & 0.6758 & 2.9595 & 0.3514 & 0.2877 & 6.9524 & 0.99\\
\hline
Dolphin community & 62 & 164 & 5.2903 & 0.0867& 0.3036 & 8 & 3.2893 & 1.3277
 &  1.7037 & 0.6951 & 2.8772 & 0.0411 & 0.0427 & 46.8571 &  $1 \times 2$\\
\hline
Florida power grid & 84 & 137 & 3.2619 & 0.0393 & 0.2112 & 13 & 5.1394 & 1.3714
 &  1.9662 & 0.8102 & 2.4685 & 0.0229 & 0.0438 & 45.6667 & $1 \times 2$\\
\hline
Glass-bead pack & 758 & 980 & 2.5858 & 0.0034 & 0.1329 & 31 & 12.6345 & 1.8103
 &  1.9853 & 0.8776 & 2.2791 & 0.0029 & 0.0214 & 93.3333 & $1 \times 80$\\
\hline
    \end{tabular}
\end{small}
    \label{tab:table}
\end{sidewaystable}

\clearpage 
\begin{table}[h]
\caption[]{
Table of $N$, $M$, $\rho$, $\lambda_1 - \lambda_{N-1}$, $K_1 - K_{N-1}$, and the 
total number of exact $w_i=0$ in the eigenvectors for the nine networks 
and the cutoff used in each case to identify the exact zeros. 
No cutoff was used for the three networks with no exact zeros. 
}
\begin{small}
\begin{tabular}{|l|c|c|c|c|c|c|c|}    
\hline
Graph type & $N$ & $M$ & $ \rho $ & $\lambda_1 - \lambda_{N-1}$ & $K_1 - K_{N-1}$ & Number exact zero & Cutoff\\
\hline
\hline
Line graph & 64 & 63 &  0.03125 & 1.99876 & 0.984127  & 0 &  \\
\hline
Line w/ shortcuts & 64 & 85 &  0.04216 & 1.88537 & 0.8  & 0 &  \\
\hline
Cayley tree & 40\ & 39 & 0.05 & 1.96825 & 0.923077  & 256 & $2.0 \times 10^{-15}$ \\
\hline
Cayley w/ shortcuts & 40 & 53 & 0.06795 & 1.84655 & 0.679245 & 168 & $1.0 \times 10^{-14}$ \\
\hline
C. elegans neurome & 277 & 1918 & 0.0502 & 1.42192 & 0.514077 & 807 & $1.0 \times 10^{-12}$ \\
\hline
St.~Mark's food web & 48 & 219 & 0.1941 & 1.28614 & 0.388128 & 0 &  \\
\hline
Dolphin community & 62 & 164 & 0.0867 & 1.66258 & 0.652439 & 116 & $1.0 \times 10^{-13}$ \\
\hline
Florida power grid & 84 & 137 & 0.0393 & 1.94327 & 0.766423 & 237 & $1.0 \times 10^{-13}$ \\
\hline
Glass-bead pack & 758 & 980 & 0.0034 & 1.98236 & 0.856122 & 56231 & $1.1 \times 10^{-12}$ \\
\hline
    \end{tabular}
\end{small}
    \label{tab:table2}
\end{table}


\begin{thebibliography}{52}%
\makeatletter
\providecommand \@ifxundefined [1]{%
 \@ifx{#1\undefined}
}%
\providecommand \@ifnum [1]{%
 \ifnum #1\expandafter \@firstoftwo
 \else \expandafter \@secondoftwo
 \fi
}%
\providecommand \@ifx [1]{%
 \ifx #1\expandafter \@firstoftwo
 \else \expandafter \@secondoftwo
 \fi
}%
\providecommand \natexlab [1]{#1}%
\providecommand \enquote  [1]{``#1''}%
\providecommand \bibnamefont  [1]{#1}%
\providecommand \bibfnamefont [1]{#1}%
\providecommand \citenamefont [1]{#1}%
\providecommand \href@noop [0]{\@secondoftwo}%
\providecommand \href [0]{\begingroup \@sanitize@url \@href}%
\providecommand \@href[1]{\@@startlink{#1}\@@href}%
\providecommand \@@href[1]{\endgroup#1\@@endlink}%
\providecommand \@sanitize@url [0]{\catcode `\\12\catcode `\$12\catcode
  `\&12\catcode `\#12\catcode `\^12\catcode `\_12\catcode `\%12\relax}%
\providecommand \@@startlink[1]{}%
\providecommand \@@endlink[0]{}%
\providecommand \url  [0]{\begingroup\@sanitize@url \@url }%
\providecommand \@url [1]{\endgroup\@href {#1}{\urlprefix }}%
\providecommand \urlprefix  [0]{URL }%
\providecommand \Eprint [0]{\href }%
\providecommand \doibase [0]{https://doi.org/}%
\providecommand \selectlanguage [0]{\@gobble}%
\providecommand \bibinfo  [0]{\@secondoftwo}%
\providecommand \bibfield  [0]{\@secondoftwo}%
\providecommand \translation [1]{[#1]}%
\providecommand \BibitemOpen [0]{}%
\providecommand \bibitemStop [0]{}%
\providecommand \bibitemNoStop [0]{.\EOS\space}%
\providecommand \EOS [0]{\spacefactor3000\relax}%
\providecommand \BibitemShut  [1]{\csname bibitem#1\endcsname}%
\let\auto@bib@innerbib\@empty
\bibitem [{\citenamefont {Monasson}(1999)}]{MONA99}%
  \BibitemOpen
  \bibfield  {author} {\bibinfo {author} {\bibfnamefont {R.}~\bibnamefont
  {Monasson}},\ }\bibfield  {title} {\bibinfo {title} {Diffusion, localization
  and dispersion relations on ``small-world{"} lattices},\ }\href@noop {}
  {\bibfield  {journal} {\bibinfo  {journal} {Eur.\ Phys.\ J.\ B}\ }\textbf
  {\bibinfo {volume} {12}},\ \bibinfo {pages} {555} (\bibinfo {year}
  {1999})}\BibitemShut {NoStop}%
\bibitem [{\citenamefont {Shuman}\ \emph {et~al.}(2013)\citenamefont {Shuman},
  \citenamefont {Narang}, \citenamefont {Frossard}, \citenamefont {Ortega},\
  and\ \citenamefont {Vandergheynst}}]{SHUM13}%
  \BibitemOpen
  \bibfield  {author} {\bibinfo {author} {\bibfnamefont {D.~I.}\ \bibnamefont
  {Shuman}}, \bibinfo {author} {\bibfnamefont {S.~K.}\ \bibnamefont {Narang}},
  \bibinfo {author} {\bibfnamefont {P.}~\bibnamefont {Frossard}}, \bibinfo
  {author} {\bibfnamefont {A.}~\bibnamefont {Ortega}},\ and\ \bibinfo {author}
  {\bibfnamefont {P.}~\bibnamefont {Vandergheynst}},\ }\bibfield  {title}
  {\bibinfo {title} {The emerging field of signal processing on graphs},\
  }\href {https://doi.org/10.1109/MSP.2012.2235192} {\bibfield  {journal}
  {\bibinfo  {journal} {IEEE Signal Processing Magazine}\ }\textbf {\bibinfo
  {volume} {May 2013}},\ \bibinfo {pages} {83} (\bibinfo {year}
  {2013})}\BibitemShut {NoStop}%
\bibitem [{\citenamefont {Shuman}\ \emph {et~al.}(2016)\citenamefont {Shuman},
  \citenamefont {Ricaud},\ and\ \citenamefont {Vandergheynst}}]{SHUM16}%
  \BibitemOpen
  \bibfield  {author} {\bibinfo {author} {\bibfnamefont {D.~I.}\ \bibnamefont
  {Shuman}}, \bibinfo {author} {\bibfnamefont {B.}~\bibnamefont {Ricaud}},\
  and\ \bibinfo {author} {\bibfnamefont {P.}~\bibnamefont {Vandergheynst}},\
  }\bibfield  {title} {\bibinfo {title} {Vertex-frequency analysis on graphs},\
  }\href {https://doi.org/10.1016/j/acha.2015.02.005} {\bibfield  {journal}
  {\bibinfo  {journal} {Appl. Comput. Harmon. Anal.}\ }\textbf {\bibinfo
  {volume} {40}},\ \bibinfo {pages} {260} (\bibinfo {year} {2016})}\BibitemShut
  {NoStop}%
\bibitem [{\citenamefont {Ricaud}\ \emph {et~al.}(2019)\citenamefont {Ricaud},
  \citenamefont {Borgnat}, \citenamefont {Tremblay}, \citenamefont
  {Goncalves},\ and\ \citenamefont {Vandergheynst}}]{RICA19}%
  \BibitemOpen
  \bibfield  {author} {\bibinfo {author} {\bibfnamefont {B.}~\bibnamefont
  {Ricaud}}, \bibinfo {author} {\bibfnamefont {P.}~\bibnamefont {Borgnat}},
  \bibinfo {author} {\bibfnamefont {N.}~\bibnamefont {Tremblay}}, \bibinfo
  {author} {\bibfnamefont {P.}~\bibnamefont {Goncalves}},\ and\ \bibinfo
  {author} {\bibfnamefont {P.}~\bibnamefont {Vandergheynst}},\ }\bibfield
  {title} {\bibinfo {title} {Fourier could be a data scientist: {F}rom graph
  {F}ourier transform to signal processing on graphs},\ }\href
  {https://doi.org/10.1016/j.crhy.2019.08.003} {\bibfield  {journal} {\bibinfo
  {journal} {Comptes Rendus Physique}\ }\textbf {\bibinfo {volume} {20}},\
  \bibinfo {pages} {474} (\bibinfo {year} {2019})}\BibitemShut {NoStop}%
\bibitem [{\citenamefont {MacMillan}\ and\ \citenamefont
  {Ouellette}(2022)}]{MACM22}%
  \BibitemOpen
  \bibfield  {author} {\bibinfo {author} {\bibfnamefont {T.}~\bibnamefont
  {MacMillan}}\ and\ \bibinfo {author} {\bibfnamefont {N.~T.}\ \bibnamefont
  {Ouellette}},\ }\bibfield  {title} {\bibinfo {title} {Lagrangian scale
  decomposition via the graph {F}ourier transform},\ }\href
  {https://doi.org/10.1103/PhysRevFluids.7.124401} {\bibfield  {journal}
  {\bibinfo  {journal} {Phys. Rev. Fluids}\ }\textbf {\bibinfo {volume} {7}},\
  \bibinfo {pages} {124401} (\bibinfo {year} {2022})}\BibitemShut {NoStop}%
\bibitem [{\citenamefont {Debye}\ \emph {et~al.}(1957)\citenamefont {Debye},
  \citenamefont {Anderson},\ and\ \citenamefont {Brumberger}}]{DEBY57}%
  \BibitemOpen
  \bibfield  {author} {\bibinfo {author} {\bibfnamefont {P.}~\bibnamefont
  {Debye}}, \bibinfo {author} {\bibfnamefont {H.~R.}\ \bibnamefont
  {Anderson}},\ and\ \bibinfo {author} {\bibfnamefont {H.}~\bibnamefont
  {Brumberger}},\ }\bibfield  {title} {\bibinfo {title} {Scattering by an
  inhomogeneous solid. {II}. {T}he correlation function and its application},\
  }\href@noop {} {\bibfield  {journal} {\bibinfo  {journal} {J.\ Appl.\ Phys.}\
  }\textbf {\bibinfo {volume} {28}},\ \bibinfo {pages} {679} (\bibinfo {year}
  {1957})}\BibitemShut {NoStop}%
\bibitem [{\citenamefont {Euler}(1736)}]{EULER1736}%
  \BibitemOpen
  \bibfield  {author} {\bibinfo {author} {\bibfnamefont {L.}~\bibnamefont
  {Euler}},\ }\bibfield  {title} {\bibinfo {title} {Solutio problematis ad
  geometriam situs pertinentis},\ }\href@noop {} {\bibfield  {journal}
  {\bibinfo  {journal} {Commentarii Acadmiae Scientiarum Imperialis
  Petropolitanae}\ }\textbf {\bibinfo {volume} {8}},\ \bibinfo {pages} {128}
  (\bibinfo {year} {1736})}\BibitemShut {NoStop}%
\bibitem [{\citenamefont {Kirchoff}(1845)}]{KIRCH1845}%
  \BibitemOpen
  \bibfield  {author} {\bibinfo {author} {\bibfnamefont {G.}~\bibnamefont
  {Kirchoff}},\ }\bibfield  {title} {\bibinfo {title} {Ueber den {D}urchgang
  eines elektrischen {S}tromes durch eine {E}bene, insbesondere durch eine
  kreisf{\"o}rmige},\ }\href@noop {} {\bibfield  {journal} {\bibinfo  {journal}
  {Ann. Phys. Chem.}\ }\textbf {\bibinfo {volume} {64}},\ \bibinfo {pages}
  {487} (\bibinfo {year} {1845})}\BibitemShut {NoStop}%
\bibitem [{\citenamefont {Kinney}\ \emph {et~al.}(2005)\citenamefont {Kinney},
  \citenamefont {Crucitti}, \citenamefont {Albert},\ and\ \citenamefont
  {Latora}}]{KINN05}%
  \BibitemOpen
  \bibfield  {author} {\bibinfo {author} {\bibfnamefont {R.}~\bibnamefont
  {Kinney}}, \bibinfo {author} {\bibfnamefont {P.}~\bibnamefont {Crucitti}},
  \bibinfo {author} {\bibfnamefont {R.}~\bibnamefont {Albert}},\ and\ \bibinfo
  {author} {\bibfnamefont {V.}~\bibnamefont {Latora}},\ }\bibfield  {title}
  {\bibinfo {title} {Modeling cascading failures in the {N}orth {A}merican
  power grid},\ }\href {https://doi.org/10.1140/epjb/e2005-00237-9} {\bibfield
  {journal} {\bibinfo  {journal} {Eur. Phys. J. B}\ }\textbf {\bibinfo {volume}
  {46}},\ \bibinfo {pages} {101} (\bibinfo {year} {2005})}\BibitemShut
  {NoStop}%
\bibitem [{\citenamefont {Zhu}\ \emph {et~al.}(2022)\citenamefont {Zhu},
  \citenamefont {Ma}, \citenamefont {Yang}, \citenamefont {Li}, \citenamefont
  {Tao},\ and\ \citenamefont {Li}}]{ZHU22}%
  \BibitemOpen
  \bibfield  {author} {\bibinfo {author} {\bibfnamefont {J.}~\bibnamefont
  {Zhu}}, \bibinfo {author} {\bibfnamefont {X.}~\bibnamefont {Ma}}, \bibinfo
  {author} {\bibfnamefont {H.}~\bibnamefont {Yang}}, \bibinfo {author}
  {\bibfnamefont {Y.}~\bibnamefont {Li}}, \bibinfo {author} {\bibfnamefont
  {C.}~\bibnamefont {Tao}},\ and\ \bibinfo {author} {\bibfnamefont
  {H.}~\bibnamefont {Li}},\ }\bibfield  {title} {\bibinfo {title}
  {Comprehensive geographic networks analysis: {S}tatistical, geometric and
  algebraic perspectives},\ }\href
  {https://doi.org/https://doi.org/10.3390/sym14040797} {\bibfield  {journal}
  {\bibinfo  {journal} {Symmetry}\ }\textbf {\bibinfo {volume} {14}},\ \bibinfo
  {pages} {797} (\bibinfo {year} {2022})}\BibitemShut {NoStop}%
\bibitem [{\citenamefont {Allheeib}\ \emph {et~al.}(2022)\citenamefont
  {Allheeib}, \citenamefont {Adhinugraha}, \citenamefont {Taniar},\ and\
  \citenamefont {Saiful~Islam}}]{ALLH22}%
  \BibitemOpen
  \bibfield  {author} {\bibinfo {author} {\bibfnamefont {N.}~\bibnamefont
  {Allheeib}}, \bibinfo {author} {\bibfnamefont {K.}~\bibnamefont
  {Adhinugraha}}, \bibinfo {author} {\bibfnamefont {D.}~\bibnamefont
  {Taniar}},\ and\ \bibinfo {author} {\bibfnamefont {M.}~\bibnamefont
  {Saiful~Islam}},\ }\bibfield  {title} {\bibinfo {title} {Computing reverse
  nearest neighbourhood on road maps},\ }\href
  {https://doi.org/https://doi.org/10.1007/s11280-021-00969-1} {\bibfield
  {journal} {\bibinfo  {journal} {World Wide Web}\ }\textbf {\bibinfo {volume}
  {25}},\ \bibinfo {pages} {99} (\bibinfo {year} {2022})}\BibitemShut {NoStop}%
\bibitem [{\citenamefont {Reis}\ and\ \citenamefont
  {M{\aa}l{\o}y}(2025)}]{REIS25}%
  \BibitemOpen
  \bibfield  {author} {\bibinfo {author} {\bibfnamefont {P.}~\bibnamefont
  {Reis}}\ and\ \bibinfo {author} {\bibfnamefont {K.~J.}\ \bibnamefont
  {M{\aa}l{\o}y}},\ }\bibfield  {title} {\bibinfo {title} {Drainage front width
  in a three-dimensional random porous medium under gravitational and capillary
  effects},\ }\href {https://doi.org/https://doi.org/10.1103/5bbz-ksds}
  {\bibfield  {journal} {\bibinfo  {journal} {Phys. Rev. Research}\ }\textbf
  {\bibinfo {volume} {7}},\ \bibinfo {pages} {033244} (\bibinfo {year}
  {2025})}\BibitemShut {NoStop}%
\bibitem [{\citenamefont {Leite}\ \emph {et~al.}(2024)\citenamefont {Leite},
  \citenamefont {Banerjee}, \citenamefont {Wei}, \citenamefont {Elowitt},\ and\
  \citenamefont {Clark}}]{LEIT24}%
  \BibitemOpen
  \bibfield  {author} {\bibinfo {author} {\bibfnamefont {L.~S.~G.}\
  \bibnamefont {Leite}}, \bibinfo {author} {\bibfnamefont {S.}~\bibnamefont
  {Banerjee}}, \bibinfo {author} {\bibfnamefont {Y.}~\bibnamefont {Wei}},
  \bibinfo {author} {\bibfnamefont {J.}~\bibnamefont {Elowitt}},\ and\ \bibinfo
  {author} {\bibfnamefont {A.~E.}\ \bibnamefont {Clark}},\ }\bibfield  {title}
  {\bibinfo {title} {Modern chemical graph theory},\ }\href
  {https://doi.org/10.1002/wcms.1729} {\bibfield  {journal} {\bibinfo
  {journal} {WIREs Comput. Mol. Sci.}\ }\textbf {\bibinfo {volume} {14}},\
  \bibinfo {pages} {1729} (\bibinfo {year} {2024})}\BibitemShut {NoStop}%
\bibitem [{\citenamefont {Arnatkeviciute}\ \emph {et~al.}(2018)\citenamefont
  {Arnatkeviciute}, \citenamefont {Fulcher}, \citenamefont {Pocock},\ and\
  \citenamefont {Fornito}}]{ARNA18}%
  \BibitemOpen
  \bibfield  {author} {\bibinfo {author} {\bibfnamefont {A.}~\bibnamefont
  {Arnatkeviciute}}, \bibinfo {author} {\bibfnamefont {B.~D.}\ \bibnamefont
  {Fulcher}}, \bibinfo {author} {\bibfnamefont {R.}~\bibnamefont {Pocock}},\
  and\ \bibinfo {author} {\bibfnamefont {A.}~\bibnamefont {Fornito}},\
  }\bibfield  {title} {\bibinfo {title} {Hub connectivity, neuronal diversity,
  and gene expression in the {C}aenorhabditis elegans connectome},\ }\href
  {https://doi.org/https://doi.org/10.1371/journal.pcbi.1005989} {\bibfield
  {journal} {\bibinfo  {journal} {{PLoS} Comput. Biol.}\ }\textbf {\bibinfo
  {volume} {14}},\ \bibinfo {pages} {e1005989} (\bibinfo {year}
  {2018})}\BibitemShut {NoStop}%
\bibitem [{\citenamefont {Rossberg}(2013)}]{ROSS13}%
  \BibitemOpen
  \bibfield  {author} {\bibinfo {author} {\bibfnamefont {A.~G.}\ \bibnamefont
  {Rossberg}},\ }\href@noop {} {\emph {\bibinfo {title} {Food Webs and
  Biodiversity}}}\ (\bibinfo  {publisher} {Wiley Blackwell},\ \bibinfo
  {address} {Oxford},\ \bibinfo {year} {2013})\BibitemShut {NoStop}%
\bibitem [{\citenamefont {Kuhlmann}\ and\ \citenamefont
  {Oepen}(2016)}]{KUHL16}%
  \BibitemOpen
  \bibfield  {author} {\bibinfo {author} {\bibfnamefont {M.}~\bibnamefont
  {Kuhlmann}}\ and\ \bibinfo {author} {\bibfnamefont {S.}~\bibnamefont
  {Oepen}},\ }\bibfield  {title} {\bibinfo {title} {Squibs: {T}owards a
  catalogue of linguistic graph banks},\ }\href
  {https://doi.org/10.1162/COLI_a_00268} {\bibfield  {journal} {\bibinfo
  {journal} {Comp. Linguistics}\ }\textbf {\bibinfo {volume} {42}},\ \bibinfo
  {pages} {819} (\bibinfo {year} {2016})}\BibitemShut {NoStop}%
\bibitem [{\citenamefont {Newman}(2010)}]{NEWM10}%
  \BibitemOpen
  \bibfield  {author} {\bibinfo {author} {\bibfnamefont {M.~E.~J.}\
  \bibnamefont {Newman}},\ }\href@noop {} {\emph {\bibinfo {title} {Networks.
  An Introduction}}}\ (\bibinfo  {publisher} {Oxford University Press},\
  \bibinfo {address} {Oxford},\ \bibinfo {year} {2010})\BibitemShut {NoStop}%
\bibitem [{\citenamefont {Masuda}\ \emph {et~al.}(2017)\citenamefont {Masuda},
  \citenamefont {Porter},\ and\ \citenamefont {Lambiotte}}]{MASU17}%
  \BibitemOpen
  \bibfield  {author} {\bibinfo {author} {\bibfnamefont {N.}~\bibnamefont
  {Masuda}}, \bibinfo {author} {\bibfnamefont {M.~A.}\ \bibnamefont {Porter}},\
  and\ \bibinfo {author} {\bibfnamefont {R.}~\bibnamefont {Lambiotte}},\
  }\bibfield  {title} {\bibinfo {title} {Random walks and diffusion on
  networks},\ }\href@noop {} {\bibfield  {journal} {\bibinfo  {journal} {Phys.\
  Rep.}\ }\textbf {\bibinfo {volume} {716-717}},\ \bibinfo {pages} {1}
  (\bibinfo {year} {2017})}\BibitemShut {NoStop}%
\bibitem [{\citenamefont {Chung}(1996)}]{CHUN96}%
  \BibitemOpen
  \bibfield  {author} {\bibinfo {author} {\bibfnamefont {F.}~\bibnamefont
  {Chung}},\ }\href@noop {} {\emph {\bibinfo {title} {Spectral graph theory}}}\
  (\bibinfo  {publisher} {American Mathematical Society},\ \bibinfo {address}
  {Providence, RI},\ \bibinfo {year} {1996})\ \bibinfo {note}
  {{C}h.~1}\BibitemShut {NoStop}%
\bibitem [{\citenamefont {Moler}\ and\ \citenamefont
  {Van~Loan}(2003)}]{MOLE03}%
  \BibitemOpen
  \bibfield  {author} {\bibinfo {author} {\bibfnamefont {C.}~\bibnamefont
  {Moler}}\ and\ \bibinfo {author} {\bibfnamefont {C.}~\bibnamefont
  {Van~Loan}},\ }\bibfield  {title} {\bibinfo {title} {Nineteen dubious ways to
  compute the exponential of a matrix, twenty-five years later},\ }\href@noop
  {} {\bibfield  {journal} {\bibinfo  {journal} {SIAM Rev.}\ }\textbf {\bibinfo
  {volume} {45}},\ \bibinfo {pages} {3} (\bibinfo {year} {2003})}\BibitemShut
  {NoStop}%
\bibitem [{\citenamefont {Arnaudon}\ \emph {et~al.}(2020)\citenamefont
  {Arnaudon}, \citenamefont {Peach},\ and\ \citenamefont {Barahona}}]{ARNA20}%
  \BibitemOpen
  \bibfield  {author} {\bibinfo {author} {\bibfnamefont {A.}~\bibnamefont
  {Arnaudon}}, \bibinfo {author} {\bibfnamefont {R.~L.}\ \bibnamefont
  {Peach}},\ and\ \bibinfo {author} {\bibfnamefont {M.}~\bibnamefont
  {Barahona}},\ }\bibfield  {title} {\bibinfo {title} {Scale-dependent measure
  of network centrality from diffusion dynamics},\ }\href@noop {} {\bibfield
  {journal} {\bibinfo  {journal} {Phys.\ Rev.\ Res.}\ }\textbf {\bibinfo
  {volume} {2}},\ \bibinfo {pages} {033104} (\bibinfo {year}
  {2020})}\BibitemShut {NoStop}%
\bibitem [{\citenamefont {Bolla}\ \emph {et~al.}(2015)\citenamefont {Bolla},
  \citenamefont {Bullins}, \citenamefont {Chaturapruek}, \citenamefont {Chen},\
  and\ \citenamefont {Friedl}}]{BOLL15}%
  \BibitemOpen
  \bibfield  {author} {\bibinfo {author} {\bibfnamefont {M.}~\bibnamefont
  {Bolla}}, \bibinfo {author} {\bibfnamefont {B.}~\bibnamefont {Bullins}},
  \bibinfo {author} {\bibfnamefont {S.}~\bibnamefont {Chaturapruek}}, \bibinfo
  {author} {\bibfnamefont {S.}~\bibnamefont {Chen}},\ and\ \bibinfo {author}
  {\bibfnamefont {K.}~\bibnamefont {Friedl}},\ }\bibfield  {title} {\bibinfo
  {title} {Spectral properties of modularity matrices},\ }\href@noop {}
  {\bibfield  {journal} {\bibinfo  {journal} {Lin. Alg. Appl.}\ }\textbf
  {\bibinfo {volume} {473}},\ \bibinfo {pages} {359} (\bibinfo {year}
  {2015})}\BibitemShut {NoStop}%
\bibitem [{\citenamefont {Banerjee}\ and\ \citenamefont
  {Jost}(2008{\natexlab{a}})}]{BANE08B}%
  \BibitemOpen
  \bibfield  {author} {\bibinfo {author} {\bibfnamefont {A.}~\bibnamefont
  {Banerjee}}\ and\ \bibinfo {author} {\bibfnamefont {J.}~\bibnamefont
  {Jost}},\ }\bibfield  {title} {\bibinfo {title} {On the spectrum of the
  normalized graph {L}aplacian},\ }\href
  {https://doi.org/https://doi.org/10.1016/j.laa.2008.01.029} {\bibfield
  {journal} {\bibinfo  {journal} {Lin. Alg. Appl.}\ }\textbf {\bibinfo {volume}
  {428}},\ \bibinfo {pages} {3015} (\bibinfo {year}
  {2008}{\natexlab{a}})}\BibitemShut {NoStop}%
\bibitem [{\citenamefont {Banerjee}\ and\ \citenamefont
  {Jost}(2008{\natexlab{b}})}]{BANE08}%
  \BibitemOpen
  \bibfield  {author} {\bibinfo {author} {\bibfnamefont {A.}~\bibnamefont
  {Banerjee}}\ and\ \bibinfo {author} {\bibfnamefont {J.}~\bibnamefont
  {Jost}},\ }\bibfield  {title} {\bibinfo {title} {Spectral plot properties:
  {T}owards a qualitative classification of networks},\ }\href
  {https://doi.org/https://doi.org/10.3934/nhm.2008.3.395} {\bibfield
  {journal} {\bibinfo  {journal} {Netw. Heterog. Media}\ }\textbf {\bibinfo
  {volume} {3}},\ \bibinfo {pages} {395} (\bibinfo {year}
  {2008}{\natexlab{b}})}\BibitemShut {NoStop}%
\bibitem [{\citenamefont {Banerjee}\ and\ \citenamefont {Jost}(2009)}]{BANE09}%
  \BibitemOpen
  \bibfield  {author} {\bibinfo {author} {\bibfnamefont {A.}~\bibnamefont
  {Banerjee}}\ and\ \bibinfo {author} {\bibfnamefont {J.}~\bibnamefont
  {Jost}},\ }\bibfield  {title} {\bibinfo {title} {Spectral characterization of
  network structure and dynamics},\ }in\ \href
  {https://doi.org/http://doi.org/10.1007/978-0-8176-4751-3-7} {\emph {\bibinfo
  {booktitle} {Dynamics on and of complex networks}}},\ \bibinfo {editor}
  {edited by\ \bibinfo {editor} {\bibfnamefont {N.}~\bibnamefont {Ganguly}}}\
  (\bibinfo  {publisher} {Birkhauser},\ \bibinfo {address} {Boston},\ \bibinfo
  {year} {2009})\ pp.\ \bibinfo {pages} {118--132}\BibitemShut {NoStop}%
\bibitem [{\citenamefont {Sciriha}(2007)}]{SCIR07}%
  \BibitemOpen
  \bibfield  {author} {\bibinfo {author} {\bibfnamefont {I.}~\bibnamefont
  {Sciriha}},\ }\bibfield  {title} {\bibinfo {title} {A characterization of
  singular graphs},\ }\href@noop {} {\bibfield  {journal} {\bibinfo  {journal}
  {El.\ J.\ Lin.\ Algebra}\ }\textbf {\bibinfo {volume} {16}},\ \bibinfo
  {pages} {451} (\bibinfo {year} {2007})}\BibitemShut {NoStop}%
\bibitem [{\citenamefont {Watts}\ and\ \citenamefont {Strogatz}(1998)}]{WS98}%
  \BibitemOpen
  \bibfield  {author} {\bibinfo {author} {\bibfnamefont {D.~J.}\ \bibnamefont
  {Watts}}\ and\ \bibinfo {author} {\bibfnamefont {S.~H.}\ \bibnamefont
  {Strogatz}},\ }\bibfield  {title} {\bibinfo {title} {Collective dynamics of
  small-world networks},\ }\href@noop {} {\bibfield  {journal} {\bibinfo
  {journal} {Nature}\ }\textbf {\bibinfo {volume} {393}},\ \bibinfo {pages}
  {440} (\bibinfo {year} {1998})}\BibitemShut {NoStop}%
\bibitem [{\citenamefont {Rosa}\ and\ \citenamefont {Ruzzene}(2022)}]{ROSA22}%
  \BibitemOpen
  \bibfield  {author} {\bibinfo {author} {\bibfnamefont {M.~I.~N.}\
  \bibnamefont {Rosa}}\ and\ \bibinfo {author} {\bibfnamefont {M.}~\bibnamefont
  {Ruzzene}},\ }\bibfield  {title} {\bibinfo {title} {Small-world disordered
  lattices: spectral gaps and diffusive transport},\ }\href@noop {} {\bibfield
  {journal} {\bibinfo  {journal} {New.\ J.\ Phys.}\ }\textbf {\bibinfo {volume}
  {24}},\ \bibinfo {pages} {073020} (\bibinfo {year} {2022})}\BibitemShut
  {NoStop}%
\bibitem [{\citenamefont {Grady}\ and\ \citenamefont
  {Polimeni}(2010)}]{GRAD10}%
  \BibitemOpen
  \bibfield  {author} {\bibinfo {author} {\bibfnamefont {J.~L.}\ \bibnamefont
  {Grady}}\ and\ \bibinfo {author} {\bibfnamefont {J.~R.}\ \bibnamefont
  {Polimeni}},\ }\href@noop {} {\emph {\bibinfo {title} {Discrete Calculus}}}\
  (\bibinfo  {publisher} {Springer-Verlag},\ \bibinfo {address} {London,
  U.K.},\ \bibinfo {year} {2010})\ \bibinfo {note} {{C}h.~8, Measuring
  Networks}\BibitemShut {NoStop}%
\bibitem [{\citenamefont {Rikvold}\ and\ \citenamefont {Stell}(1985)}]{RIKV85}%
  \BibitemOpen
  \bibfield  {author} {\bibinfo {author} {\bibfnamefont {P.~A.}\ \bibnamefont
  {Rikvold}}\ and\ \bibinfo {author} {\bibfnamefont {G.}~\bibnamefont
  {Stell}},\ }\bibfield  {title} {\bibinfo {title} {D-dimensional
  interpenetrable-sphere models of random two-phase media: Microstructure and
  an application to chromatography},\ }\href@noop {} {\bibfield  {journal}
  {\bibinfo  {journal} {J. Coll. Int. Sci.}\ }\textbf {\bibinfo {volume}
  {108}},\ \bibinfo {pages} {158} (\bibinfo {year} {1985})}\BibitemShut
  {NoStop}%
\bibitem [{\citenamefont {Armstrong}\ \emph {et~al.}(2019)\citenamefont
  {Armstrong}, \citenamefont {McClure}, \citenamefont {Robins}, \citenamefont
  {Liu}, \citenamefont {Arns}, \citenamefont {Schl{\"u}ter},\ and\
  \citenamefont {Berg}}]{ARMS19}%
  \BibitemOpen
  \bibfield  {author} {\bibinfo {author} {\bibfnamefont {R.~T.}\ \bibnamefont
  {Armstrong}}, \bibinfo {author} {\bibfnamefont {J.~E.}\ \bibnamefont
  {McClure}}, \bibinfo {author} {\bibfnamefont {V.}~\bibnamefont {Robins}},
  \bibinfo {author} {\bibfnamefont {Z.}~\bibnamefont {Liu}}, \bibinfo {author}
  {\bibfnamefont {C.~H.}\ \bibnamefont {Arns}}, \bibinfo {author}
  {\bibfnamefont {S.}~\bibnamefont {Schl{\"u}ter}},\ and\ \bibinfo {author}
  {\bibfnamefont {S.}~\bibnamefont {Berg}},\ }\bibfield  {title} {\bibinfo
  {title} {Porous media characterization using {M}inkowski functionals:
  Theories, applications and future directions},\ }\href
  {https://doi.org/10.1007/s11242-018-1201-4} {\bibfield  {journal} {\bibinfo
  {journal} {Transp. Por. Media}\ }\textbf {\bibinfo {volume} {130}},\ \bibinfo
  {pages} {305} (\bibinfo {year} {2019})}\BibitemShut {NoStop}%
\bibitem [{\citenamefont {Newman}(2000)}]{NEWM00}%
  \BibitemOpen
  \bibfield  {author} {\bibinfo {author} {\bibfnamefont {M.~E.~J.}\
  \bibnamefont {Newman}},\ }\bibfield  {title} {\bibinfo {title} {Models of the
  {S}mall {W}orld},\ }\href@noop {} {\bibfield  {journal} {\bibinfo  {journal}
  {J. Stat. Phys.}\ }\textbf {\bibinfo {volume} {101}},\ \bibinfo {pages} {819}
  (\bibinfo {year} {2000})}\BibitemShut {NoStop}%
\bibitem [{\citenamefont {Newman}\ and\ \citenamefont {Watts}(1999)}]{NEWM99}%
  \BibitemOpen
  \bibfield  {author} {\bibinfo {author} {\bibfnamefont {M.~E.~J.}\
  \bibnamefont {Newman}}\ and\ \bibinfo {author} {\bibfnamefont {D.~J.}\
  \bibnamefont {Watts}},\ }\bibfield  {title} {\bibinfo {title}
  {Renormalization group analysis of the small-world network model},\
  }\href@noop {} {\bibfield  {journal} {\bibinfo  {journal} {Phys. Lett. A}\
  }\textbf {\bibinfo {volume} {263}},\ \bibinfo {pages} {341} (\bibinfo {year}
  {1999})}\BibitemShut {NoStop}%
\bibitem [{\citenamefont {Tuncer}\ and\ \citenamefont {Erzan}(2015)}]{TUNC15}%
  \BibitemOpen
  \bibfield  {author} {\bibinfo {author} {\bibfnamefont {A.}~\bibnamefont
  {Tuncer}}\ and\ \bibinfo {author} {\bibfnamefont {A.}~\bibnamefont {Erzan}},\
  }\bibfield  {title} {\bibinfo {title} {Spectral renormalization group for the
  {G}aussian model and {$\psi^4$} theory on nonspatial networks},\ }\href
  {https://doi.org/https://doi.org/101103/PhysRevE.92.022106} {\bibfield
  {journal} {\bibinfo  {journal} {Phys.\ Rev.\ E}\ }\textbf {\bibinfo {volume}
  {92}},\ \bibinfo {pages} {022106} (\bibinfo {year} {2015})}\BibitemShut
  {NoStop}%
\bibitem [{\citenamefont {Tuncer}\ and\ \citenamefont {Erzan}(2020)}]{TUNC20}%
  \BibitemOpen
  \bibfield  {author} {\bibinfo {author} {\bibfnamefont {A.}~\bibnamefont
  {Tuncer}}\ and\ \bibinfo {author} {\bibfnamefont {A.}~\bibnamefont {Erzan}},\
  }\bibfield  {title} {\bibinfo {title} {Explicit construction of the
  eigenvectors and eigenvalues of the graph {L}aplacian on the {C}ayley tree},\
  }\href {https://doi.org/https://doi.org/10.1016/j.laa.2019.10.023} {\bibfield
   {journal} {\bibinfo  {journal} {Linear Algebra Appl.}\ }\textbf {\bibinfo
  {volume} {586}},\ \bibinfo {pages} {111} (\bibinfo {year}
  {2020})}\BibitemShut {NoStop}%
\bibitem [{CE2()}]{CE277}%
  \BibitemOpen
  \href@noop {} {\ }\bibinfo {note}
  {{https://sites.google.com/view/dynamicconnectomelab/resources}}\BibitemShut
  {NoStop}%
\bibitem [{\citenamefont {Kaiser}\ and\ \citenamefont
  {Hilgetag}(2006)}]{KAIS06}%
  \BibitemOpen
  \bibfield  {author} {\bibinfo {author} {\bibfnamefont {M.}~\bibnamefont
  {Kaiser}}\ and\ \bibinfo {author} {\bibfnamefont {C.~C.}\ \bibnamefont
  {Hilgetag}},\ }\bibfield  {title} {\bibinfo {title} {Non-optimal component
  placement, but short processing paths, due to long-distance projections in
  neural systems},\ }\href
  {https://doi.org/https://doi.org/10.1371/journal.pcbi.0020095} {\bibfield
  {journal} {\bibinfo  {journal} {{PLoS} Comput. Biol.}\ }\textbf {\bibinfo
  {volume} {2}},\ \bibinfo {pages} {e95} (\bibinfo {year} {2006})}\BibitemShut
  {NoStop}%
\bibitem [{\citenamefont {White}\ \emph {et~al.}(1986)\citenamefont {White},
  \citenamefont {Southgate}, \citenamefont {Thompson},\ and\ \citenamefont
  {Brenner}}]{WHIT86}%
  \BibitemOpen
  \bibfield  {author} {\bibinfo {author} {\bibfnamefont {J.~G.}\ \bibnamefont
  {White}}, \bibinfo {author} {\bibfnamefont {E.}~\bibnamefont {Southgate}},
  \bibinfo {author} {\bibfnamefont {J.~N.}\ \bibnamefont {Thompson}},\ and\
  \bibinfo {author} {\bibfnamefont {S.}~\bibnamefont {Brenner}},\ }\bibfield
  {title} {\bibinfo {title} {The structure of the nervous system of the
  nematode {C}aenorhabditis {E}legans},\ }\href@noop {} {\bibfield  {journal}
  {\bibinfo  {journal} {Phil. Trans. R. Soc. London}\ }\textbf {\bibinfo
  {volume} {314}},\ \bibinfo {pages} {1} (\bibinfo {year} {1986})}\BibitemShut
  {NoStop}%
\bibitem [{\citenamefont {Christian}\ and\ \citenamefont
  {Luczkovich}(1999)}]{CHRI99}%
  \BibitemOpen
  \bibfield  {author} {\bibinfo {author} {\bibfnamefont {R.~R.}\ \bibnamefont
  {Christian}}\ and\ \bibinfo {author} {\bibfnamefont {J.~J.}\ \bibnamefont
  {Luczkovich}},\ }\bibfield  {title} {\bibinfo {title} {Organizing and
  understanding a winter’s seagrass foodweb network through effective trophic
  levels},\ }\href@noop {} {\bibfield  {journal} {\bibinfo  {journal} {Ecol.
  Model.}\ }\textbf {\bibinfo {volume} {117}},\ \bibinfo {pages} {99} (\bibinfo
  {year} {1999})}\BibitemShut {NoStop}%
\bibitem [{\citenamefont {Williams}\ \emph {et~al.}(2002)\citenamefont
  {Williams}, \citenamefont {Berlow}, \citenamefont {Dunne}, \citenamefont
  {Barab{\'a}si},\ and\ \citenamefont {Martinez}}]{WILL02}%
  \BibitemOpen
  \bibfield  {author} {\bibinfo {author} {\bibfnamefont {R.~J.}\ \bibnamefont
  {Williams}}, \bibinfo {author} {\bibfnamefont {E.~L.}\ \bibnamefont
  {Berlow}}, \bibinfo {author} {\bibfnamefont {J.~A.}\ \bibnamefont {Dunne}},
  \bibinfo {author} {\bibfnamefont {A.-L.}\ \bibnamefont {Barab{\'a}si}},\ and\
  \bibinfo {author} {\bibfnamefont {N.~D.}\ \bibnamefont {Martinez}},\
  }\bibfield  {title} {\bibinfo {title} {Two degrees of separation in complex
  food webs},\ }\href
  {https://doi.org/www.pnas.org/cgi/doi/10.1073/pnas.192448799} {\bibfield
  {journal} {\bibinfo  {journal} {Proc. Natl. Acad. Sci. U.S.A.}\ }\textbf
  {\bibinfo {volume} {99}},\ \bibinfo {pages} {12913} (\bibinfo {year}
  {2002})}\BibitemShut {NoStop}%
\bibitem [{\citenamefont {Rikvold}\ and\ \citenamefont {Sevim}(2007)}]{PAR07}%
  \BibitemOpen
  \bibfield  {author} {\bibinfo {author} {\bibfnamefont {P.~A.}\ \bibnamefont
  {Rikvold}}\ and\ \bibinfo {author} {\bibfnamefont {V.}~\bibnamefont
  {Sevim}},\ }\bibfield  {title} {\bibinfo {title} {Individual-based
  predator-prey model for biological coevolution: {F}luctuations, stability,
  and community structure},\ }\href
  {https://doi.org/10.1103/PhysRevE.75.051920} {\bibfield  {journal} {\bibinfo
  {journal} {Phys. Rev. E}\ }\textbf {\bibinfo {volume} {75}},\ \bibinfo
  {pages} {051920} (\bibinfo {year} {2007})}\BibitemShut {NoStop}%
\bibitem [{\citenamefont {Lusseau}(2003)}]{LUSS03B}%
  \BibitemOpen
  \bibfield  {author} {\bibinfo {author} {\bibfnamefont {D.}~\bibnamefont
  {Lusseau}},\ }\bibfield  {title} {\bibinfo {title} {The emergent properties
  of a dolphin social network},\ }\href
  {https://doi.org/10.1098/rsbl.2003.0057} {\bibfield  {journal} {\bibinfo
  {journal} {Proc. R. Soc. London B (suppl.)}\ }\textbf {\bibinfo {volume}
  {270}},\ \bibinfo {pages} {S186} (\bibinfo {year} {2003})}\BibitemShut
  {NoStop}%
\bibitem [{\citenamefont {Lusseau}\ \emph {et~al.}(2003)\citenamefont
  {Lusseau}, \citenamefont {Schneider}, \citenamefont {Boisseau}, \citenamefont
  {Haase}, \citenamefont {Slooten},\ and\ \citenamefont {Dawson}}]{LUSS03}%
  \BibitemOpen
  \bibfield  {author} {\bibinfo {author} {\bibfnamefont {D.}~\bibnamefont
  {Lusseau}}, \bibinfo {author} {\bibfnamefont {K.}~\bibnamefont {Schneider}},
  \bibinfo {author} {\bibfnamefont {O.~J.}\ \bibnamefont {Boisseau}}, \bibinfo
  {author} {\bibfnamefont {P.}~\bibnamefont {Haase}}, \bibinfo {author}
  {\bibfnamefont {E.}~\bibnamefont {Slooten}},\ and\ \bibinfo {author}
  {\bibfnamefont {S.~M.}\ \bibnamefont {Dawson}},\ }\bibfield  {title}
  {\bibinfo {title} {The bottlenose dolphin community of {D}oubtful {S}ound
  features a large proportion of long-lasting associations},\ }\href
  {https://doi.org/10.1007/s00265-003-0651-y} {\bibfield  {journal} {\bibinfo
  {journal} {Behavioral Ecol. Sociobiol.}\ }\textbf {\bibinfo {volume} {54}},\
  \bibinfo {pages} {396} (\bibinfo {year} {2003})}\BibitemShut {NoStop}%
\bibitem [{DNO()}]{DNOTE}%
  \BibitemOpen
  \href@noop {} {\ }\bibinfo {note} {Data with 62 individuals, as entered by
  M.\ Newman, were obtained from
  https://web.archive.org/web/20170822170651/http://networkdata.ics.uci.edu/data/dolphins/dolphins.gml}\BibitemShut
  {NoStop}%
\bibitem [{\citenamefont {Dale}\ \emph {et~al.}(2009)\citenamefont {Dale},
  \citenamefont {Alquthami}, \citenamefont {Baldwin}, \citenamefont {Faruque},
  \citenamefont {Langston}, \citenamefont {McLaren}, \citenamefont {Meeker},
  \citenamefont {Steurer},\ and\ \citenamefont {Schoder}}]{FLmap}%
  \BibitemOpen
  \bibfield  {author} {\bibinfo {author} {\bibfnamefont {S.}~\bibnamefont
  {Dale}}, \bibinfo {author} {\bibfnamefont {T.}~\bibnamefont {Alquthami}},
  \bibinfo {author} {\bibfnamefont {T.}~\bibnamefont {Baldwin}}, \bibinfo
  {author} {\bibfnamefont {O.}~\bibnamefont {Faruque}}, \bibinfo {author}
  {\bibfnamefont {J.}~\bibnamefont {Langston}}, \bibinfo {author}
  {\bibfnamefont {P.}~\bibnamefont {McLaren}}, \bibinfo {author} {\bibfnamefont
  {R.}~\bibnamefont {Meeker}}, \bibinfo {author} {\bibfnamefont
  {M.}~\bibnamefont {Steurer}},\ and\ \bibinfo {author} {\bibfnamefont
  {K.}~\bibnamefont {Schoder}},\ }\href@noop {} {\emph {\bibinfo {title}
  {Progress Report for the Institute for Energy Systems, Economics and
  Sustainability and the Florida Energy Systems Consortium}}},\ \bibinfo {type}
  {Tech. Rep.}\ (\bibinfo  {institution} {Center for Advanced Power Systems,
  Florida State University},\ \bibinfo {address} {Tallahassee, FL},\ \bibinfo
  {year} {2009})\BibitemShut {NoStop}%
\bibitem [{\citenamefont {Abou~Hamad}\ \emph {et~al.}(2010)\citenamefont
  {Abou~Hamad}, \citenamefont {Israels}, \citenamefont {Rikvold},\ and\
  \citenamefont {Poroseva}}]{ABOU10B}%
  \BibitemOpen
  \bibfield  {author} {\bibinfo {author} {\bibfnamefont {I.}~\bibnamefont
  {Abou~Hamad}}, \bibinfo {author} {\bibfnamefont {B.}~\bibnamefont {Israels}},
  \bibinfo {author} {\bibfnamefont {P.~A.}\ \bibnamefont {Rikvold}},\ and\
  \bibinfo {author} {\bibfnamefont {S.~V.}\ \bibnamefont {Poroseva}},\
  }\bibfield  {title} {\bibinfo {title} {Spectral matrix methods for
  partitioning power grids: Applications to the {I}talian and {F}loridian
  high-voltage networks},\ }\href@noop {} {\bibfield  {journal} {\bibinfo
  {journal} {Phys. Proc.}\ }\textbf {\bibinfo {volume} {4}},\ \bibinfo {pages}
  {125} (\bibinfo {year} {2010})}\BibitemShut {NoStop}%
\bibitem [{\citenamefont {Abou~Hamad}\ \emph {et~al.}(2011)\citenamefont
  {Abou~Hamad}, \citenamefont {Rikvold},\ and\ \citenamefont
  {Poroseva}}]{ABOU11}%
  \BibitemOpen
  \bibfield  {author} {\bibinfo {author} {\bibfnamefont {I.}~\bibnamefont
  {Abou~Hamad}}, \bibinfo {author} {\bibfnamefont {P.~A.}\ \bibnamefont
  {Rikvold}},\ and\ \bibinfo {author} {\bibfnamefont {S.~V.}\ \bibnamefont
  {Poroseva}},\ }\bibfield  {title} {\bibinfo {title} {Floridian high-voltage
  power-grid network partitioning and cluster optimization using simulated
  annealing.},\ }\href@noop {} {\bibfield  {journal} {\bibinfo  {journal}
  {Phys. Proc.}\ }\textbf {\bibinfo {volume} {15}},\ \bibinfo {pages} {2}
  (\bibinfo {year} {2011})}\BibitemShut {NoStop}%
\bibitem [{\citenamefont {Rikvold}\ \emph {et~al.}(2012)\citenamefont
  {Rikvold}, \citenamefont {{Abou Hamad}}, \citenamefont {Israels},\ and\
  \citenamefont {Poroseva}}]{Rikvold}%
  \BibitemOpen
  \bibfield  {author} {\bibinfo {author} {\bibfnamefont {P.~A.}\ \bibnamefont
  {Rikvold}}, \bibinfo {author} {\bibfnamefont {I.}~\bibnamefont {{Abou
  Hamad}}}, \bibinfo {author} {\bibfnamefont {B.}~\bibnamefont {Israels}},\
  and\ \bibinfo {author} {\bibfnamefont {S.~V.}\ \bibnamefont {Poroseva}},\
  }\bibfield  {title} {\bibinfo {title} {Modeling power grids},\ }\href@noop {}
  {\bibfield  {journal} {\bibinfo  {journal} {Phys. Proc.}\ }\textbf {\bibinfo
  {volume} {34}},\ \bibinfo {pages} {119} (\bibinfo {year} {2012})}\BibitemShut
  {NoStop}%
\bibitem [{\citenamefont {Xu}\ \emph {et~al.}(2014)\citenamefont {Xu},
  \citenamefont {Gurfinkel},\ and\ \citenamefont {Rikvold}}]{XU14}%
  \BibitemOpen
  \bibfield  {author} {\bibinfo {author} {\bibfnamefont {Y.}~\bibnamefont
  {Xu}}, \bibinfo {author} {\bibfnamefont {A.~J.}\ \bibnamefont {Gurfinkel}},\
  and\ \bibinfo {author} {\bibfnamefont {P.~A.}\ \bibnamefont {Rikvold}},\
  }\bibfield  {title} {\bibinfo {title} {Architecture of the {F}lorida power
  grid as a complex network},\ }\href
  {https://doi.org/http://dx.doi.org/10.1016/j.physa.2014.01.035} {\bibfield
  {journal} {\bibinfo  {journal} {Physica A}\ }\textbf {\bibinfo {volume}
  {401}},\ \bibinfo {pages} {130} (\bibinfo {year} {2014})}\BibitemShut
  {NoStop}%
\bibitem [{\citenamefont {Brodin}\ \emph {et~al.}(2025)\citenamefont {Brodin},
  \citenamefont {Pierce}, \citenamefont {Reis}, \citenamefont {Rikvold},
  \citenamefont {Moura}, \citenamefont {Jankov},\ and\ \citenamefont
  {M{\aa}l{\o}y}}]{BROD25}%
  \BibitemOpen
  \bibfield  {author} {\bibinfo {author} {\bibfnamefont {J.~F.}\ \bibnamefont
  {Brodin}}, \bibinfo {author} {\bibfnamefont {K.}~\bibnamefont {Pierce}},
  \bibinfo {author} {\bibfnamefont {P.}~\bibnamefont {Reis}}, \bibinfo {author}
  {\bibfnamefont {P.~A.}\ \bibnamefont {Rikvold}}, \bibinfo {author}
  {\bibfnamefont {M.}~\bibnamefont {Moura}}, \bibinfo {author} {\bibfnamefont
  {M.}~\bibnamefont {Jankov}},\ and\ \bibinfo {author} {\bibfnamefont {K.~J.}\
  \bibnamefont {M{\aa}l{\o}y}},\ }\bibfield  {title} {\bibinfo {title}
  {Interface instability of two-phase flow in a three-dimensional porous
  medium},\ }\href {https://doi.org/https://doi.org/10.1103/hm82-167x}
  {\bibfield  {journal} {\bibinfo  {journal} {Phys. Rev. Fluids}\ }\textbf
  {\bibinfo {volume} {10}},\ \bibinfo {pages} {064003} (\bibinfo {year}
  {2025})}\BibitemShut {NoStop}%
\bibitem [{\citenamefont {Gostick}(2017)}]{GOST17}%
  \BibitemOpen
  \bibfield  {author} {\bibinfo {author} {\bibfnamefont {J.~T.}\ \bibnamefont
  {Gostick}},\ }\bibfield  {title} {\bibinfo {title} {Versatile and efficient
  pore network extraction method using marker-based watershed segmentation},\
  }\href {https://doi.org/https://doi.org/10.1103/PhysRevE.96.023307}
  {\bibfield  {journal} {\bibinfo  {journal} {Phys. Rev. E}\ }\textbf {\bibinfo
  {volume} {96}},\ \bibinfo {pages} {023307} (\bibinfo {year}
  {2017})}\BibitemShut {NoStop}%
\bibitem [{\citenamefont {Gostick}\ \emph {et~al.}(2019)\citenamefont
  {Gostick}, \citenamefont {Khan}, \citenamefont {Tranter}, \citenamefont
  {Kok}, \citenamefont {Agnaou}, \citenamefont {Sadeghi},\ and\ \citenamefont
  {Jervis}}]{GOST19}%
  \BibitemOpen
  \bibfield  {author} {\bibinfo {author} {\bibfnamefont {J.~T.}\ \bibnamefont
  {Gostick}}, \bibinfo {author} {\bibfnamefont {Z.~A.}\ \bibnamefont {Khan}},
  \bibinfo {author} {\bibfnamefont {T.~G.}\ \bibnamefont {Tranter}}, \bibinfo
  {author} {\bibfnamefont {M.~D.~R.}\ \bibnamefont {Kok}}, \bibinfo {author}
  {\bibfnamefont {M.}~\bibnamefont {Agnaou}}, \bibinfo {author} {\bibfnamefont
  {M.}~\bibnamefont {Sadeghi}},\ and\ \bibinfo {author} {\bibfnamefont
  {R.}~\bibnamefont {Jervis}},\ }\bibfield  {title} {\bibinfo {title}
  {Pore{S}py: A {P}ython toolkit for quantitative analysis of porous media
  images},\ }\href {https://doi.org/https://doi.org/10.21105/joss.01296}
  {\bibfield  {journal} {\bibinfo  {journal} {J. Open Source Software}\
  }\textbf {\bibinfo {volume} {4}},\ \bibinfo {pages} {1296} (\bibinfo {year}
  {2019})}\BibitemShut {NoStop}%
\end{thebibliography}
%


\end{document}